\documentclass[journal]{IEEEtran}

\ifCLASSINFOpdf
\else
\fi

\usepackage[compact]{titlesec}
\usepackage{subfig}
\usepackage{graphicx} 
\usepackage[noend]{algorithm2e}
\usepackage{algorithmic}
\usepackage{setspace}
\usepackage{url}
\usepackage{amsmath,amsthm}
\usepackage{amssymb,wasysym}
\usepackage{mathrsfs}
\usepackage{cite}
\usepackage{xcolor}
\usepackage{wrapfig}
\usepackage{array}
\usepackage{tabularray}
\usepackage{verbatim}
\usepackage{dsfont}
\usepackage{multirow}
\usepackage{breqn,xspace}
\usepackage[T1]{fontenc}
\usepackage{cmap}
\usepackage{amsmath} 
\usepackage{amssymb} 
\usepackage{amsfonts} 
\usepackage{wasysym}
\usepackage{pifont}        
\usepackage[table]{xcolor} 
\usepackage{hhline}        

\usepackage{multirow}
\usepackage{booktabs}
\usepackage{graphicx}
\usepackage[compact]{titlesec}
\newcounter{workflowstep}

\ifodd 1

\else

\fi

\begin{document}

\title{Satellite Intelligence: Distributed Training, Collaborative Inference, and LLM Adaptation in LEO Constellations}

\author{Zheng Lin, Zhanwei Wang, Yuxin Zhang, Yafei Wang, Ons Aouedi, Zihan Fang, Jinfeng Xu, Jing Yang, Vu Nguyen Ha, Yuanwei Liu,~\IEEEmembership{Fellow,~IEEE}, Mohamed-Slim Alouini,~\IEEEmembership{Fellow,~IEEE}, Symeon Chatzinotas,~\IEEEmembership{Fellow,~IEEE}


\thanks{Z. Lin, O. Aouedi, V. Ha, and S. Chatzinotas are with the Interdisciplinary Centre for Security, Reliability and Trust (SnT), University of Luxembourg, L-1855 Luxembourg, Luxembourg (e-mail: zhenglin@ieee.org; ons.aouedi@uni.lu; vu-nguyen.ha@uni.lu; symeon.chatzinotas@uni.lu).}
\thanks{Z. Wang and Y. Liu are with the Department of Computer and Electronic Engineering, University of Hong Kong, Pok Fu Lam, Hong Kong, China (e-mail: zhanwei.wang@hku.hk; yuanwei@hku.hk).}
\thanks{Y. Zhang is with the Institute of Space
Internet, Fudan University, Shanghai, China (e-mail: yuxinzhang22@m.fudan.edu.cn).}
\thanks{Y. Wang is with the National Mobile
Communications Research Laboratory, Southeast University, Nanjing 210096, China, and also with Purple Mountain Laboratories, Nanjing 211100, China (e-mail: wangyf@seu.edu.cn).}
\thanks{Z. Fang is with the Department of Computer Science, City University of Hong Kong, Kowloon, Hong Kong SAR, China (e-mail: zihanfang3-c@my.cityu.edu.hk).}
\thanks{J. Xu is with the Department of Electrical and Computer Engineering, The University of British Columbia, Vancouver, Canada (e-mail: jinfeng.xu@ubc.ca).}
\thanks{J. Yang is with the Center of Research for Cyber Security and Network (CSNET), Faculty of Computer Science and Information Technology, Universiti Malaya, 50603 Kuala Lumpur, Malaysia (e-mail: s2147529@siswa.um.edu.my).}
\thanks{M. S. Alouini is with CEMSE Division, King Abdullah University of
Science and Technology (KAUST), Thuwal, 239556900, Kingdom of Saudi
Arabia (KSA) (e-mail: slim.alouini@kaust.edu.sa).}
}




\maketitle

\begin{abstract}
The proliferation of orbital sensing data and the real-time demands of emerging global services are pushing low Earth orbit (LEO) satellite networks from bent-pipe relays toward intelligent onboard processing. This shift confronts a fundamental tension, where modern artificial intelligence (AI) training paradigms require intensive computation, abundant memory, and stable connectivity, all of which are scarce under the dynamic topologies, intermittent links, and severe size, weight, and power (SWaP) constraints of satellite networks. Addressing this tension demands rethinking AI workflows, from training and inference to the emerging deployment of large language models (LLMs), under a constraint-aware design philosophy. To the best of our knowledge, this paper provides the first survey that jointly covers all three, whereas prior surveys examine these aspects in isolation. We begin by identifying the deployment challenge of satellite intelligence (SI) and deriving design principles for SI. We then examine three advancing paradigms: i) distributed training, where federated learning (FL) and split learning (SL) enable satellite-ground co-training under resource limitations; ii) collaborative inference, which exploits single-satellite, multi-satellite, and satellite-ground architectures to meet real-time service demands; and iii) on-orbit LLM adaptation and serving, which tackles the frontier challenge of deploying LLMs under severe orbital constraints and resource-aware serving. Finally, we identify open challenges and future directions. By bridging the AI and satellite networks, this paper provides a research roadmap for the implementation of space-native AI technologies.
\end{abstract}


\IEEEpeerreviewmaketitle


\section{Introduction}
\label{sec:introduction}

\begin{table*}[t]
\centering
\caption{Comparison of related surveys and this survey, where \CIRCLE\ fully addressed, \LEFTcircle\ partially covered, \Circle\ not addressed.}
\label{tab:related}
\small

\resizebox{0.86\textwidth}{!}{%
\begin{tblr}{
  width=\textwidth,
  colspec={X[5,c,m] X[7,c,m] X[44,l,m] X[11,c,m] X[11,c,m] X[12,c,m] X[10,c,m]},
  colsep=3pt, rowsep=2.2pt, stretch=1.2,
  row{1}={font=\bfseries, bg=black!8, halign=c, m},
  column{1}={font=\bfseries}, 
  row{15}={bg=blue!10}, 
  hline{1,Z}={0.8pt},
  vline{1,Z}={0.8pt},
  vline{2-7}={0.4pt},
  hline{2}={0.6pt},
  hline{3-5,7,9,14,15}={wd=0.3pt, fg=black}, 
  hline{6,8,10,11,12,13}={2-Z}{wd=0.3pt, fg=black}, 
}
Year & Ref. & Introduction & {Orbital Constraint\\Analysis} & {Satellite Distributed\\Training} & {Satellite Collaborative\\Inference} & {On-Orbit LLM\\Adaptation} \\
2020 & \cite{kodheli2020satellite} & Present an overview of satellite communication systems in the New Space era, and discuss their applications, enabling technologies, and future challenges & \CIRCLE & \Circle & \Circle & \Circle \\
2021 & \cite{fourati2021artificial} & Introduce major AI techniques for satellite communication systems, and discuss their applications in satellite network optimization and management & \CIRCLE & \Circle & \Circle & \Circle \\
2022 & \cite{azari2022evolution} & Present the evolution of non-terrestrial networks from 5G to 6G, and discuss their architectures, enabling technologies, potential applications, and open challenges & \CIRCLE & \Circle & \Circle & \Circle \\
\SetCell[r=2]{c,m} 2023
& \cite{shen2023survey} & Provide a comprehensive overview of next-generation computing technologies for space-air-ground integrated networks, including their potential applications and open research challenges & \CIRCLE & \LEFTcircle & \Circle & \Circle \\
& \cite{yue2023low} & Investigate security and reliability issues in LEO satellite communication systems, and discuss corresponding solutions, design guidelines, and future research directions & \CIRCLE & \Circle & \Circle & \Circle \\
\SetCell[r=2]{c,m} 2024
& \cite{mahboob2024revolutionizing} & Review AI-empowered satellite-based non-terrestrial networks for 6G and discuss enabling technologies and research challenges & \CIRCLE & \LEFTcircle & \Circle & \Circle \\
& \cite{bakambekova2024interplay} & Examine AI-enabled communication and network optimization across space-air-ground segments, including edge intelligence and federated learning & \LEFTcircle & \LEFTcircle & \Circle & \Circle \\
\SetCell[r=5]{c,m} 2025
& \cite{fontanesi2025artificial} & Survey AI/ML applications and implementation challenges across satellite communication systems & \LEFTcircle & \LEFTcircle & \Circle & \Circle \\
& \cite{ntontin2025vision} & Present a vision and roadmap for space communications in the 6G-and-beyond era, covering use cases, system architectures, and enabling technologies & \LEFTcircle & \Circle & \Circle & \Circle \\
& \cite{chen2025survey} & Survey resource management models and optimization methods for communication- and computing-integrated SAGINs & \CIRCLE & \Circle & \LEFTcircle & \Circle \\
& \cite{shi2025satellite} & Review architectures and enabling technologies for deploying large AI models in satellite edge networks & \CIRCLE & \LEFTcircle & \LEFTcircle & \LEFTcircle \\
& \cite{yin2025comprehensive} & Summarize applications and algorithms for orbital edge computing, including computation and resource management & \LEFTcircle & \LEFTcircle & \LEFTcircle & \Circle \\
2026 & \cite{guimaraes2026survey} & Discuss satellite computing from networking, computing, and application perspectives & \LEFTcircle & \Circle & \LEFTcircle & \Circle \\
& Ours & Survey constraint-aware satellite intelligence in LEO constellations, covering distributed training, collaborative inference, and on-orbit LLM adaptation and serving & \CIRCLE & \CIRCLE & \CIRCLE & \CIRCLE \\
\end{tblr}%
}
\end{table*}

\subsection{Background}

The proliferation of low Earth orbit (LEO) mega-constellations, combined with advances in high-throughput satellite technologies and multi-modal onboard sensors, is fundamentally reshaping non-terrestrial networks~\cite{mao2024intelligent, huang2025robust, chen2024shortest, wang2024ultra, du2025multi,zhang2025satfed,luo2026high}. Modern satellites continuously acquire massive volumes of remote sensing imagery and positioning data~\cite{zhou2024mobility, tang2024joint, fan2025satellite, zhai2024seco, zhu2023collaborative}, demanding intelligent and timely onboard processing unseen in earlier satellite generations. This demand has catalyzed growing interest in deploying artificial intelligence (AI) in space-borne systems~\cite{zhang2024generative, asheralieva2023efficient, bhandari2024user, zhang2024resource, hassan2024spaceris,teng2026communication,yuan2025satsense,cao2026robust,zhang2026towards}. However, the prevailing terrestrial-centric processing paradigm, where raw data is downlinked to ground stations (GSs) for centralized AI processing, faces significant barriers. 
First, satellite-ground link (SGL) spectrum is shared among mega-constellations and other satellite services, making it a key bottleneck~\cite{al2022survey,kim2024feasibility}. Downlinking raw high-fidelity sensor streams over these links therefore imposes a prohibitive bandwidth burden, even though such measurements, unlike ground-originated traffic, are compressible~\cite{ma2023network, cao2023satcp, chen2025game, lin2025fedsn}. Second, LEO satellites can access a given GS only during short and intermittent contact windows. Mitigating the resulting latency requires a geographically dense ground segment, whose deployment cost is prohibitive for small operators and compels them to rely on third-party ground-station-as-a-service providers~\cite{vasisht2021l2d2}. Together, these two barriers render terrestrial-centric processing incompatible with satellite applications requiring real-time decision-making, such as disaster response and planetary monitoring~\cite{zhai2023fedleo, elmahallawy2024communication, zhao2025alanine, zhang2025ground}.

\subsection{Motivation: From Terrestrial-Centric Processing to Satellite Intelligence}

To combat these challenges, satellite intelligence (SI) has emerged as a transformative paradigm, shifting the computational locus from GSs to the individual satellites~\cite{zhang2024energy, xia2024efficient, yao2025leoedge, xing2024deciphering}. Rather than transmitting raw data, SI enables onboard feature extraction and real-time analysis, thereby shrinking the downlink transmission overhead required for both model training and inference. The recent advances of commercial off-the-shelf (COTS) AI accelerators, such as embedded graphics processing units (GPUs) and vision processing units, whose nominal throughput reaches the tera-operations-per-second level, demonstrate that the hardware foundation for on-orbit AI is beginning to materialize~\cite{xing2024deciphering,giuffrida2022phisat1}. However, realizing SI still remains severely bottlenecked. Unlike terrestrial data centers, LEO satellites operate under stringent size, weight, and power (SWaP) constraints that cap individual computing and energy budgets to a small fraction of their ground counterparts~\cite{zhou2024latency, luo2024energy, hu2025joint, huang2025utility, dai2025energy}. Concurrently, orbital dynamics induce rapidly evolving network topologies and frequent link handovers, which may be partially mitigated by extensive GS, albeit at a cost prohibitive for most operators~\cite{vasisht2021l2d2}.
These constraints give rise to a core tension, where contemporary AI models demand intensive computation and stable, high-bandwidth connectivity, but the orbital environment available to most operators can rarely provide either. Such severe physical and network limitations preclude the direct transplantation of resource-intensive terrestrial AI workflows, spanning both training and inference, into orbit~\cite{sun2024efficient, zhou2024decomposition, yang2024communication, wu2024accelerating, tang2024opportunistic}.
Since relying on an isolated satellite is insufficient to handle the explosive computing demands of modern AI tasks, there is a critical need to design an efficient and collaborative SI framework.

\begin{figure*}[t!]
\centering
\includegraphics[width=0.83\textwidth]{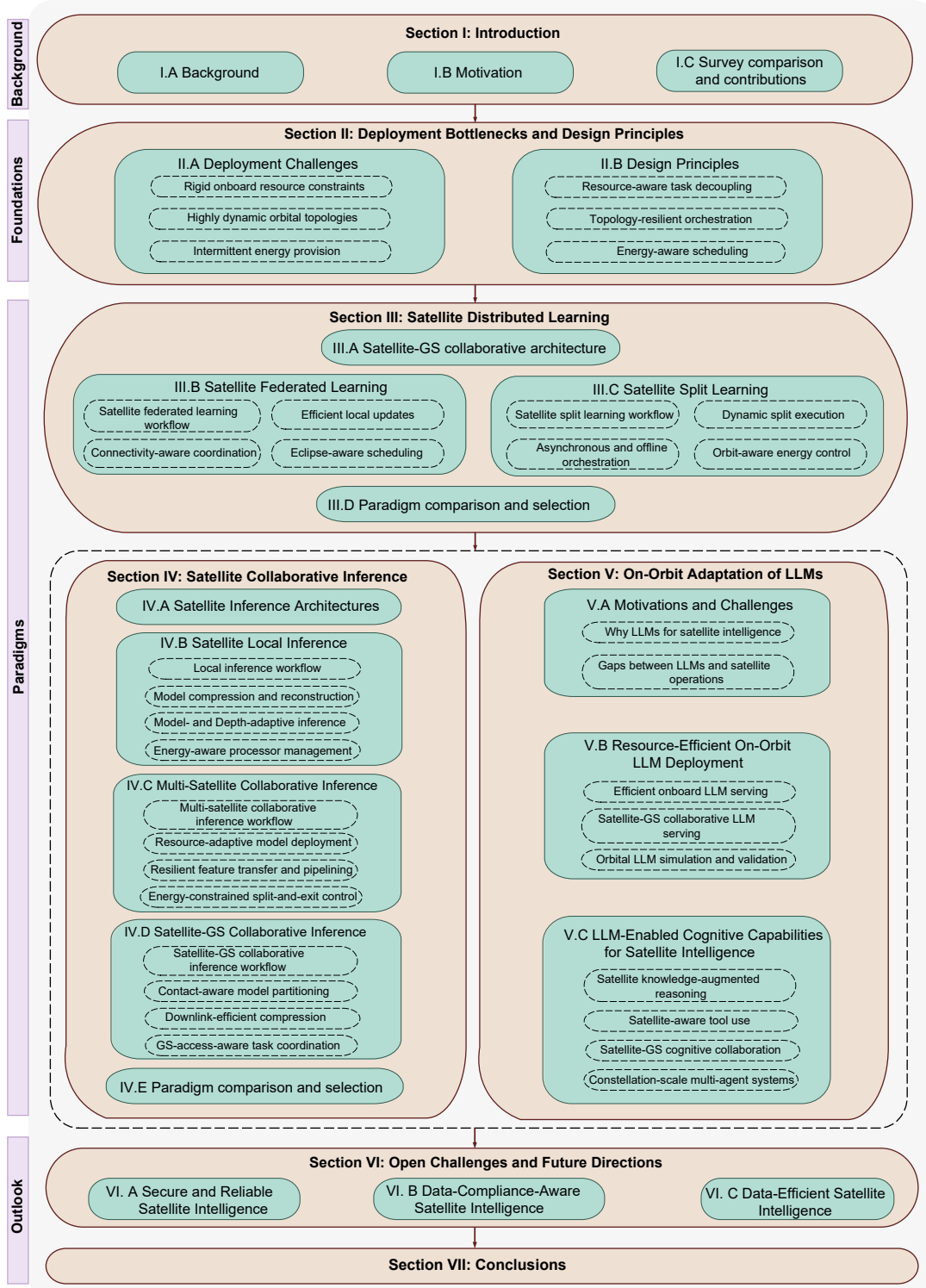}
\vspace{-0.5cm}
\centering\caption{Organization of this survey.}
\label{fig_survey_outiline}
\end{figure*}

\subsection{Comparisons with Prior Surveys and Our Contributions}

While existing surveys have explored various facets of satellite-AI integration, they exhibit critical limitations across four key dimensions. First, in terms of orbital constraints, several works~\cite{bakambekova2024interplay,fontanesi2025artificial,ntontin2025vision,yin2025comprehensive,guimaraes2026survey} only partially account for the rigid SWaP budgets, highly dynamic topologies, and intermittent energy provision of LEO satellites, treating them as peripheral considerations rather than first-class drivers of AI system design. Second, regarding distributed training, many surveys~\cite{kodheli2020satellite,fourati2021artificial,azari2022evolution,yue2023low,ntontin2025vision,chen2025survey,guimaraes2026survey} do not address on-orbit model training at all. Some works~\cite{shen2023survey,mahboob2024revolutionizing,bakambekova2024interplay,fontanesi2025artificial,shi2025satellite,yin2025comprehensive} mainly adopt generic federated learning frameworks or examine isolated architectures, without systematically tailoring federated and split learning to LEO-specific resource limits, contact-window-driven connectivity, and sunlit-eclipse energy cycles. Third, satellite collaborative inference is absent from most existing surveys~\cite{kodheli2020satellite,fourati2021artificial,azari2022evolution,shen2023survey,yue2023low,mahboob2024revolutionizing,bakambekova2024interplay,fontanesi2025artificial,ntontin2025vision}. The few that discuss it~\cite{chen2025survey,shi2025satellite,yin2025comprehensive,guimaraes2026survey} largely frame inference as a task offloading or resource allocation problem, or focus on a single execution architecture. They lack a unified treatment of local, multi-satellite, and satellite-GS inference under energy periodicity and stringent onboard constraints. Finally, on-orbit large language model (LLM) adaptation remains largely unexplored. Only~\cite{shi2025satellite} considers large AI models in satellite edge networks, and it focuses on vision foundation models rather than LLMs. Critical challenges, such as memory-constrained autoregressive serving, KV-cache management under fixed onboard budgets, satellite-GS collaborative LLM serving, and the integration of LLM-enabled reasoning and tool use into satellite operations, have not been systematically investigated. As summarized in Table~\ref{tab:related}, no existing survey provides comprehensive coverage across these four dimensions. In contrast, this survey adopts a constraint-aware perspective that investigates distributed training, collaborative inference, and on-orbit LLM adaptation, thereby offering a unified roadmap toward space-native AI.

To bridge these gaps, this article presents a comprehensive survey on the integration of AI and LEO satellite networks, organized along a principled progression: from the physical constraints that govern on-satellite processing, through distributed training and collaborative inference, to the emerging frontier of on-orbit LLM adaptation. We begin by analyzing the challenge of deploying SI, including rigid onboard resource constraints, highly dynamic orbital topologies, and intermittent energy provision, and present the constraint-driven design principles for SI. After that, we examine how distributed learning frameworks, federated learning (FL)~\cite{mcmahan2017communication,fang2026aggregation} and split learning (SL)~\cite{thapa2022splitfed,vepakomma2018split,lin2024split}, enable satellite-ground co-training under resource limitations. We then turn to collaborative inference across single-satellite, multi-satellite, and satellite-ground architectures for meeting real-time service demands. We further investigate resource-efficient on-orbit LLM deployment, as well as how onboard LLMs can support advanced capabilities for SI. Finally, we identify open challenges and future directions. The main contributions of this paper are summarized as follows:

\begin{itemize} 
\item We delineate the bottlenecks that arise from deploying AI models within LEO constellations and present the constraint-driven design principles for SI architectures.
\item We investigate the potential of distributed training and collaborative inference in satellite networks, tailoring strategies to the resource, connectivity, and energy constraints of orbital environments.
\item We explore resource-efficient deployment of onboard LLMs under severe orbital constraints, and the advanced LLM-enabled capabilities for SI. 
\item We identify several critical open challenges and future research directions, providing a foundational blueprint for the evolution of SI.
\end{itemize}

\begin{figure}[t!]
\centering
\includegraphics[width=0.5\textwidth]{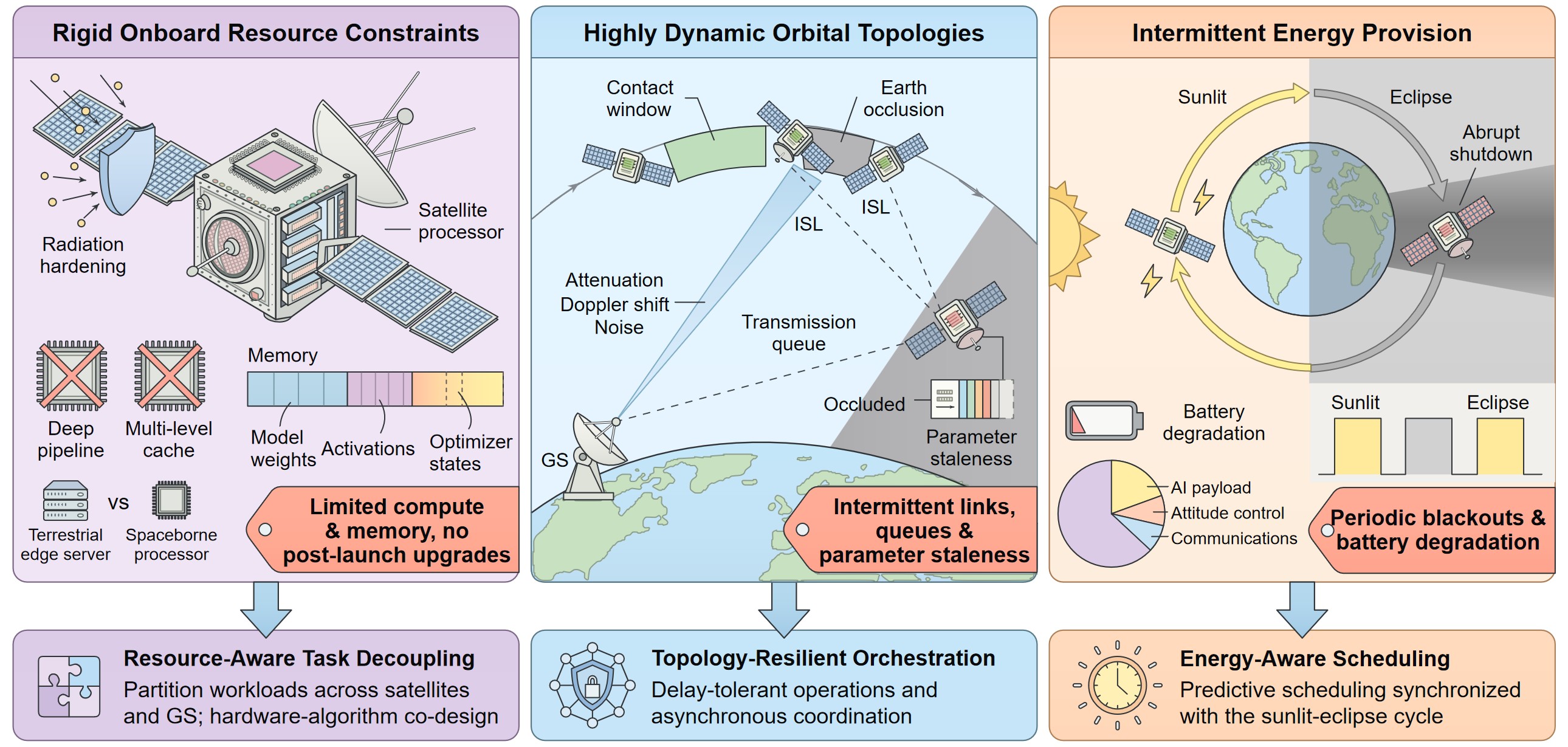}
\centering\caption{Deployment challenges of SI and the corresponding design principles.}
\label{fig_SL_challenge_priciple}
\end{figure}

As outlined in Fig.~\ref{fig_survey_outiline}, the survey is organized as follows. Section~\ref{sec:characteristics_constraints} presents the deployment challenges of SI and analyzes its design principle. Sections~\ref{sec:distributed_training} and~\ref{sec:collaborative_inference} investigate distributed training and collaborative inference paradigms tailored for satellite networks. Section~\ref{sec:LLM_adaptation} delves into resource-efficient adaptation and serving for onboard LLMs, and Section~\ref{sec:challenges} outlines critical open challenges and future research directions. Finally, conclusions are presented in Section~\ref{sec:conclusion}.

\section{Deployment Challenges and Design Principles}
\label{sec:characteristics_constraints}

Unlike terrestrial edge intelligence, built on static infrastructure with abundant computing resources, stable network topology, and continuous energy supply, SI must operate under rigid onboard SWaP budgets, highly dynamic orbital topologies, and intermittent energy provision. These three constraints jointly reshape how AI workloads can be designed, scheduled, and executed in orbit. In this section, we first delineate these deployment challenges and then present the constraint-driven design principles for SI architectures, as shown in Fig.~\ref{fig_SL_challenge_priciple}.

\subsection{Deployment Challenges of SI}\label{subsec:SI_constraints}

{\textit{Rigid Onboard Resource Constraints.}} The implementation of SI faces a mismatch between the computing demands of modern AI and the stringent physical constraints of satellites~\cite{xing2024deciphering,denby2020orbital,giuffrida2022phisat1}. Constrained by radiation hardening and power budgets, spaceborne processors typically deliver computing capabilities ranging from sub-giga floating-point operations per second (GFLOPS) on radiation-hardened processors to a few tera-FLOPS (TFLOPS) on commercial off-the-shelf (COTS) GPU-class accelerators, up to five orders of magnitude lower than terrestrial edge servers~\cite{liu2025computing,wang2023satellite}. Radiation-hardened architectures further omit advanced microarchitectural features, such as deep pipelines and multi-level caches, to ensure reliability~\cite{wang2026radshield}. This omission degrades arithmetic logic unit utilization and induces severe memory access bottlenecks. Alternatively, COTS processors can rely on redundant or resilient designs, which in turn incur considerable computation and power overhead. Under either approach, billion-parameter models can easily exhaust the limited onboard memory (from about 1 GB on CubeSat-class platforms~\cite{marszk2020ops} to 8–32 GB on GPU-equipped payloads), leaving negligible headroom for activation and optimizer states. Moreover, satellite resource budgets are fixed at launch and immutable throughout the operational lifespan~\cite{liu2024inorbit}, whereas ground infrastructure can be upgraded far more cheaply and frequently to keep pace with rapidly evolving AI hardware. Consequently, any mismatch between onboard AI demands and available compute or memory cannot be resolved through post-launch upgrades, necessitating the design of resource-aware SI architectures.

{\textit{Highly Dynamic Orbital Topologies.}} Realizing SI demands the timely exchange of substantial data volumes, including high-dimensional feature maps (several kilobytes (KB) to several megabytes (MB) per layer) and model parameters (hundreds of MB to several gigabytes (GB)). However, the highly dynamic orbital topology hinders this exchange~\cite{cao2023satcp}. As LEO satellites traverse at velocities approaching 7.5 km/s, Earth occlusion restricts SGL availability with a given GS to only a few contact windows per day, each lasting up to about 10 minutes~\cite{denby2020orbital,ma2023network,tao2024serval}. Even within these narrow visibility windows, communication quality is further degraded by long-distance propagation attenuation exceeding 200 decibels (dB), extreme Doppler shifts of up to 48 kilohertz (kHz)~\cite{azari2022evolution}, and cosmic and ionospheric noise, which leads to bit error rates several orders of magnitude higher than typical of terrestrial wired networks~\cite{saeed2020cubesat,saradhi2009pliar}. Compounded by onboard power constraints and limited antenna gains, these channel degradations typically confine radio-frequency SGLs on small satellites to 30–100 Mbps~\cite{zhang2024resource,xing2024deciphering}. Higher throughput is attainable, e.g., via gateways or optical links, but incurs substantial cost or new challenges such as cloud blockage~\cite{kaushal2016optical}, so SGL capacity is easily saturated by the transmission of intermediate features or model parameters. Consequently, the mismatch between massive AI data exchanges and intermittent connectivity gives rise to persistent transmission queues and parameter staleness that must be explicitly accounted for in SI design.

{\textit{Intermittent Energy Provision.}} LEO satellites rely exclusively on solar energy harvesting, supplemented by battery storage during eclipse periods~\cite{xing2024deciphering,cheng2024energy}. Typically, a small LEO satellite (e.g., a microsatellite of tens of kilograms) allocates a mere 10–50 W to its computing payload from a total platform power budget that rarely exceeds 200 W, a strict envelope shared with mission-critical system components such as attitude determination and communications. This energy profile is also inherently intermittent, where the approximately 30-minute eclipse within a typical 90-minute orbit completely suspends solar generation, compelling all active computing tasks to draw from finite battery reserves~\cite{abdi2013emb}. Since the dedicated battery capacity for AI workloads is severely limited (typically 10–100 Wh), continuous onboard AI execution is unsustainable. Sustaining compute-intensive workloads throughout each eclipse rapidly depletes this reserve, and prolonged power overdraw not only risks forcing abrupt shutdowns but also accelerates irreversible battery degradation. This degradation compounds over the satellite's operational lifespan, with cumulative charge-discharge cycling causing up to 25\% capacity loss within about six months of LEO cycling~\cite{cook2021impact}, which progressively tightens an already narrow energy budget. Such failures interrupt long-running iterative processes, invalidate accumulated intermediate states, and create periodic service blackouts~\cite{wang2026temperature}, ultimately rendering onboard AI systems incompatible with the continuity that time-sensitive decision-making requires.

\subsection{Design Principles for SI}\label{subsec:design_principle}

Beyond the above three bottlenecks, satellite platforms face other engineering challenges, such as thermal management and spacecraft-level system integration, which lie outside the scope of this survey and are discussed only where they directly affect AI execution. To overcome the fundamental bottlenecks outlined in Section~\ref{subsec:SI_constraints}, the deployment of SI cannot rely on a direct transplantation of terrestrial AI frameworks. Instead, it necessitates a constraint-aware design governed by three core principles:

{\textit{Resource-Aware Task Decoupling.}} Given the rigid SWaP constraints that severely cap individual satellite capabilities, SI architectures must transition from monolithic, compute-intensive execution to collaborative training paradigms~\cite{zhang2024leader}. Complex AI workloads should be partitioned or flexibly scheduled across inter-satellite and satellite-GS resources. Furthermore, hardware-algorithm co-design is imperative to tailor AI models to the tight computing and memory limits of satellite payloads, thereby maximizing onboard processing efficiency without exhausting individual satellite capacities.

{\textit{Topology-Resilient Orchestration.}} To counteract highly dynamic orbital topologies and intermittent connectivity, SI systems must be intrinsically resilient to network disruptions, rather than relying on the continuous and synchronous data exchanges characteristic of terrestrial networks~\cite{jiang2026csafl}. This requires the SI architecture to support delay-tolerant operations and asynchronous coordination among satellites at the algorithmic level, while the orchestration layer must dynamically adapt to frequent link handovers and transient satellite outages at the system level. Together, these mechanisms sustain AI workflows despite the time-varying topology of LEO constellations.

{\textit{Energy-Aware Scheduling.}} Since LEO satellites rely exclusively on cyclical solar energy harvesting and finite battery storage, a fixed execution cannot sustain continuous AI operation. SI workflows therefore demand a dynamic and energy-driven orchestration synchronized with the satellite's sunlit-eclipse cycle, customizing computational intensity based on instantaneous energy availability to prevent critical battery depletion~\cite{li2024battery}. Beyond reacting to the current energy state, such orchestration should be inherently predictive, exploiting the predictability of orbital dynamics to anticipate upcoming sunlit-eclipse transitions, thereby sustaining service continuity across the full orbital cycle.

\begin{table*}[t!]
\centering
\caption{Summary of representative techniques for satellite FL.}
\label{tab:satellite_fl_summary}
\footnotesize
\resizebox{0.92\textwidth}{!}{
\begin{tblr}{
  width=\textwidth,
  colspec={Q[c,m,wd=2.2cm] Q[c,m,wd=0.9cm] X[2.2,l,m] X[2.4,l,m] X[3.4,l,m]},
  colsep=4pt, rowsep=2.2pt, stretch=1.1,
  row{1}={font=\bfseries, bg=black!8, halign=c},
  hline{1,Z}={0.8pt},
  vline{1,Z}={0.8pt},
  vline{2-5}={0.4pt},
  hline{2}={0.6pt},
  hline{8,15}={0.5pt},
  hline{3-7,9-14,16-17}={2-Z}{wd=0.3pt, fg=black},
}
Approach & Ref. & Performance Metrics & Control Variables & Key Contribution \\
\SetCell[r=6]{c,m} {\textbf{Efficient Local}\\\textbf{Updates}} 
& \cite{hu2021lora}
& Accuracy; Trainable parameters; Training memory
& Adaptation rank; Adapted linear layers
& Low-rank factorized updates of large linear layers on a frozen backbone \\
& \cite{lin2025fedsn}
& Test accuracy; Convergence speed
& Per-satellite sub-structure combination; Staleness-based aggregation switching
& Sub-structure training matched to heterogeneous satellite computing, memory, and uplink budgets, with pseudo-synchronous aggregation compensating staleness \\
& \cite{luo2021cost}
& Total time and energy cost; Target accuracy
& Participating clients; Local iterations
& Local-epoch selection balancing model contribution against computing overhead \\
& \cite{li2020federated}
& Convergence; Accuracy under heterogeneous data and resources
& Proximal coefficient; Variable local work
& Proximal penalty bounding local deviation under heterogeneous epochs \\
& \cite{stich2018sparsified}
& Convergence rate; Communication volume
& Top-$k$ sparsity level; Residual memory
& Sparsified updates with residual accumulation compensating discarded elements \\
& \cite{yang2024communication}
& Accuracy; Training time
& Block-wise bit-width; Per-window training workload
& Progressive block-wise quantization ensuring each satellite completes per-round training within its connection window \\
\SetCell[r=7]{c,m} {\textbf{Connectivity-Aware}\\\textbf{Coordination}}
& \cite{rodio2023federated}
& Accuracy; Convergence time; Model bias
& Aggregation weights; Client exclusion
& Bias-aware aggregation under correlated, intermittent client availability \\
& \cite{zhou2024decomposition}
& Communication-training efficiency; Local training accuracy
& Device scheduling within LEO service time
& Decomposition and meta-DRL-based multi-objective optimization for LEO-assisted asynchronous FL \\
& \cite{liu2024fedasmu}
& Accuracy; Training time
& Staleness-aware aggregation weights; Local model refresh
& Delay-based downweighting of stale updates in asynchronous aggregation \\
& \cite{gu2021fast}
& Convergence; Accuracy under device unavailability
& Stored latest updates of unavailable clients
& Bias correction of missing and delayed updates via memorized updates \\
& \cite{nguyen2022federated}
& Time-to-accuracy; Communication trips
& Buffer size; Client concurrency
& Buffered asynchronous aggregation tolerating stragglers \\
& \cite{razmi2024board}
& Training time; Communication cost; Accuracy
& Intra-orbit ISL relaying; Partial aggregation
& ISL relaying exploiting predictable satellite visits to bridge intermittent GS contacts \\
& \cite{lei2025joint}
& Training time; Network traffic
& Data partitioning; Resource allocation; Aggregation transmission
& In-network aggregation transmission reducing traffic toward the GS \\
\SetCell[r=3]{c,m} {\textbf{Eclipse-Aware}\\\textbf{Scheduling}}
& \cite{li2024battery}
& Battery depth of discharge; Energy consumption
& Onboard energy scheduling; Inter-satellite collaboration
& Energy budgeting exploiting periodic harvesting to preserve battery reserves \\
& \cite{luo2024energy}
& Energy consumption; Convergence; Accuracy
& Aggregation frequency; Model compression ratio
& Energy-efficient hierarchical training jointly adapting aggregation frequency and compression under onboard resource and energy limits \\
& \cite{xu2020client}
& Selected clients over time; Long-term energy consumption
& Client selection; Bandwidth allocation
& Long-term client selection keeping energy use within sustainable budgets \\
\end{tblr}%
}
\end{table*}

\begin{figure}[t!]
\centering
\includegraphics[width=0.5\textwidth]{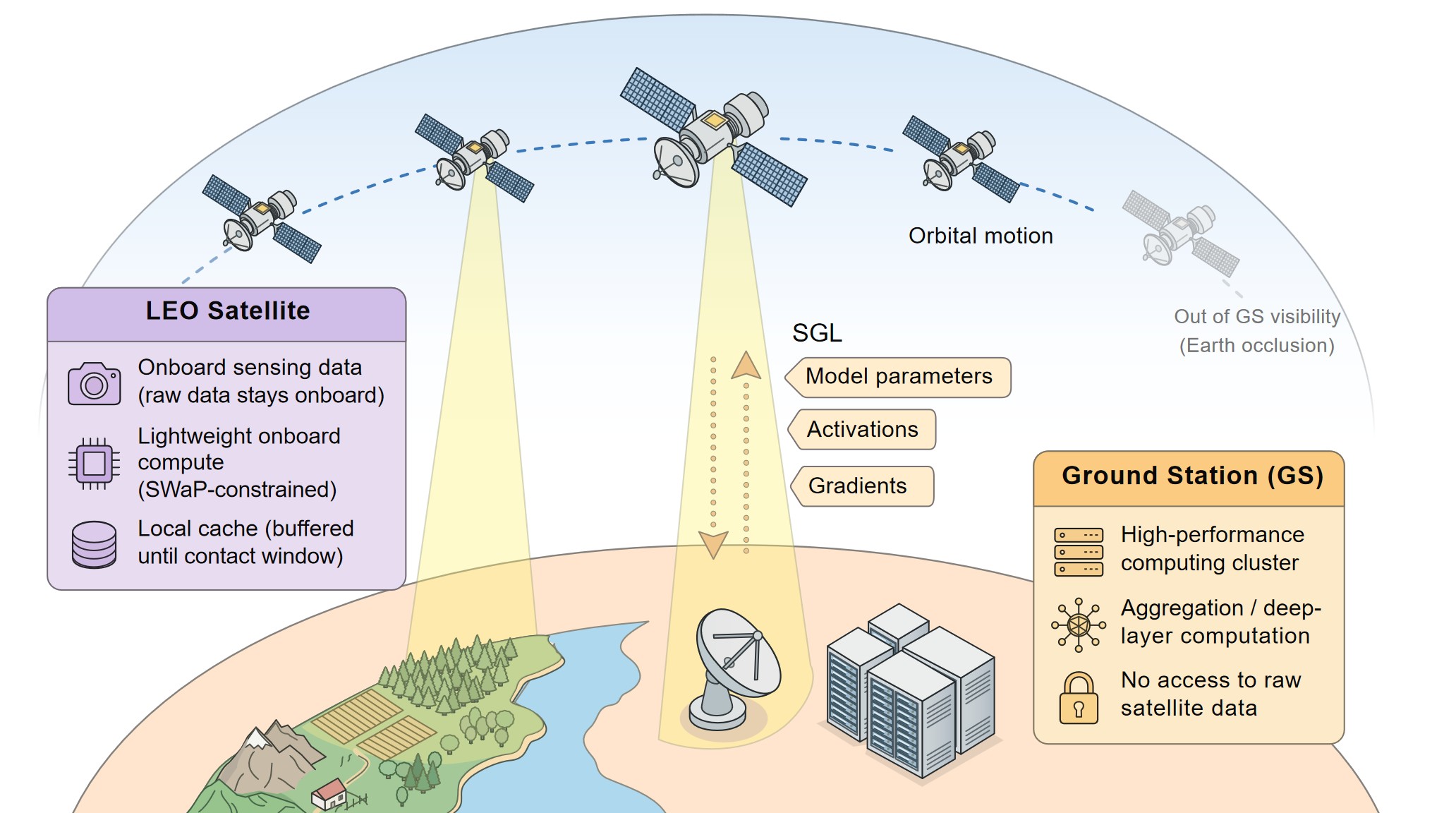}
\centering\caption{Satellite-GS collaborative architecture for distributed training.}
\label{fig_sat_GS_archi}
\end{figure}

\section{Satellite Distributed Training Frameworks}
\label{sec:distributed_training}

On-orbit model updating is increasingly necessary for satellite missions, as static models cannot account for seasonal land cover changes such as vegetation growth or snow retreat~\cite{koh2021wilds}, nor for fault signatures that vary across individual satellites and mission phases~\cite{hundman2018detecting}. Realizing such updates under the deployment challenges in Section~\ref{subsec:SI_constraints} calls for the constraint-aware design principles established in Section~\ref{subsec:design_principle}. This section investigates distributed training frameworks for SI, wherein the resource-aware, topology-resilient, and energy-aware principles are instantiated through satellite-GS collaboration mechanisms. We first present the satellite-GS collaborative architecture (Section~\ref{subsec:satellite_GS_archi}). Building on this architecture, we then explore two representative distributed training paradigms, namely satellite FL (Section~\ref{subsec:FL}) and satellite SL (Section~\ref{subsec:SL}).



\subsection{Satellite-GS Collaborative Architecture}\label{subsec:satellite_GS_archi}

As illustrated in Fig.~\ref{fig_sat_GS_archi}, the collaborative architecture comprises two primary components: the satellite and the GS, which exchange intermediate representations such as model parameters, activations, and gradients to enable collaborative computing. This architecture serves as the common foundation for both FL and SL, as detailed below. 
\begin{itemize} 
\item {\textit{Satellite:}} In the satellite-ground collaborative architecture, LEO satellites serve as both primary data acquisition terminals and proximal processing units. Constrained by stringent SWaP limitations, these satellites are restricted to lightweight computation, updating either the full model (in FL) or only its shallow layers (in SL). To accommodate intermittent satellite-GS connectivity, intermediate results, such as model parameters or smashed data\footnote{Smashed data is a term specific to SL. Smashed data refers collectively to the activations transmitted from the satellite to the GS during the forward propagation and the gradients transmitted back from the GS to the satellite during backpropagation, both exchanged at the cut layer.}, are cached locally and transmitted only during visible windows.
\item  {\textit{GS:}} Equipped with high-performance computing clusters, the GS compensates for spaceborne hardware limitations by taking over model aggregation or the compute-intensive workload, namely aggregating local model updates (in FL) or updating the deep model layers (in SL). By operating exclusively on these received intermediate results without accessing raw satellite data, the GS simultaneously balances the computing load across the satellite-GS network and preserves data privacy by design.
\end{itemize}

\subsection{Satellite FL}\label{subsec:FL}

As a prominent distributed machine learning framework, FL enables multi-party collaborative training by exchanging model updates rather than raw data~\cite{mcmahan2017communication,fang2025automated,hong2026conflict}. However, its deployment over satellite-GS collaborative architectures is shaped by the three challenges identified in Section~\ref{subsec:SI_constraints}. This subsection first describes the workflow pipeline of satellite FL  (Section~\ref{subsec:FL_pipeline}), and subsequently presents efficient local updates (Section~\ref{subsec:FL_1}), connectivity-aware coordination (Section~\ref{subsec:FL_2}), and eclipse-aware scheduling (Section~\ref{subsec:FL_3}) optimization strategies. Table~\ref{tab:satellite_fl_summary} summarizes the key techniques for satellite FL reviewed below.

\subsubsection{FL Workflow in Satellite Networks}\label{subsec:FL_pipeline}

\begin{figure}[t!]
\centering
\includegraphics[width=0.5\textwidth]{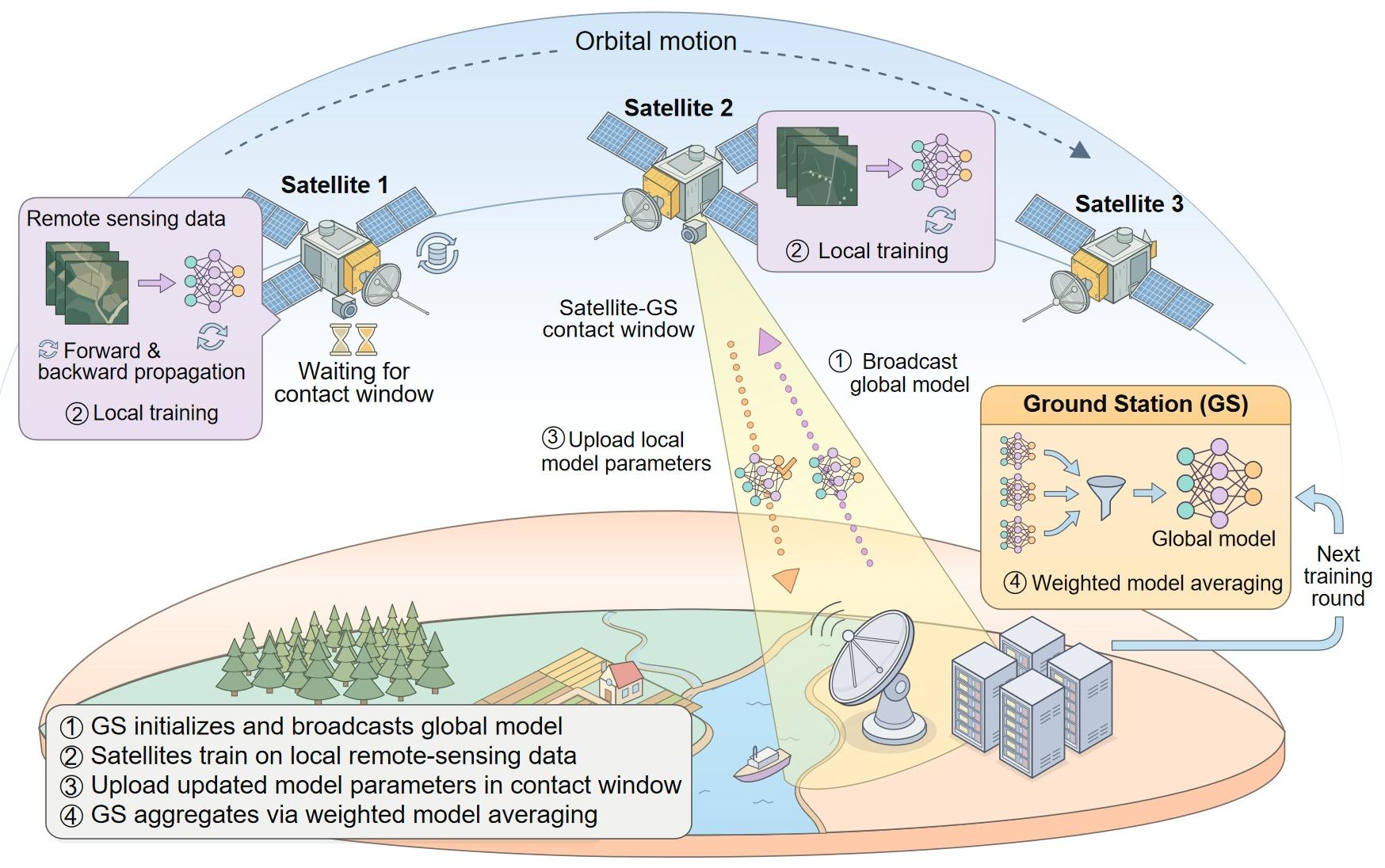}
\centering\caption{Training procedure of satellite FL with satellite-GS model exchange during contact windows.}
\label{fig_FL}
\end{figure}

Building upon conventional FL, satellite FL performs collaborative training through a repeated satellite-GS interaction loop, as illustrated in Fig.~\ref{fig_FL}. At the beginning of each training round, the GS initializes a global model and broadcasts it to the participating satellites. Each satellite then uses its locally collected remote-sensing data to perform forward and backward propagation, updating its local model parameters. After that, each satellite transmits its updated model parameters to the GS during an available satellite-GS contact window. Upon receiving the updates, the GS aggregates them via weighted model averaging to produce a new global model, which is then redistributed in the next training round.

\subsubsection{Efficient Local Updates}\label{subsec:FL_1}

Satellite FL requires each participating satellite to perform local model training on its local dataset. While this paradigm eliminates raw data transmission, it imposes substantial memory and computational overhead on spaceborne processors operating under stringent SWaP constraints. Addressing this bottleneck necessitates strategies that target distinct stages of the training pipeline. Reducing the volume of trainable parameters being optimized, adapting the amount of local computation to available resources, and compressing the resulting local updates prior to transmission. Motivated by this perspective, we organize the following discussion around three complementary strategies for enabling efficient local updates: parameter-efficient fine-tuning, adaptive local epoch control, and sparse and quantized model training.

{\textit{a) Parameter-Efficient Fine-Tuning}.} In satellite FL, on-satellite training requires updating all trainable parameters, generating optimizer states and gradients whose memory footprint exceeds satellite hardware capacities. Parameter-efficient fine-tuning (PEFT)~\cite{han2024parameter} mitigates this bottleneck by freezing the pre-trained backbone and restricting optimization to a lightweight subset of parameters, substantially reducing computing and memory overhead to enable on-satellite training of large-scale models. Selecting an appropriate PEFT mechanism requires jointly weighing data heterogeneity, downstream task complexity, and available satellite resources, as representative methods trade off representational capacity against resource efficiency. Among higher-capacity methods, Adapter tuning~\cite{houlsby2019parameter} embeds trainable modules to capture domain-specific features, providing effective adaptation under highly non-IID data, while low-rank adaptation (LoRA)~\cite{hu2021lora} updates large linear layers via low-rank matrix factorization, substantially reducing the trainable parameter count. Lighter alternatives exist for more severe onboard memory and compute constraints: prompt-based tuning~\cite{lester2021power} prepends trainable virtual tokens to the input, and bias-only updating~\cite{zaken2022bitfit} restricts optimization to existing additive bias vectors, both trading representational capacity for a reduced resource footprint.

{\textit{b) Adaptive Local Epoch Control}.} The number of local training epochs governs a trade-off: more local epochs improve each satellite's contribution to the global model, but increase its onboard computing and memory overhead~\cite{luo2021cost}. Satellite FL typically assigns a fixed number of epochs across all satellites; this configuration is ill-suited to satellite networks, where available computing resources fluctuate due to heterogeneous processors and concurrent mission priorities~\cite{lin2025fedsn}. A straightforward solution treats epoch allocation as a per-satellite decision, simply allocating more epochs to resource-rich satellites. However, its impacts are not confined to any single satellite, where a satellite performing substantially more local updates than its peers moves further away from the current round's global model, an effect worsened by the data heterogeneity across satellites. Effective epoch control must therefore go beyond matching each satellite's epoch count to its available computing budget, which can be estimated from onboard telemetry such as processor utilization and remaining memory~\cite{wang2024latte}. It must additionally bound how much any two satellites' local updates are allowed to differ, for instance, via a proximal penalty on local deviation~\cite{li2020federated} or early termination once deviation exceeds a threshold~\cite{niu2024flrce}; otherwise, unbounded divergence can cause satellites to converge to inconsistent local models.

{\textit{c) Sparse and Quantized Model Training}.} FL requires retaining gradients, activations, and optimizer states at full precision throughout the training process, imposing substantial computing and memory overhead~\cite{rajbhandari2020zero}. To combat this issue, two complementary techniques can be adopted. The first is sparsification, which retains only the most informative elements, for instance via Top-k gradient selection~\cite{aji2017sparse}, while discarding the remainder; rather than discarding these elements entirely, they can be accumulated as residuals and incorporated into subsequent updates, so that information lost to sparsification at one step is compensated over later steps~\cite{stich2018sparsified}. The second is quantization, which reduces the numerical precision to low-bit formats~\cite{chen2024mixed,yang2024communication}, with the precision adjusted according to available onboard resources. Other techniques, such as low-rank factorization~\cite{vogels2019powersgd} and sketch-based compression~\cite{rothchild2020fetchsgd}, can be combined with sparsification and quantization to further reduce the training overhead. However, sparsifying or quantizing too aggressively introduces optimization noise that can outweigh the resulting resource savings, particularly under highly heterogeneous data, and these techniques must therefore be calibrated to strike a good balance between resource savings and training performance.

\subsubsection{Connectivity-Aware Coordination}\label{subsec:FL_2}

FL assumes continuous connectivity. In practice, however, most operators can afford only a few GSs, so each LEO satellite can reach the GS only during short and intermittent contact windows. Unlike terrestrial wireless channels, satellite-GS visibility in LEO networks follows a predictable pattern, since contact windows can be computed in advance from ephemeris data~\cite{tao2023transmitting}. FL coordination should therefore move from accommodating intermittent connectivity toward scheduling client participation, model aggregation, and inter-satellite relaying according to predicted satellite visibility.

{\textit{a) Orbital-Aware FL Client Selection}.} Conventional FL typically relies on random client sampling~\cite{mcmahan2017communication,zhang2025lcfed,fang2026hfedmoe,zhang2025satfed}, an approach that inevitably selects satellites lacking GS visibility or approaching imminent link handovers, resulting in severe participation failures and straggler effects~\cite{chen2026fedhra} in satellite networks.  To adapt to the time-varying topology, the GS must replace random sampling with contact-window-aware selection, prioritizing satellites with predicted SGL windows sufficient to complete an entire FL round without interruption. However, optimizing solely for connectivity risks spatial bias, overfitting the global model to the orbital trajectories most frequently in contact with the GS~\cite{ribero2022federated,rodio2023federated}. Client selection must therefore be formulated as a joint optimization problem that considers link availability against model performance. This balance can be achieved through algorithmic frameworks such as Lyapunov optimization~\cite{perazzone2025communication}, multi-armed bandits~\cite{xia2020multi}, or deep reinforcement learning (DRL)~\cite{zhang2022multiagent} for dynamic scheduling~\cite{zhou2024decomposition}, allowing contact-window-aware selection to reduce participation failures without compromising model performance.

{\textit{b) Staleness-Aware Asynchronous Aggregation}.} Conventional FL requires the GS to collect updates from all selected clients before global aggregation. This synchronization is impractical in LEO networks, as short SGL windows and long gaps in GS coverage frequently create stragglers. The GS therefore faces a dilemma: waiting for delayed updates increases training latency, while discarding them wastes onboard computation~\cite{sun2024staleness,jiang2026csafl}. To mitigate this, the GS can adopt asynchronous FL, updating the global model sequentially upon receiving individual updates~\cite{xie2019asynchronous,wang2025tackling}. However, this asynchrony induces severe model staleness, as delayed updates are computed on outdated global models; aggregating them directly amplifies client drift under the highly non-independent and identically distributed (non-IID) satellite data, degrading model convergence~\cite{sun2024staleness,elmahallawy2024communication}. Therefore, the GS must dynamically compensate for the impact of stale updates using mechanisms like downweighting older updates via delay-based decay~\cite{liu2024fedasmu,jiang2026csafl}, correcting the gradient bias through variance reduction~\cite{gu2021fast}, or aggregating updates via asynchronous buffering~\cite{nguyen2022federated}. By applying these techniques, the GS can incorporate delayed updates based on delay and link reliability while maintaining robust training performance.

\begin{figure}[t!]
\centering
\includegraphics[width=0.48\textwidth]{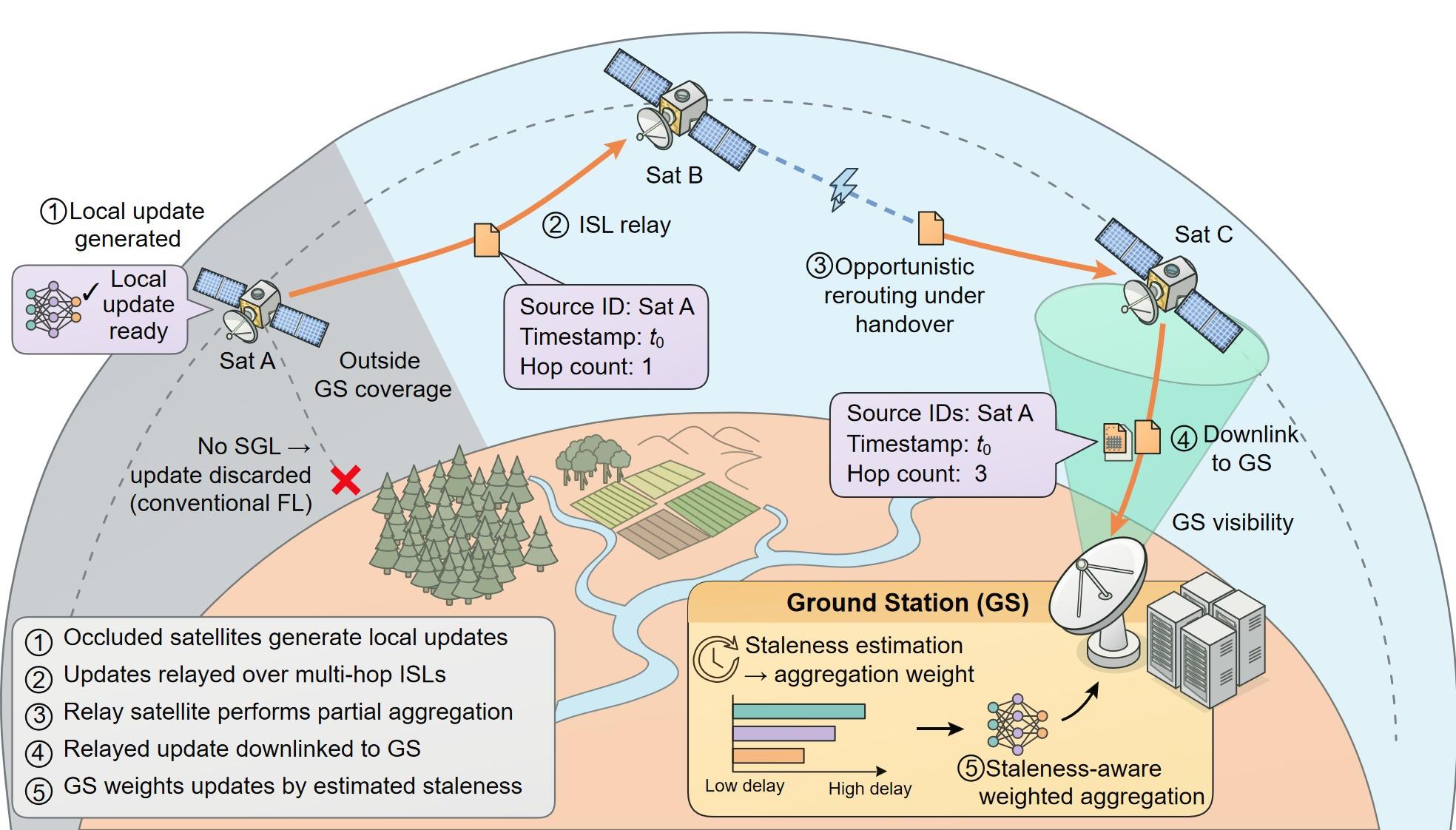}
\centering\caption{ISL-assisted update relay for satellites without GS visibility in satellite FL.}
\label{fig_relay_FL}
\end{figure}

{\textit{c) ISL-Assisted Update Relay}.} Since GSs can be deployed only at a limited number of suitable sites~\cite{baeza2023gateway}, most LEO satellites lie outside direct GS coverage at any given time. In satellite FL, satellites without GS visibility cannot upload their local updates, resulting in discarded updates and underutilized onboard resources. Inter-satellite links (ISLs) address this limitation by extending SGL coverage through multi-hop relaying. As illustrated in Fig.~\ref{fig_relay_FL}, an isolated satellite forwards its local update over ISLs to a neighboring satellite, which then relays the payload onward to the GS~\cite{razmi2024board}. By decoupling model transmission from the satellite-GS contact window, this scheme enables disconnected satellites to remain involved in FL. However, relaying every update risks saturating the limited SGL capacity; thus, the ISL-assisted relaying process should incorporate communication-efficient strategies, including opportunistic routing to accommodate dynamic link handovers~\cite{tang2024opportunistic}, in-network partial aggregation to reduce downlink traffic~\cite{lei2025joint}, and decentralized coordination to avoid redundant transmission~\cite{zehtabi2026resource,zhai2023fedleo,zhao2025alanine,chen2026fedhra}. Each relayed update should further carry source identifiers, generation timestamps, and hop counts, enabling the GS to estimate its staleness and assign an appropriate aggregation weight. Together with staleness-aware aggregation, ISL-assisted relaying thus recovers updates that would otherwise be lost due to the lack of GS coverage, sustaining FL participation under dynamic orbital connectivity.

   \begin{table*}[t!]
\centering
\caption{Summary of representative techniques for satellite SL}
\label{tab:satellite_sl_summary}
\footnotesize

\resizebox{0.92\textwidth}{!}{%
\begin{minipage}{\textwidth}
\setlength{\tabcolsep}{3pt}
\newlength{\satSLTableWidth}
\setlength{\satSLTableWidth}{\dimexpr\textwidth-10\tabcolsep-6\arrayrulewidth\relax}

\begin{tblr}{
  colspec={Q[c,m,wd=0.14\satSLTableWidth]
           Q[c,m,wd=0.05\satSLTableWidth]
           Q[l,m,wd=0.22\satSLTableWidth]
           Q[l,m,wd=0.25\satSLTableWidth]
           Q[l,m,wd=0.33\satSLTableWidth]},
  colsep=3pt, rowsep=2pt, stretch=1.15,
  row{1}={font=\bfseries, bg=black!8, halign=c, m},
  hline{1,Z}={0.8pt},
  vline{1,Z}={0.8pt},
  vline{2-5}={0.4pt},
  hline{2}={0.6pt},
  hline{8,14}={0.5pt},
  hline{3-7,9-13,15-16}={2-Z}{wd=0.3pt, fg=black},
}
\textbf{Approach} & \textbf{Ref.} & \textbf{Performance Metrics} & \textbf{Control Variables} & \textbf{Key Contribution} \\
\SetCell[r=6]{c} {\bf{Dynamic Split Execution}}
& \cite{lin2025adaptsfl}
& Training latency; Accuracy
& Cut layer; Client-side model aggregation interval
& Convergence-guided cut-layer adaptation to limited client-side resources \\
& \cite{lin2024efficient}
& Per-round latency; Accuracy
& Cut layer; Subchannel allocation; Transmit power
& Joint cut-layer and resource allocation with last-layer gradient aggregation lightening client-side computation \\
& \cite{ilhan2023scalefl}
& Accuracy; Computation cost
& Sub-model width; Sub-model depth
& Capacity-matched nested sub-models for heterogeneous client hardware \\
& \cite{imanishi2023fast}
& Peak memory; Recomputation overhead
& Retained activations; Recomputation order
& Activation recomputation trading extra computation for reduced peak memory \\
& \cite{lin2026sl}
& Time-to-accuracy; Communication volume
& Channel grouping; Group-wise compression level
& Entropy-guided channel-wise compression shrinking smashed-data volume \\
& \cite{oh2025communication}
& Accuracy; Communication overhead
& Feature-wise dropout probability; Quantization level
& Dispersion-based feature-wise dropout and quantization shrinking smashed-data volume \\
\SetCell[r=6]{c} {\bf{Pipeline and Offline Orchestration}}
& \cite{lin2026hasfl}
& Convergence; Training latency
& Batch size; Cut layer
& Adaptive batch size and cut layer shortening per-round exchange to curb stragglers \\
& \cite{qin2021simigrad}
& Training time; Accuracy
& Batch size; Target gradient similarity
& Mini-batch-level batch-size adaptation enabling per-round workload adjustment \\
& \cite{narayanan2019pipedream}
& Time-to-accuracy; Throughput
& Stage partitioning; Stage replication
& Asynchronous forward-backward pipelining with weight stashing tolerating delayed gradients \\
& \cite{zheng2017asynchronous}
& Accuracy; Speedup
& Compensation coefficient; Moving-average rate
& Taylor-expansion compensation correcting the staleness of delayed gradients \\
& \cite{lin2026leo}
& Accuracy; Training time
& Auxiliary model; Pseudo-labels; Activation interpolation
& Auxiliary model sustaining client-side training during satellite-GS non-contact time \\
& \cite{wu2025sfl}
& Accuracy; Training time
& Auxiliary network; Splitting network size
& Auxiliary-network local updates continuing training under intermittent connectivity \\
\SetCell[r=3]{c} {\bf{Orbit-Aware Energy Control}}
& \cite{meng2026asfl}
& Convergence rate; Delay; Energy consumption
& Cut layer; Resource allocation
& Online splitting and resource allocation under long-term energy and delay budgets \\
& \cite{lee2025lyapunov}
& Time-averaged energy; Round latency; Privacy
& Per-round cut layer; Drift-plus-cost trade-off weight
& Lyapunov-based round-wise cut-layer control minimizing time-averaged energy \\
& \cite{kim2023bargaining}
& Energy; Training time; Privacy
& Cut layer; Device-side model personalization
& Bargaining-game cut-layer selection balancing energy consumption against time and privacy \\
\end{tblr}
\end{minipage}
}
\end{table*}

 \begin{figure}[t!]
\centering
\includegraphics[width=0.5\textwidth]{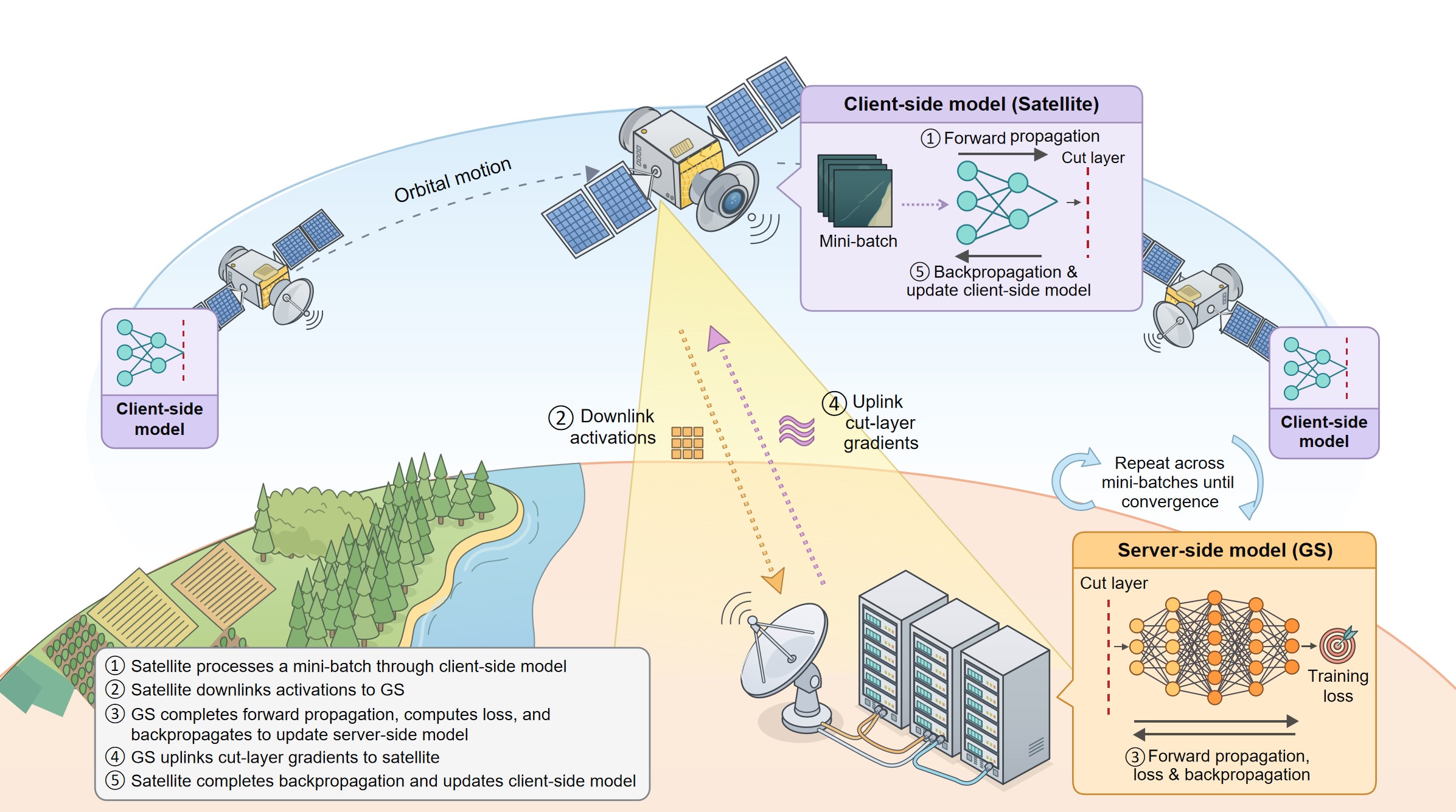}
\centering\caption{Training procedure of satellite SL with activation and gradient exchange at the cut layer.}
\label{fig_SL}
\end{figure}

\subsubsection{Eclipse-Aware Scheduling}\label{subsec:FL_3}

FL depends on repeated rounds of local training, but the energy available for this training fluctuates between sunlit and eclipse phases. During eclipse, satellites rely solely on limited battery reserves, competing for this same supply with mission-critical spacecraft functions such as attitude control and communications. Sustaining FL training under these conditions is therefore inherently difficult, and satellite FL cannot simply assume energy is always available. Its scheduling must instead follow the sunlit-eclipse cycle, as detailed in the following.

{\textit{a) Energy-Adaptive Local Training}.}
Since solar power generation and battery states vary over the orbit, a fixed training configuration may waste energy during sunlit periods or cause energy shortages during eclipses~\cite{chen2026adaptive}. Satellite FL therefore requires an energy-adaptive training scheme that scales the training workload to the available energy. Before each round, the controller estimates the FL energy budget from predicted solar power, current battery energy, and the reserve required for the next eclipse~\cite{li2024battery}. This budget, together with the available training time, determines how much local training the satellite can perform, and the controller then jointly configures the number of local epochs and mini-batch size~\cite{xiao2024time}, to maximize learning performance without depleting the battery below its required eclipse reserve. Because actual solar supply and platform power consumption may deviate from their predictions, the controller should calibrate the training configuration as updated battery and power measurements become available. For instance, model predictive control~\cite{majumdar2017dynamic} can adjust the training load based on such short-term predictions, while Lyapunov-based optimization~\cite{mao2016dynamic} and reinforcement learning~\cite{park2026joint} support online decisions under uncertain energy conditions.

{\textit{b) Energy-Driven Cross-Plane Scheduling}.}
Deploying FL across LEO mega-constellations introduces a macro-level scheduling challenge that goes beyond adapting training at the level of a single satellite: since satellites in different orbital planes enter eclipse at different times, energy availability across the constellation is highly heterogeneous~\cite{chen2026satsfc}. Standard FL requires the GS to wait for every selected satellite to upload its update before aggregating; energy-depleted satellites in eclipse cannot continue training and therefore bottleneck model aggregation. To address this, the GS can prioritize satellites with sufficient remaining energy over those currently in eclipse when selecting satellites for each round. This selection cannot rely only on each satellite's energy level at the moment of selection, however: selecting the same energy-rich satellites round after round can drain their reserves ahead of their next eclipse. Therefore, the selection must also consider how each satellite's energy will evolve over subsequent rounds. Decision-making frameworks such as Lyapunov optimization~\cite{xu2020client} and DRL~\cite{wang2020optimizing} address this by using predictable ephemeris data to select participants round by round in a way that keeps every satellite's battery above a safe threshold over the long run, even as the constellation's topology continues to change~\cite{yang2026energy,li2023networking}.

\subsection{Satellite SL}\label{subsec:SL}

While satellite FL avoids transmitting raw data by retaining it on-orbit, it requires each satellite to train the full model locally, exhausting the limited satellite memory and compute budgets and rendering large-scale model training impractical. SL~\cite{thapa2022splitfed,vepakomma2018split,lin2024split} addresses this limitation via model partitioning, where it splits a deep neural network across the satellite-GS collaborative architecture, so that satellites execute only lightweight shallow-layer computations while the compute-intensive deep layers are offloaded to the resource-rich GS~\cite{lin2026leo}. This subsection first presents the execution pipelines of satellite SL (Section~\ref{subsec:SL_pipeline}) and then explores dynamic split execution (Section~\ref{subsec:SL_resource_partition}), pipeline and offline orchestration (Section~\ref{subsec:SL_intermittency_connect}), and orbit-aware energy control (Section~\ref{subsec:SL_energy_control}). Table~\ref{tab:satellite_sl_summary} summarizes the key techniques for satellite SL reviewed below.

\subsubsection{SL Workflow in Satellite Networks}\label{subsec:SL_pipeline}

Unlike satellite FL, which requires full-model on-satellite training, SL partitions a deep neural network at an intermediate layer, referred to as the cut layer, into a client-side model deployed on the satellite and a server-side model hosted at the GS. Building on this partition, satellite SL establishes a collaborative training loop between satellites and the GS, as illustrated in Fig.~\ref{fig_SL}. In each round, every satellite processes a mini-batch of data through its client-side model to produce activations, and downlinks them to the GS. The GS then completes the forward propagation through its server-side model, computes the training loss, and backpropagates to update its server-side parameters and derives the gradient at the cut layer. Finally, the GS uplinks these intermediate gradients back to the satellite, allowing it to complete the remaining backpropagation and update the client-side model on the satellite. The process repeats across mini-batches until training converges.

\subsubsection{Dynamic Split Execution}\label{subsec:SL_resource_partition}

While SL offloads the deep layers to the GS, caching smashed data and updating the client-side model still remain constrained by satellite compute and memory budgets, especially for immense constellations or large-scale models. Addressing these constraints requires determining the cut layer based on a satellite's available resources, adapting the client-side model to the heterogeneous hardware across a constellation, and managing the memory demands of caching activations and gradients within each round. The details are as follows.

{\textit{a) Dynamic Cut Layer Adaptation}.} A static cut layer cannot accommodate the fluctuating satellite resources, particularly when concurrent tasks such as remote-sensing acquisition or attitude control compete for compute and memory, causing a fixed-depth client-side model to exceed the available budget and delay training~\cite{wu2025split,lin2025adaptsfl}. To address this, a resource-aware adaptive partitioning strategy instead treats the cut layer as an adjustable variable, updated from real-time telemetry such as processor utilization, remaining memory, and buffer occupancy~\cite{you2026adaptive}. If the cut layer is adjusted solely based on computation, it moves toward the input when onboard resources are constrained, offloading most computation to the GS, and deeper when resources are abundant, allowing the satellite to take over compute-intensive load. However, such an adjustment disregards a strict constraint on memory: for architectures such as convolutional neural networks, activations in shallower layers can be substantially larger than those in deeper layers~\cite{kang2017neurosurgeon}, and the satellite must retain the activations of every client-side layer until the corresponding gradients return; shifting the cut layer deeper therefore accumulates a larger activation footprint, which becomes infeasible whenever it exceeds the satellite's available memory, irrespective of any reduction in computation time. Joint optimization of the cut layer and resource allocation~\cite{lin2024efficient,lin2025hierarchical} can identify the cut layer that minimizes onboard computation time while satisfying the satellite's memory capacity.

\begin{figure}[t!]
\centering
\includegraphics[width=0.5\textwidth]{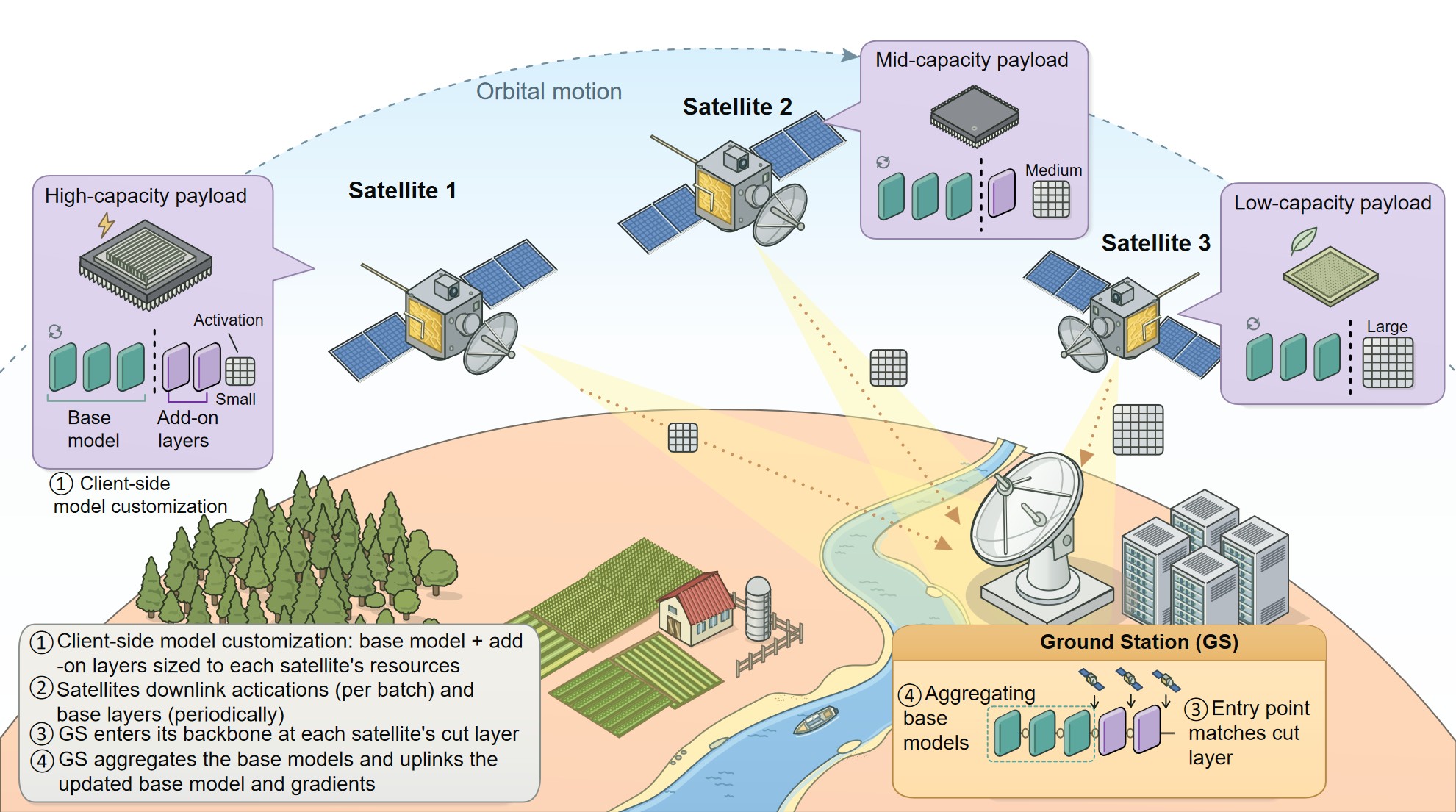}
\centering\caption{Heterogeneous client-side model customization with a shared base model in satellite SL.}
\label{fig_Hetero_model_cus}
\end{figure}

{\textit{b) Heterogeneous Client-Side Model Customization}.} Satellites launched across different batches and hardware generations exhibit substantial heterogeneity within a constellation, with onboard payloads, such as AI accelerators versus low-power FPGAs, varying by orders of magnitude in capacity~\cite{lin2025fedsn,zhang2026sleon}. No uniform client-side model size can accommodate this heterogeneity. For example, a model sized for the most capable hardware would exceed the budget of resource-constrained satellites, whereas a conservative model for the weakest satellite would leave the computing capacity of high-performance satellites largely idle~\cite{ilhan2023scalefl}.  As shown in Fig.~\ref{fig_Hetero_model_cus}, the client-side model customization design inspired by nested and slimmable networks~\cite{kim2018nestednet,yu2019universally} resolves this by assigning each satellite a client-side model aligned with its own resources, where all satellites share several initial layers as a common base model; satellites with sufficient capacity execute additional layers to produce more compact smashed data, while resource-constrained satellites execute only the base model. Every few training rounds, satellites additionally downlink their base-layer parameters alongside the smashed data, and the GS aggregates them and uplinks the updated base model, ensuring consistent low-level feature extraction and preventing shallow-representation divergence. Correspondingly, the GS maintains a backbone modularized into layer-wise components: as smashed data arrive from satellites with different client-side models, the GS locates the module entry point matching each satellite's cut layer and invokes the subsequent layers to complete forward and backward propagation, achieving flexible per-satellite model customization.

{\textit{c) Activation-Gradient Memory Coordination}.} 
The forward-backward cycle unique to SL introduces a dual memory bottleneck absent in FL: satellites must cache activations after the forward propagation and store the GS's returned gradients before backpropagation~\cite{lin2024split}. Since these two footprints coexist within a single training round, their combined peak can exceed the onboard memory budget, particularly for high-resolution remote-sensing tasks~\cite{xing2024deciphering}. Addressing this peak requires the satellite and the GS to jointly reduce the memory footprint of activations and gradients. On the satellite side, a gradient checkpointing mechanism can be utilized to retain only the input sample and a few initial-layer checkpoints~\cite{imanishi2023fast}, recomputing the remaining activations once gradients arrive, trading additional computation for reduced memory. On the GS side, quantizing~\cite{lin2026sl,oh2025communication} and sparsifying~\cite{zheng2023reducing} the cut layer gradients before transmission reduces both the downlink volume and the satellite's storage overhead, since gradients used immediately for a parameter update need not be retained. The compression ratio, however, must be customized to the model and task, as overly aggressive compression introduces noise that degrades training accuracy. Staggered batch scheduling~\cite{fan2021dapple,narayanan2021memory} can further help by timing activation caching and gradient updates so that their memory footprints do not overlap, thereby keeping the peak usage within budget.

\subsubsection{Pipeline and
Offline
Orchestration} \label{subsec:SL_intermittency_connect}

Satellite SL couples the client-side and server-side computations throughout model training, demanding continuous forward-backward interactions between satellites and the GS. Consequently, any SGL interruption directly hinders the training pipeline, rather than simply delaying it. This vulnerability is further exacerbated by the short, intermittent SGL windows of LEO networks. Therefore, orchestrating satellite SL must address three problems: contact-window-aware scheduling, delay-tolerant computation, and continued training during GS unavailability.

\begin{table*}[!t]
\centering
\caption{Comparison of satellite FL and satellite SL}
\label{tab:training_comparison}
\small

\resizebox{0.8\textwidth}{!}{%
\begin{tblr}{
  width=\textwidth,
  colspec={X[1.1,c,m] X[1.4,c,m] X[1.4,c,m]}, 
  colsep=4pt, rowsep=2.2pt, stretch=1.15,
  row{1}={font=\bfseries, bg=black!8, halign=c, m},
  column{1}={font=\bfseries}, 
  hline{1,Z}={0.8pt},
  vline{1,Z}={0.8pt},
  vline{2,3}={0.4pt},
  hline{2}={0.6pt},
  hline{3-Z}={wd=0.3pt, fg=black},
}
Dimension & Satellite FL & Satellite SL \\
Onboard computation & Full model, forward and backward & Shallow client-side layers \\
Onboard memory demand & Model, optimizer states, gradients & Client model, cached activations/gradients \\
Communication pattern & One update per round & Smashed data per batch \\
Communication frequency & Low, round-level & High, batch-level \\
Link interruption tolerance & High, rounds can be delayed & Low, stalls the training loop \\
Convergence dependency & Cross-satellite aggregation & Continuous satellite-GS interaction \\
Suited satellite profile & Adequate onboard compute/memory & Severe compute/memory constraints \\
\end{tblr}%
}
\end{table*}

{\textit{a) Adaptive Batch-Size Scheduling.}} Since each SL training round requires a complete forward-backward interaction between the satellite and the GS, adopting a uniform batch size across all rounds is highly inefficient~\cite{wei2026pipelining,lin2026hasfl}. Excessively small batches necessitate frequent interactions, repeatedly incurring GS-side queuing delays that consume a substantial portion of the window and reduce effective data transmission~\cite{tirana2025minimization}. Conversely, overly large batches risk extending a single round beyond the remaining window duration, triggering a timeout that discards the data already transmitted. To optimize window utilization, the batch size can be adjusted dynamically to avoid both extremes. Before each round, the satellite estimates a batch size sufficient to complete the exchange within the available window, based on real-time channel bandwidth and the predictable duration of upcoming satellite passes~\cite{vasisht2021l2d2} via dynamic batch-size regulation~\cite{qin2021simigrad,lin2026hasfl}. This is achieved by decoupling local computation from communication, where the satellite executes forward propagation batch by batch, accumulating outputs locally until reaching the target size before transmission, while the GS performs the corresponding server-side computation and returns the resulting gradients as a single batch. By amortizing the queuing overhead across multiple batches rather than incurring it for each, this scheme reduces idle time and improves training efficiency.

\begin{figure}[t!]
\centering
\includegraphics[width=0.5\textwidth]{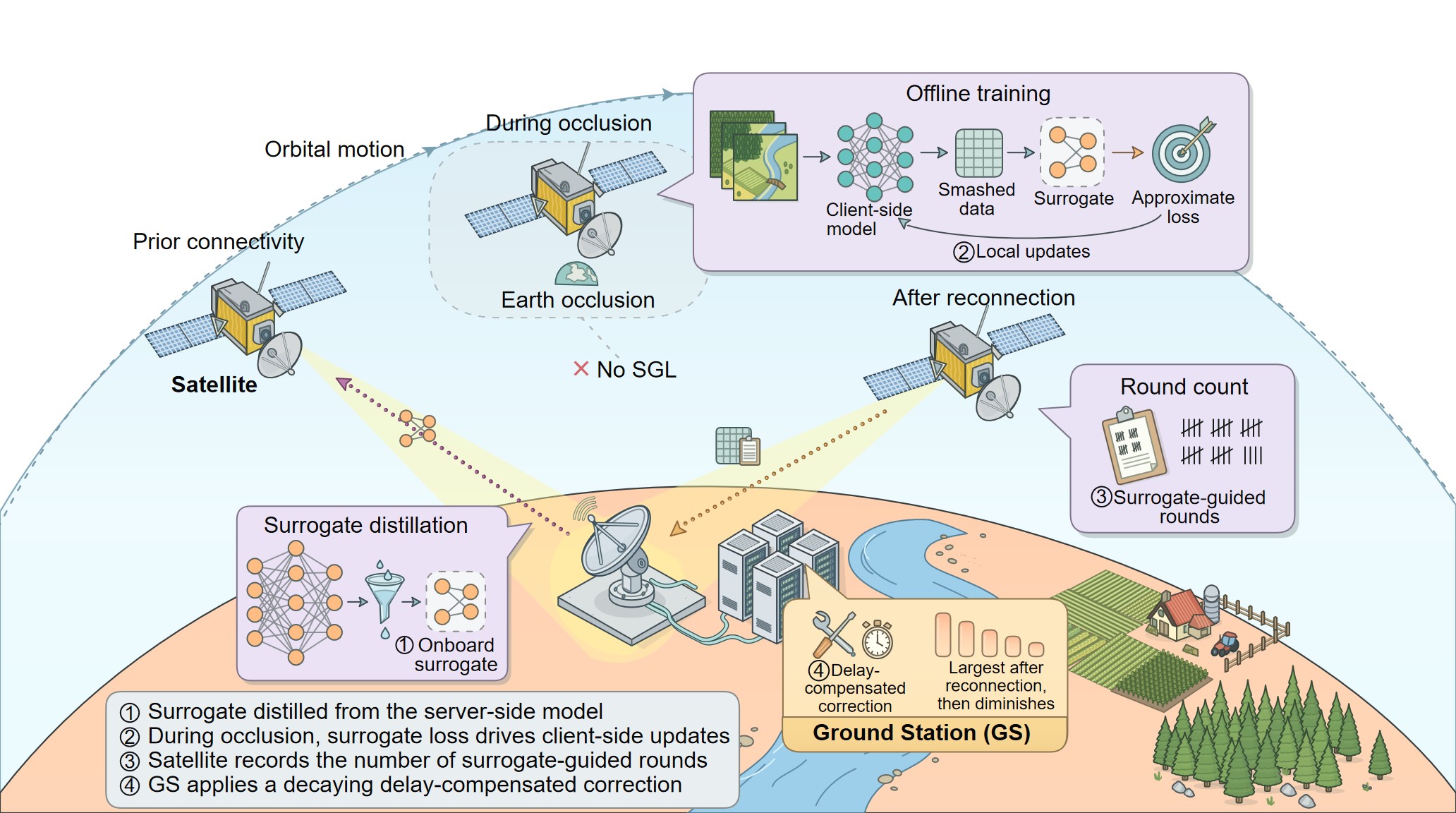}
\centering\caption{Surrogate-assisted offline training in satellite SL.}
\label{fig_surroga_assist_off}
\end{figure}

{\textit{b) Asynchronous Forward-Backward Pipelining}.} SGLs are occasionally subject to unforeseen disruptions, such as sudden channel degradation or the steering and tracking limitations of GS antennas~\cite{vennam2026satellites}, leaving a satellite's forward propagation complete before the corresponding gradients from the GS arrive. Suspending onboard computation to wait for gradients renders processing capacity idle. This waste can be eliminated by decoupling the forward propagation and backpropagation into concurrent and asynchronous pipelines. The forward pipeline continues processing subsequent batches using the client-side model associated with the most recently received gradient, buffering the resulting activations locally until their corresponding gradients arrive~\cite{narayanan2021memory}. The backward pipeline operates independently: upon receiving a delayed gradient, it matches each gradient to its buffered activation via round identifiers and completes the corresponding parameter update without interrupting the forward pipeline. This design sustains continuous utilization of onboard compute resources, preventing a delay in any single gradient from stalling the entire training process. However, since delayed gradients stem from an outdated client-side model, applying them directly may steer the parameter update away from the direction appropriate for the current model, thus introducing a staleness issue that undermines training stability~\cite{narayanan2019pipedream}. This can be mitigated via delay-compensated gradient correction, which estimates and corrects gradient drift over the delay period (e.g., using Taylor expansion)~\cite{zheng2017asynchronous}, or through momentum correction techniques~\cite{ajanthan2025nesterov}, which mitigate the adverse impact of stale gradients on parameter update directions.

{\textit{c) Surrogate-Assisted Offline Training.}} During prolonged periods without GS coverage, which may span multiple orbital periods, a satellite may lose SGL connectivity entirely, leaving SL's forward-backward interaction unable to proceed and halting model training~\cite{denby2020orbital}. To make use of this idle period, the satellite can retain a lightweight onboard surrogate of the server-side model, distilled from the server-side model during prior connectivity to approximate the mapping from smashed data to task loss~\cite{lin2025esl}, as shown in Fig.~\ref{fig_surroga_assist_off}. During occlusion, the satellite feeds smashed data into this surrogate to compute an approximate loss and backpropagates it through the surrogate and the client-side model to obtain local parameter updates~\cite{lin2026leo,wu2025sfl}. Because the surrogate only approximates the actual server-side model, these updates inevitably diverge from the gradients the server-side model would produce. The satellite therefore records how many surrogate-guided rounds it performs during occlusion; upon reconnection, the GS treats these updates as stale, using this count as a measure of their staleness, and mitigates the accumulated deviation through delay-compensated correction~\cite{zheng2017asynchronous}, whose magnitude is largest immediately after reconnection and diminishes as the surrogate-induced deviation is resolved. This mechanism enables client-side training throughout occlusion, at the cost of a temporary accuracy drop while the correction takes effect after reconnection.

\subsubsection{Orbit-Aware
Energy Control}\label{subsec:SL_energy_control}

Unlike onboard compute and memory, which are allocated at the start of each training round and released once it completes, battery energy accumulates its consumption across rounds and is replenished only during sunlit phases. This cyclical replenishment, combined with the irreversible capacity degradation from repeated charge-discharge cycling, distinguishes energy control from compute and memory resource management. Therefore, energy coordination for satellite SL must account for both the immediate energy cost and its cumulative effect on the battery's long-term capacity.

{\textit{a) Sunlit-Phase Energy Budget Scheduling}.} To prevent premature battery depletion and ensure a safe transition into the upcoming eclipse, a dynamic and closed-loop energy budget scheduling scheme can be adopted during the sunlit phase. At the start of this phase, the controller establishes an initial budget by predicting solar generation from orbital geometry and subtracting the battery reserve necessary to survive the subsequent eclipse~\cite{liu2024inorbit}. Because the energy consumed by each SL training round depends on its batch size and cut layer, the controller estimates the cost of each candidate round from these two factors, admitting the round only if the sum of this cost and the cumulative consumption remains within the current budget~\cite{meng2026asfl,lee2025lyapunov,lee2024exploring}. Since actual power generation and consumption typically deviate from their predictions, the controller periodically recalibrates the remaining budget using real-time battery state-of-charge telemetry~\cite{li2024exploring}. If this recalibration indicates a more constrained budget than expected, the controller reduces the energy allocated to subsequent rounds by decreasing the batch size or adopting a shallower cut layer~\cite{liao2024mergesfl,kim2023bargaining}. This closed-loop regulation aligns energy consumption with the energy harvesting remaining for the rest of the sunlit phase, preserving the battery reserve required for the subsequent eclipse.

    {\textit{b) Overlap Power Regulation}.}  The satellite's shared power supply imposes a hard limit on the instantaneous power it can deliver, independent of a training round's total energy budget~\cite{lee2013design}. In SL, the concurrent nature of processor computation and radio transmission yields overlapping power demands. Because the AI payload shares this supply with other spacecraft components, their combined instantaneous draw can easily exceed the hard limit~\cite{xing2024deciphering}. Exceeding this threshold triggers overcurrent protection, forcing a payload reboot and discarding the round's intermediate results. This constraint becomes significantly tighter during the eclipse phase; without solar supplementation, the battery alone must satisfy the combined demand, exacerbating the risk of power overloads. To prevent this, the controller predicts the joint instantaneous power required by computation and transmission, factoring in the telemetry reported by other system components~\cite{wang2026temperature}.  It intervenes only when this prediction exceeds the limit, e.g., delaying transmission or downscaling the processor frequency. By applying a tighter margin during eclipses, this mechanism preserves achievable throughput while safely confining the satellite within its peak power limits.

\begin{figure*}[t!]
\centering
\includegraphics[width=0.85\textwidth]{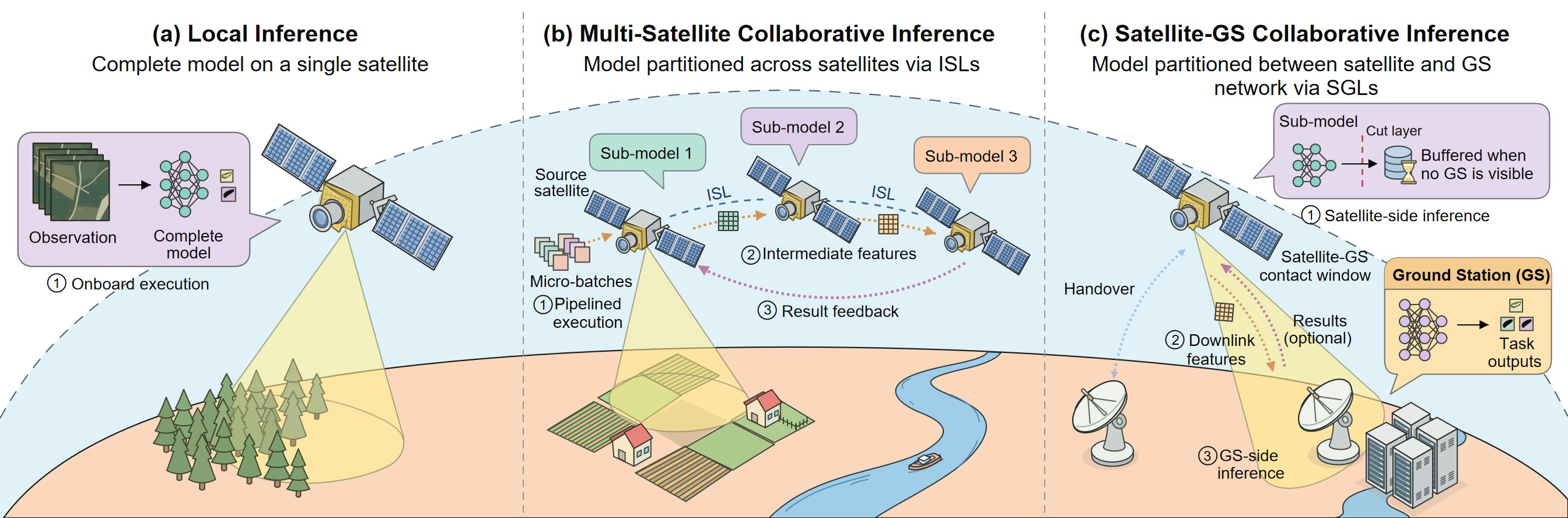}
\centering\caption{An illustration of local inference, multi-satellite collaborative inference, and satellite-GS collaborative inference.}
\label{fig_three_infer}
\end{figure*}

\subsection{Paradigm Comparison and Selection}\label{subsec:SL_FL_compare}

FL and SL both enable satellite-GS co-training under stringent satellite resource constraints, but Table~\ref{tab:training_comparison} reveals a fundamental difference in how they achieve it. FL confines computation to the satellite and communicates only occasionally, while SL offloads major computation-intensive tasks but relies on frequent exchange with the GS. Thus, the optimal paradigm is not fixed but depends on each satellite's resources and the connectivity it can sustain during training. Satellite FL is preferable when a satellite retains sufficient compute and memory to complete full-model training locally, since it then only needs to exchange updates once per round, effectively tolerating the short and infrequent contact windows characteristic of LEO orbits. Conversely, satellite SL becomes advantageous when a satellite's compute and memory are insufficient for full-model training, offloading the deep layers to the GS at the expense of the more frequent satellite-GS interaction. This tradeoff is worthwhile only when the SGL can sustain the resulting per-batch communication without becoming the dominant bottleneck. However, switching a satellite from one paradigm to the other requires re-partitioning the model, re-establishing the corresponding satellite-GS protocol, and, for a switch to SL, negotiating a new split layer. Therefore, such transitions must be driven by a sustained mismatch between a satellite's capacity and its assigned paradigm, rather than by transient resource fluctuations.

This selection follows the design principles established in Section~\ref{subsec:design_principle}. Consequently, a constellation can support a hybrid architecture where FL and SL operate concurrently across different satellites, or permit individual satellites to transition between paradigms as their hardware availability, mission phases, or connectivity profiles evolve.

\section{Satellite Collaborative Inference}
\label{sec:collaborative_inference}

Distributed training in Section~\ref{sec:distributed_training} optimizes model parameters to satellite data, whereas inference applies the resulting model to new observations, shifting the primary constraint from convergence across repeated rounds to the timely and accurate execution of a single task. Depending on a task's latency requirements and the resources available at the time, inference can be executed onboard a single satellite, distributed across cooperating satellites, or offloaded to the GS. This section examines how the design principles of Section~\ref{subsec:design_principle} translate into each of these inference architectures for SI.

\subsection{Inference Architectures for SI}
\label{subsec:inference_metrics}

Executing trained AI models onboard, i.e., inference, is expected to support a variety of satellite missions, including Earth
observation, satellite communications, and spacecraft health monitoring. For instance, cloud screening can filter cloud-obscured observations before downlink, object detection can identify targets of interest for remote sensing analysis~\cite{pitonak2022cloudsatnet,gu2023learning}, and telemetry anomaly detection can flag abnormal subsystem behavior during spacecraft operations~\cite{hundman2018detecting}. Supporting such a range of tasks under varying latency requirements and available resources motivates more than one way of executing inference. As shown in Fig.~\ref{fig_three_infer}, we organize three inference architectures for SI along where model execution takes place. Local inference~\cite{yang2024toward} confines execution to a single satellite, while distributed inference (comprising multi-satellite~\cite{fang2026nonintrusive} and satellite-GS collaborative inference~\cite{zhao2026osgs}) expands the available resources by involving additional participants, either other satellites or the GS:
\begin{itemize}
\item \textit{Local Inference}: executes the complete model on the satellite that captured the observation, relying solely on its own onboard compute and memory budget.
\item \textit{Multi-Satellite Collaborative Inference}: distributes model execution across satellites connected by ISLs, pooling their onboard resources within the constellation.
\item \textit{Satellite-GS Collaborative Inference}: partitions model execution between the satellite and the GS over SGLs, combining satellite onboard processing with GS's computing resources.
\end{itemize}

The following subsections review the architecture of the three inference paradigms, then systematically analyze and compare the corresponding methods for addressing the deployment challenges of SI identified in Section~\ref{subsec:SI_constraints}.

\subsection{Local Inference for SI}

Local inference executes the complete AI model on the satellite without exchanging data over wireless links. This eliminates link availability as a constraint on inference feasibility, enabling autonomous and time-critical decisions independent of SGLs and ISLs. However, communication independence does not imply resource abundance, where the complete model and its runtime workspace must be accommodated within a single satellite's computing, memory, and energy budget. This subsection first describes the local inference workflow and then examines the methodologies developed to operate within this onboard resource envelope. Table~\ref{tab:local_inference_summary} summarizes the key techniques for local inference reviewed below.

\begin{table*}[t!]
\centering
\caption{Summary of representative techniques for local inference}
\label{tab:local_inference_summary}
\footnotesize 

\resizebox{0.92\textwidth}{!}{%
\begin{minipage}{\textwidth} 
\setlength{\tabcolsep}{3pt}
\newlength{\localInferenceTableWidth}
\setlength{\localInferenceTableWidth}{\dimexpr\textwidth-10\tabcolsep-6\arrayrulewidth\relax}

\begin{tblr}{
  colspec={Q[c,m,wd=0.14\localInferenceTableWidth]
           Q[c,m,wd=0.05\localInferenceTableWidth]
           Q[l,m,wd=0.22\localInferenceTableWidth]
           Q[l,m,wd=0.25\localInferenceTableWidth]
           Q[l,m,wd=0.33\localInferenceTableWidth]},
  colsep=3pt, rowsep=2pt, stretch=1.15,
  row{1}={font=\bfseries, bg=black!8, halign=c, m},
  hline{1,Z}={0.8pt},
  vline{1,Z}={0.8pt},
  vline{2-5}={0.4pt},
  hline{2}={0.6pt},
  hline{8,11}={0.5pt},
  hline{3-7,9-10,12}={2-Z}{wd=0.3pt, fg=black},
}
\textbf{Approach} & \textbf{Ref.} & \textbf{Performance Metrics} & \textbf{Control Variables} & \textbf{Key Contribution} \\
\SetCell[r=6]{c,m} {\textbf{Model Compression and}\\\textbf{ Reconstruction}}
& \cite{cratere2025efficient}
& Accuracy; Latency; Model size
& Pruning ratio; Numerical precision
& Channel pruning; INT8 quantization-aware training \\
& \cite{pitonak2022cloudsatnet}
& Accuracy; Throughput; Board power
& Weight and activation bit widths
& Quantized CNN; FPGA mapping \\
& \cite{grabowski2024squeezing}
& Segmentation IoU; Model size; Latency
& Teacher/student architectures
& nnU-Net-to-U-Net knowledge distillation \\
& \cite{gu2023learning}
& Detection mAP; Parameter count
& Student backbone; Teacher loss weights
& Contextual feature transfer; Multi-teacher distillation \\
& \cite{ali2026lightweight}
& Cloud-masking accuracy; Parameter count
& Tensor structure; Bond dimensions
& Low-rank tensor factorization \\
& \cite{capogrosso2026tinyml}
& Accuracy; Latency; Energy; Memory
& Pruning ratio; Precision; Operator mapping
& Compression-accelerator co-design \\
\SetCell[r=3]{c,m} {\textbf{Model- and Depth-}\\\textbf{Adaptive Inference}}
& \cite{denby2023kodan}
& Filtering precision; Processing time; Data value
& Specialized model; Tile processing
& Geospatial specialization; Context-based model selection \\
& \cite{wang2025lightweight}
& Accuracy-energy efficiency
& Model choice per time slot
& DQN-based model selection \\
& \cite{liu2025spaceexit}
& Detection quality; Throughput
& Exit depth; Uncertainty thresholds
& Uncertainty-guided early exits; Geospatial priors \\
\SetCell[r=2]{c,m} {\textbf{Energy-Aware}\\\textbf{Processor Management}}
& \cite{lyholm2026statistical}
& Modeled processing energy; E2E deadline success probability
& GPU clock frequency
& GPU frequency matched to remaining processing time \\
& \cite{kim2019temperature}
& Modeled mean time to failure; power
& Core mapping; DVFS; Temperature thresholds; Dummy tasks
& Thermal-cycle control for processor reliability \\
\end{tblr}

\par\vspace{3pt}
\scriptsize
IoU: intersection over union; mAP: mean average precision; DQN: deep Q-network;
DVFS: dynamic voltage and frequency scaling.
\end{minipage} 
} 
\end{table*}

\subsubsection{Workflow of Local Inference}

As illustrated in Fig.~\ref{fig_three_infer}, the local inference workflow proceeds through three distinct stages. First, during the pre-deployment phase, candidate models are trained, compressed, and compiled for the target onboard hardware. They are subsequently profiled for operator compatibility, peak memory footprint, accuracy, latency, and energy consumption; only configurations satisfying the task requirements within the onboard resource budget are installed with their execution policies. Second, in the data acquisition stage, the satellite captures sensor observations triggered by mission schedules or dynamic events. These raw observations are preprocessed into structured model inputs (e.g., patch extraction, resizing, and normalization), and routed to the onboard inference runtime. Finally, during the execution stage, the onboard controller dynamically selects a deployed model and its optimal execution configuration based on immediate task requirements, observation context, and real-time resource availability. The satellite executes the forward propagation and post-processes the predictions into task outputs (e.g., class labels, detections, or segmentation masks). Within this workflow, we next examine key methodologies for operating within the onboard resource envelope and for improving local inference efficiency.


\subsubsection{Model Compression and Reconstruction}

Model compression reduces a model's parameter count or numerical precision before deployment to lower its onboard inference workload, trading inference cost against accuracy. In particular, aggressive compression can reduce computation latency, energy consumption, and temperature rise, but may degrade inference accuracy. Four representative compression approaches for local inference in SI address this trade-off, organized below according to whether they modify the original model directly or reconstruct it through auxiliary knowledge.

{\textit{a) Structural and Precision Compression}.} Model pruning reduces onboard computational and memory demands by removing network components that contribute marginally to task performance. Structured pruning~\cite{fang2023depgraph,he2024structured}, which discards entire channels or filters rather than individual weights, is generally preferred for satellite deployment; it yields regular computation patterns that map efficiently onto embedded accelerators, whereas unstructured sparsity often provides negligible practical speedup without dedicated hardware support~\cite{fang2023depgraph}. Quantization instead lowers storage and arithmetic costs by representing weights and activations with lower-bit formats. The achievable bit-width depends jointly on the target hardware's native support for low-precision arithmetic and the specific task's tolerance for reduced numerical resolution. Quantization-aware training is therefore typically used to adapt the model to this reduced precision prior to deployment, rather than quantizing an already-trained model afterward~\cite{liu2025unfairness,chen2025efficientqat}. Because pruning targets structural redundancy while quantization reduces numerical precision, the two techniques address complementary sources of overhead. They are commonly applied in sequence, pruning the network first to establish a compact architecture before quantizing the remaining weights, rather than optimizing both simultaneously from scratch~\cite{qu2025automatic,yu2023boost}. Notably, both techniques must be calibrated against the task's error tolerance. For instance, in CubeSat cloud screening, pruning strength is chosen to limit false cloud detections that would discard useful observations~\cite{cratere2025efficient}, while for onboard cloud classification, quantization bit-widths are tailored to balance classification accuracy against available FPGA resources~\cite{pitonak2022cloudsatnet}.

{\textit{b) Knowledge-Based Model Reconstruction}.} Knowledge distillation transfers knowledge from a larger teacher network to a compact student network during offline training. The computationally intensive burden, i.e., training the teacher and generating its supervision signal~\cite{grabowski2024squeezing}, is strictly confined to the GS. Consequently, onboard execution is restricted entirely to the lightweight student model, decoupling training overhead from inference overhead~\cite{gu2023learning}. The transferred knowledge can take several forms, ranging from the teacher's output logits to its intermediate feature representations. While intermediate representations typically convey richer supervisory information, they incur the cost of requiring structural compatibility between the teacher and student~\cite{hu2023discriminator,lin2025perspective}. As an alternative approach to model compactness, low-rank factorization reduces model storage by approximating large weight tensors as products of lower-dimensional matrices~\cite{ali2026lightweight}. This decomposition substantially curtails both the parameter count and the arithmetic complexity. The degree of compression is directly governed by the rank retained in the approximation~\cite{capogrosso2026tinyml}. Because a lower rank yields a more compact representation at the expense of greater deviation from the original weights, the rank is typically calibrated on a per-layer basis according to each layer's sensitivity to approximation error~\cite{guo2025integrating,xiao2023haloc}. Notably, unlike pruning and quantization, which are cascaded within a single compression pipeline, knowledge distillation and low-rank factorization are generally treated as distinct alternative pathways. Both a distilled student and a factorized network serve as standalone replacements for the original model, rather than to complement one another's compression.

Across the two above categories, compression need not follow a single route. Structural and precision compression modify the deployed model directly, whereas knowledge-based reconstruction produces an independent compact replacement. In practice, a lightweight model obtained through knowledge distillation or factorization can subsequently be further pruned or quantized, cascading these methodologies rather than treating them as mutually exclusive alternatives.

\subsubsection{Model- and Depth-Adaptive Inference}

Adaptive model execution dynamically modulates the runtime behavior of deployed models according to task requirements, observation content, and available resources. This adaptation is achieved either by selecting among multiple candidate models or by adjusting the execution depth of a single model. We systematically investigate both paradigms below.

{\textit{a) Model Selection}.} Model selection dynamically routes tasks among complete networks prepared offline, balancing inference accuracy, latency, and energy consumption. This routing decision is primarily governed by two aspects. First, observation content can guide the selection, since distinct scene types impose varying demands on model capacity. For instance, context-specialized models can be assigned to individual image tiles based on scene content, balancing filtering precision against processing time to maximize the informational value of downlinked observations within constrained onboard budgets~\cite{denby2023kodan}. Such specialization enables domain-specific models, such as those tuned for coastal or urban imagery, to outperform general-purpose networks, at the cost of requiring a distinct model per scene category. Second, the satellite's evolving energy state and task load over the orbit can also govern the choice, as the affordable computation varies with the orbital position: a deep Q-network can select among candidate models of distinct sizes and predictive capabilities at each orbital time slot to optimize a joint accuracy-energy objective~\cite{wang2025lightweight}, treating model selection as a sequential decision in which the choice made at one slot affects the energy available for subsequent ones. A robust system may need to consider these two criteria, favoring a smaller model under scarce resources even for scenes that would otherwise merit a more capable one. Model selection nonetheless incurs a storage penalty, as maintaining multiple candidates onboard renders the candidate pool's size an additional deployment constraint: a larger pool improves coverage across scene types and resource conditions but consumes memory otherwise available to a single, more capable model.

\begin{figure}[t!]
\centering
\includegraphics[width=0.48\textwidth]{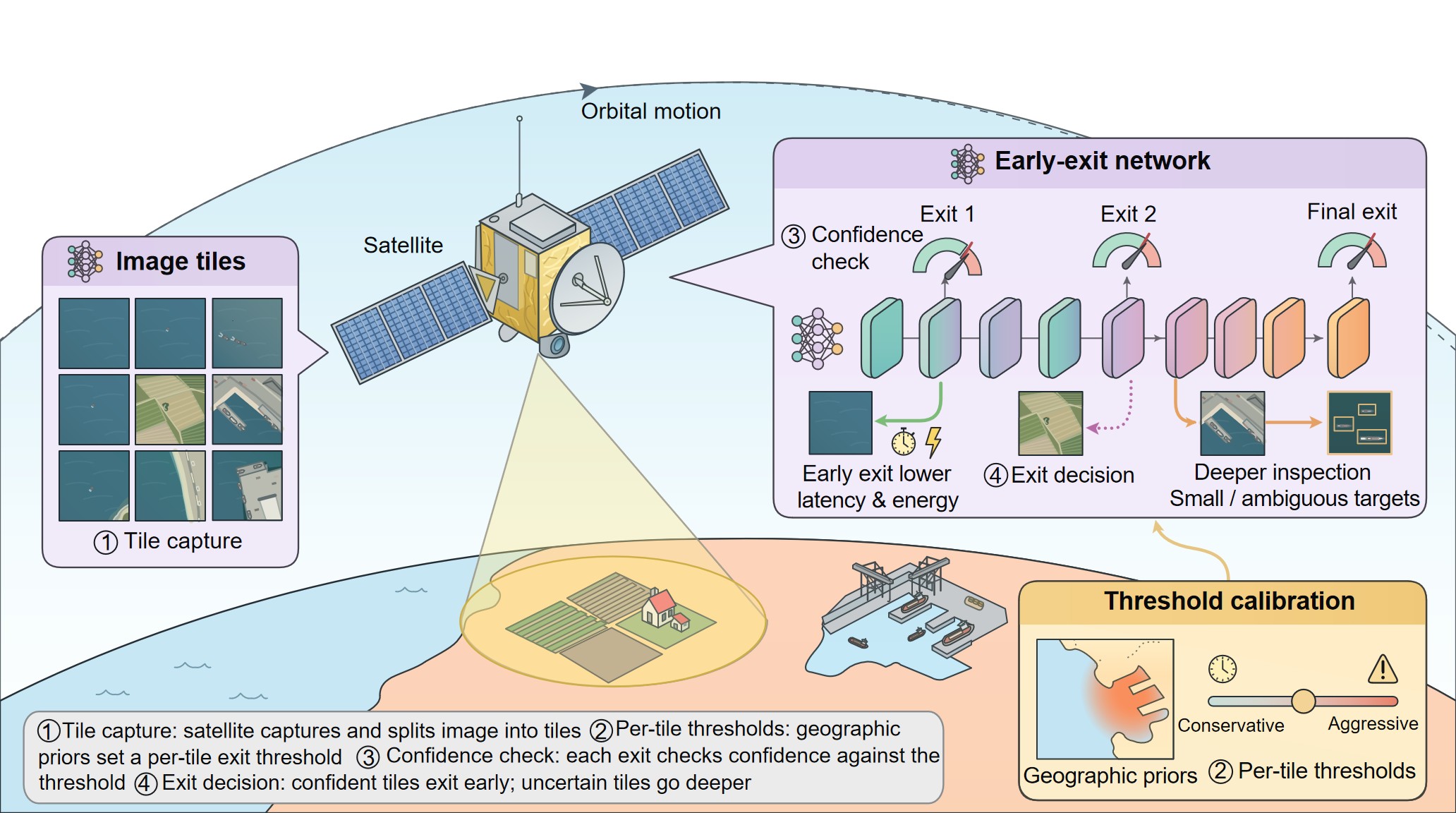}
\centering\caption{Early-exit inference with confidence-based exit decisions for onboard processing.}
\label{fig_early_exit}
\end{figure}

{\textit{b) Early-Exit Inference}.} As shown in Fig.~\ref{fig_early_exit}, early-exit inference~\cite{han2022dynamic,li2023predictive} adaptively controls the execution depth of a deployed model. Lightweight auxiliary classifiers attached at intermediate layers produce a prediction once further computation is judged unnecessary for the current input. Earlier exits reduce processing latency and energy consumption, but may compromise accuracy for small or ambiguous targets requiring deeper features~\cite{li2023predictive,lin2023dynamicdet}. The placement of each exit, together with the confidence threshold required to invoke it, is therefore critical to this approach. These thresholds are calibrated on held-out data, trading the proportion of inputs exited early against the resulting drop in accuracy. Overly conservative thresholds forfeit most of the achievable latency savings, whereas overly aggressive thresholds risk exiting on inputs the network has not yet classified reliably~\cite{tan2021empowering,han2022dynamic}. For onboard object detection, learned uncertainty estimates determine whether an image tile proceeds to deeper detector stages, while visual features and geographic priors refine the exit thresholds. Certain regions or object classes are inherently more likely to warrant deeper inspection than others~\cite{liu2025spaceexit}, and calibrating the thresholds to reflect this heterogeneity directs the network's deeper layers where they are most needed.

\begin{table*}[!t]
\centering
\caption{Summary of representative techniques for multi-satellite collaborative inference}
\label{tab:multi_satellite_inference_summary}
\footnotesize
\setlength{\tabcolsep}{3pt}
\newlength{\multiSatelliteInferenceTableWidth}
\setlength{\multiSatelliteInferenceTableWidth}{\dimexpr\textwidth-10\tabcolsep-6\arrayrulewidth\relax}

\resizebox{0.92\textwidth}{!}{%
\begin{minipage}{\textwidth}
\begin{tblr}{
  colspec={Q[c,m,wd=0.13\multiSatelliteInferenceTableWidth]
           Q[c,m,wd=0.05\multiSatelliteInferenceTableWidth]
           Q[l,m,wd=0.21\multiSatelliteInferenceTableWidth]
           Q[l,m,wd=0.22\multiSatelliteInferenceTableWidth]
           Q[l,m,wd=0.34\multiSatelliteInferenceTableWidth]},
  colsep=3pt, rowsep=2pt, stretch=1.15,
  row{1}={font=\bfseries, bg=black!8, halign=c, m},
  hline{1,Z}={0.8pt},
  vline{1,Z}={0.8pt},
  vline{2-5}={0.4pt},
  hline{2}={0.6pt},
  hline{7,11}={0.5pt},
  hline{3-6,8-10}={2-Z}{wd=0.3pt, fg=black},
  hline{12}={2,5}{wd=0.3pt, fg=black},
}
\textbf{Approach}
& \textbf{Ref.}
& \textbf{Performance Metrics}
& \textbf{Control Variables}
& \textbf{Key Contribution} \\

\SetCell[r=5]{c,m} {\textbf{Resource-Adaptive Model}\\\textbf{Deployment}}
& \cite{xu2023collaborative}
& Throughput; Completion time; Network traffic
& Satellite group; Layer split points
& Neighborhood selection; Runtime prediction; PPO-based partitioning \\

& \cite{peng2026aptsat}
& Task completion rate; Delay; Energy consumption; Resource utilization
& DNN slice boundaries; Offloading destinations; Routes
& Workload-balanced partitioning; SAC-based routing and offloading \\

& \cite{yu2026microservice}
& Successfully processed requests; Deployed microservice count
& Microservice placement; Per-satellite replica counts
& Dependency-aware deployment; Multi-step robust adversarial reinforcement learning \\

& \cite{chen2026hierarchical}
& Throughput; Average task response time
& Task-to-swarm assignment; Partition count; Cut points; Subtask placement
& Swarm-level assignment; Runtime-aware partitioning and node-level offloading \\

& \cite{wang2026spacemoe}
& Token-generation latency under top-$K$ expert activation
& Layer-to-subnet mapping; Expert-to-satellite placement
& Distributed MoE inference architecture;
Ring-based layer placement; Activation-aware expert placement \\

\SetCell[r=4]{c,m} {\textbf{Resilient Feature}\\\textbf{Transfer and Pipelining}}
& \cite{zhang2026communication}
& Inference delay; Activation traffic; Accuracy
& Layer split points; Activation compression ratios
& Learned activation masks; Joint split/compression search on a DAG \\

& \cite{chen2026iapetus}
& Task completion ratio; Latency; Battery draw; Quality loss
& Satellite swarm; Layer placement; Per-layer token retention
& Content-aware quality prediction; Lyapunov-guided compression/offloading planning \\

& \cite{hu2022pipeedge}
& Inference throughput; Memory utilization; Pipeline idle time
& Micro-batch size; Model split points; Worker allocation
& Micro-batch pipelining with communication overlap for heterogeneous edge nodes \\

& \cite{borzunov2023distributed}
& Inference latency; Fault recovery time; Network traffic
& Checkpointing frequency; Routing topology; Activation buffer
& Fault-tolerant distributed inference; Stateful checkpointing for seamless task resumption \\

\SetCell[r=2]{c,m} {\textbf{Energy-Constrained}\\\textbf{Split-and-Exit Control}}
& \cite{guan2024collaborative}
& \SetCell[r=2]{l,m} Task completion rate; Average task latency
& \SetCell[r=2]{l,m} Per-task exit depth; Model split point
& Dynamic programming for joint exit/split selection in multi-satellite systems \\

& \cite{zhang2025satcooper}
&
&
& Offline dynamic programming; Online DRL-based decisions in collaboration \\
\end{tblr}

\par\vspace{3pt}
\scriptsize
PPO: proximal policy optimization; 
MoE: Mixture-of-experts;
DRL: deep reinforcement learning;
DAG: directed acyclic graph;
SAC: soft actor-critic.
\end{minipage}
}
\end{table*}

\subsubsection{Energy-Aware Processor Management}

The intermittent energy provision dictates that a satellite cannot sustainably execute inference workloads at peak computational capacity. Reducing processor frequency conserves energy but exacerbates inference latency, and this tradeoff cannot be resolved through naive frequency reduction. Most inference tasks enforce strict execution deadlines, so an indiscriminately conservative frequency policy risks deadline violations, rendering the computation operationally void regardless of the energy conserved~\cite{bin2025neurobalancer,zhang2024dvfo}. Energy-aware inference control therefore formulates frequency selection as a constrained optimization problem, minimizing energy cost subject to a strict execution deadline rather than treating latency and power as a tradeoff decoupled from task urgency.

Execution time is inherently stochastic, fluctuating with input complexity and instantaneous hardware conditions such that no fixed frequency can offer a deterministic deadline guarantee.
Empirical latency distributions, profiled offline for each candidate frequency, instead characterize how execution time varies across all settings, enabling the selection of the lowest frequency that satisfies the deadline with high probability and minimizing power consumption under this probabilistic constraint~\cite{lyholm2026statistical}. Because this decision must be revisited continuously as task queues and remaining energy evolve over the orbit, frameworks such as Lyapunov optimization or DRL adjust frequency online rather than relying on a single offline computation. Frequency adjustment, however, introduces a longer-term cost that per-task energy minimization alone fails to capture. Cycling a processor repeatedly between high and low operating temperatures accelerates its physical degradation; optimizing exclusively for instantaneous energy efficiency can shorten the payload's operational lifespan. Multicore task mapping and dynamic voltage and frequency scaling (DVFS) can be co-orchestrated to bound the thermal excursions experienced across compute-intensive and idle intervals, rather than letting frequency alone dictate temperature~\cite{kim2019temperature}, substantially improving modeled processor reliability under a fixed power budget.


\subsection{Multi-Satellite Collaborative Inference}\label{subsec:multi_satellite}

The rigid onboard resource constraints identified in Section~\ref{subsec:SI_constraints} impose a fundamental cap on the computing and memory available to any individual satellite. However, as mission complexity escalates, the resource requirements of modern neural network architectures increasingly surpass these limits. Rather than overly compressing a model to fit a single satellite's budget, which risks substantial accuracy degradation, the inference workload can be partitioned across multiple collaborating satellites interconnected via ISLs. This collaborative paradigm effectively combines distributed memory and computational resources to accommodate large-scale models that exceed the capacity of any individual payload. This subsection first introduces the workflow of multi-satellite collaborative inference, as depicted in Fig.~\ref{fig_three_infer}, and subsequently investigates optimization methodologies to enhance its distributed execution efficiency. Table~\ref{tab:multi_satellite_inference_summary} summarizes the key techniques for multi-satellite collaborative inference reviewed below.

\subsubsection{Workflow of Multi-Satellite Collaborative Inference}

Realizing AI model inference across multiple networked satellites proceeds through three steps. First, during model deployment, a trained AI model is partitioned into several sub-models, i.e., partial layers, each mapped to a satellite along an ISL-connected path; model partitioning and placement are jointly determined (e.g., accounting for layer workloads, intermediate feature sizes, onboard computing and memory capacities, and expected link and energy availability), and the resulting sub-models are placed on their assigned satellites together with the corresponding execution and forwarding policies. Second, during pipelined inference, the source satellite preprocesses sensor observations and executes the initial sub-model, forwarding its intermediate features over ISLs to the satellite hosting the subsequent sub-model, a process that repeats until the terminal sub-model yields the final predictions; for a sustained input stream, inputs are grouped into micro-batches, enabling distinct satellites to process successive micro-batches concurrently while preserving the sub-model dependencies within each. Finally, during result feedback, the satellite executing the terminal sub-model returns the predictions to the source satellite for supporting onboard decision-making, or to be stored for later downlink.

Building on this workflow, multi-satellite collaborative inference offers three principal advantages over single-satellite execution. First, aggregating distributed memory and computing resources supports models exceeding individual satellite capacities. This cooperative execution concurrently distributes the computation and energy burden across multiple satellites. Second, confining the forward propagation to the constellation enables inference during GS non-visibility. This orbital isolation effectively eliminates the bandwidth overhead associated with downlinking raw sensor observations. Third, micro-batch pipelining significantly enhances system throughput for continuous data streams, allowing different satellites to process successive inputs concurrently while preserving the sub-model dependencies within each.

\subsubsection{Bottlenecks in Multi-Satellite Collaborative Inference}

Partitioning a model into a pipeline of satellites tightly couples their execution, so that the three SI challenges identified in Section~\ref{subsec:SI_constraints}, each isolated to a single satellite in prior sections, evolve into systemic bottlenecks once inference spans multiple dependent satellites:
\begin{itemize}
\item \textit{Rigid onboard resource constraints} extend beyond any single satellite into a load-balancing problem across the constellation. Each model partition must fit within the computing and memory limits of the satellite hosting it; thus, an uneven partition leaves the faster satellites idle while the slowest one caps the throughput of the entire pipeline.
\item \textit{Dynamic orbital topologies}  expose a vulnerability absent in single-satellite execution. As in NFV service function chain placement~\cite{xia2024efficient}, hosting each sub-model on a single satellite forces features through long multi-hop ISL paths, while replicating it shortens these paths at the cost of onboard memory~\cite{yu2026microservice}. Since ISLs suffer from frequent handovers and pointing errors, a transient link disruption stalls all downstream satellites, turning a single failure into a loss of availability for the whole pipeline.
\item \textit{Intermittent energy provision} constrains the pipeline as a whole rather than any individual satellite. Sustained execution requires every participating satellite to remain powered, but satellites along a multi-hop path do not enter eclipse at the same time; the moment any single satellite exhausts its battery or enters eclipse halts the pipeline, irrespective of the energy still held by the remaining satellites.
\end{itemize}

We categorize representative studies into three research directions targeting these coupled constraints, which are reviewed in the following subsections.


\subsubsection{Resource-Adaptive Model Deployment}

Model deployment mitigates rigid onboard resource constraints by jointly optimizing model partitioning and placement. These coupled decisions determine both hardware feasibility and pipeline execution efficiency. The memory limits constraining this process are practically severe. For instance, ViT-Huge contains 632 million parameters~\cite{dosovitskiy2021image} and requires approximately 1.26 GB for its 16-bit weights alone, exceeding the 1 GB CPU memory limit on experimental platforms like OPS-SAT~\cite{marszk2020ops}. A large-scale model must therefore be partitioned to fit onboard, and each resulting sub-model must in turn be placed on a satellite capable of hosting it without leaving faster satellites idle or slower ones bottlenecking the pipeline.

The joint optimization grows increasingly complex as system dynamics are considered. The baseline scheme reduces this problem to matching sub-model execution times against available satellite compute capacity. Predicted runtimes based on CPU and memory occupancy guide layer partitioning within neighboring satellite groups~\cite{xu2023collaborative}. Hierarchical allocation similarly coordinates task assignment across swarms based on dynamic workload and network conditions~\cite{chen2026hierarchical}. Since model placement determines the physical transmission distance between sub-models, the problem therefore extends to jointly coordinating model partitioning and placement with network routing and offloading decisions. This minimizes latency and energy consumption under strict computing and communication limits~\cite{peng2026aptsat}. Furthermore, modern deployments frequently comprise interdependent microservices rather than a single pipeline; model placement must account for temporally fluctuating task demand. Dependency-aware placement and service replica counts can be optimized via multi-step adversarial reinforcement learning~\cite{yu2026microservice} to maximize the request volume each satellite successfully processes despite demand volatility. Across these diverse methods, the underlying objective remains consistent: keeping every satellite in the pipeline productively occupied rather than idle.

Beyond layer-wise model partitioning, SpaceMoE introduces the first framework for distributed mixture-of-experts (MoE) inference in satellite networks by exploiting expert placement as an additional deployment dimension~\cite{wang2026spacemoe}.
An MoE model retains a sequential layer structure, with each MoE layer comprising multiple parallel expert modules and a gating network that activates only a subset of experts for each token.
Building on this structure, SpaceMoE maps MoE layers to satellite subnets arranged along an orbital ring and distributes each layer's experts across satellites within its subnet.
Its expert-placement strategy assigns frequently activated experts to satellites with lower expected path latency, thereby reducing token-generation latency under time-varying inter-satellite connectivity.

\begin{figure}[t!]
\centering
\includegraphics[width=0.48\textwidth]{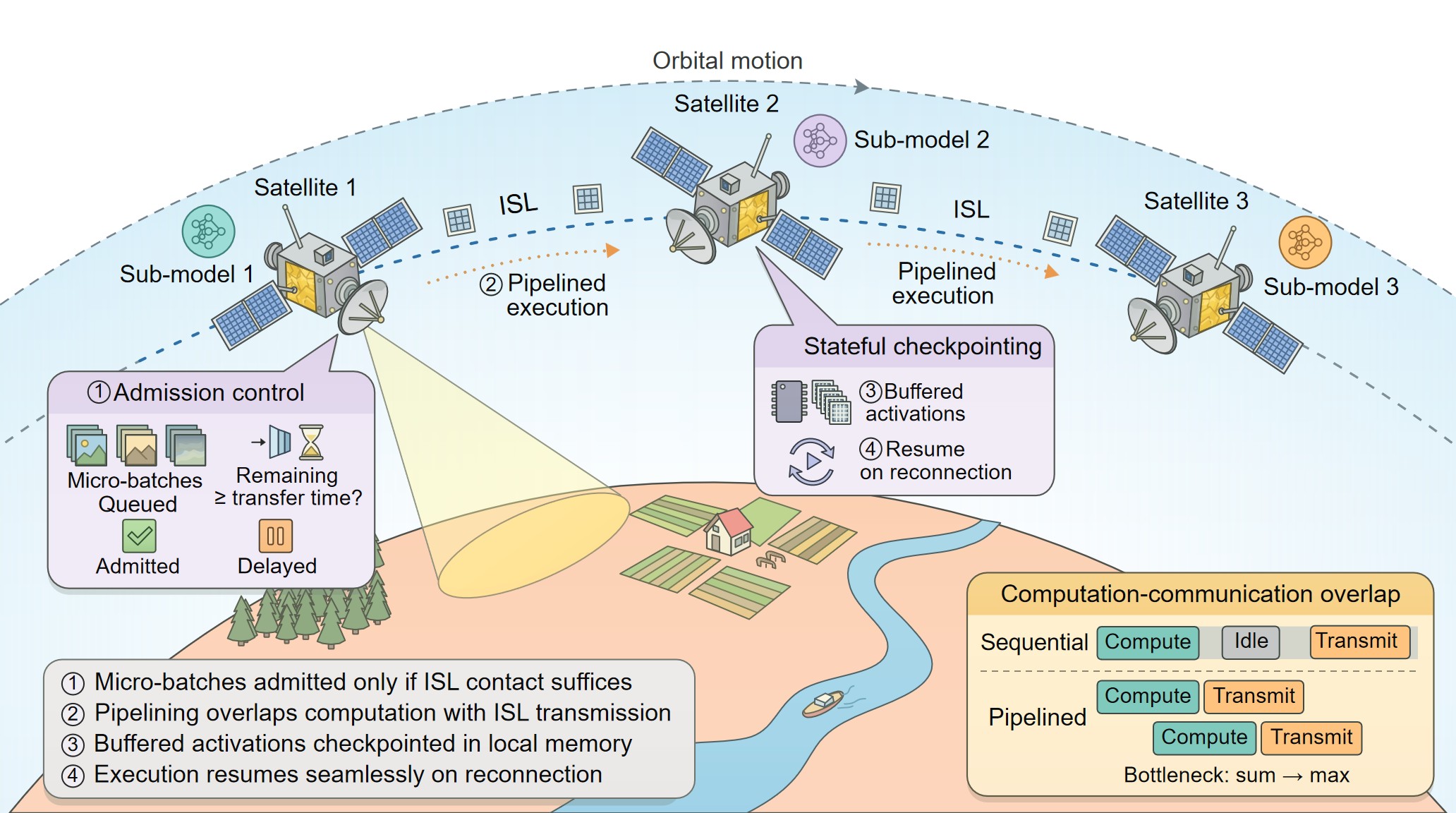}
\centering\caption{Micro-batch pipelining with admission control and checkpointing in multi-satellite collaborative inference.}
\label{fig_pipeline}
\end{figure}

\begin{table*}[!t]
\centering
\caption{Summary of representative techniques for satellite-GS collaborative inference}
\label{tab:satellite_ground_inference_summary}
\footnotesize
\setlength{\tabcolsep}{3pt}
\newlength{\satelliteGroundInferenceTableWidth}
\setlength{\satelliteGroundInferenceTableWidth}{\dimexpr\textwidth-10\tabcolsep-6\arrayrulewidth\relax}

\resizebox{0.92\textwidth}{!}{%
\begin{minipage}{\textwidth}
\begin{tblr}{
  colspec={Q[c,m,wd=0.15\satelliteGroundInferenceTableWidth]
           Q[c,m,wd=0.05\satelliteGroundInferenceTableWidth]
           Q[l,m,wd=0.20\satelliteGroundInferenceTableWidth]
           Q[l,m,wd=0.22\satelliteGroundInferenceTableWidth]
           Q[l,m,wd=0.38\satelliteGroundInferenceTableWidth]},
  colsep=3pt, rowsep=2pt, stretch=1.15,
  row{1}={font=\bfseries, bg=black!8, halign=c, m},
  hline{1,Z}={0.8pt},
  vline{1,Z}={0.8pt},
  vline{2-5}={0.4pt},
  hline{2}={0.6pt},
  hline{4,7}={0.5pt}, 
  hline{3,5,6,8,9}={2-Z}{wd=0.3pt, fg=black}, 
}
\textbf{Approach} & \textbf{Ref.} & \textbf{Performance Metrics} & \textbf{Control Variables} & \textbf{Key Contribution} \\
\SetCell[r=2]{c,m} {\textbf{Contact-Aware}\\\textbf{Model Partitioning}}
& \cite{chen2024energyaware}
& End-to-end latency; Resource estimation error
& Layer execution locations; Processing schedule
& Waiting between contacts enters the deadline constraint for layer placement \\
& \cite{song2026semantic}
& Weighted task success rate; Weighted average latency
& Offloading destination; Semantic split point; Computing allocation
& Visibility masks exclude unavailable GSs; Congestion penalties guide splitting and offloading \\
\SetCell[r=3]{c,m} {\textbf{Downlink-Efficient}\\\textbf{Compression}}
& \cite{zhu2023collaborative}
& HSI classification accuracy; Band-selection time; Energy consumption
& Selected spectral bands; Band count
& Band selection balances task quality and energy while fitting processing and transmission into a contact \\
& \cite{gadre2024adapting}
& Classification accuracy; Reconstruction PSNR; First-contact latency
& Latent size and precision; Receiver configuration
& Channel-aware encoding limits payloads; Doppler-aware GS reception improves packet recovery \\
& \cite{furutanpey2025fool}
& Bitrate; Detection quality; Encoder throughput
& Codec configuration; Tiling; Rate-distortion operating point
& Codec profiling balances encoding before a contact against coded volume; Inter-tile coding lowers bitrate \\
\SetCell[r=3]{c,m} {\textbf{GS-Access-Aware}\\\textbf{Task Coordination}}
& \cite{yao2025leoedge}
& Inference latency; Data-delivery latency; Model accuracy
& Task placement; Direct or relayed delivery; Result-return path
& Remaining contact time guides direct transfer, Neighbor relaying, and local fallback \\
& \cite{tao2024serval}
& Query response latency; Image-delivery latency
& Query-filter placement; Processing and downlink priorities
& Ground precomputation prepares query filtering; Priority queues allocate each contact to latency-sensitive results first \\
& \cite{zhang2024resource}
& Normalized counting error; Downlink volume
& Tile size; Similarity and confidence thresholds; Downlink tile set
& Deduplication and tile selection under a downlink budget limit ground reprocessing traffic during each contact \\
\end{tblr}

\par\vspace{3pt}
\scriptsize
HSI: hyperspectral image; PSNR: peak signal-to-noise ratio.
First-contact latency measures waiting for the next satellite--ground contact rather than end-to-end inference.
\end{minipage}
}
\end{table*}

\subsubsection{Resilient Feature Transfer and Pipelining}

Highly dynamic orbital topologies severely constrain the tightly coupled forward-passing sequences of pipeline execution. Since ISLs connecting adjacent satellites suffer from frequent handovers and pointing errors, these vulnerabilities can interrupt intermediate feature transfers, inevitably causing cascading pipeline stalls and severely degrading service availability.

{\textit{a) Feature Compression}.} Dynamic orbital topologies impose severe bandwidth and temporal constraints on feature transmission. Specifically, the capacity and remaining duration of ISLs fluctuate continuously as orbital geometry evolves~\cite{nam2026prediction,li2026spatiotemporal}. Since high-dimensional intermediate features easily saturate these fragile links, features must be compressed before being forwarded to the next satellite in the pipeline. However, discarding task-relevant information inherently risks degrading overall inference accuracy. A fixed compression ratio cannot consistently resolve this dynamic tradeoff: when the link is stable with sufficient capacity, compression should be relaxed to preserve more information, whereas as capacity tightens or the link is disrupted, more aggressive compression becomes necessary to ensure successful delivery within the shrinking contact window. Therefore, recent methods leverage learned activation masks to sparsify transmitted features, while a shortest-path search over a directed acyclic graph jointly determines layer split points and compression ratios~\cite{zhang2026communication}. Furthermore, the compression level also depends on input content and previously discarded information. Content-aware quality prediction captures these cumulative effects, enabling the joint optimization of token retention and model placement, and the cooperating satellite swarm through Lyapunov-guided planning under the constellation's contact windows~\cite{chen2026iapetus}.

{\textit{b) Micro-batch Pipelining}.} The sequential dependencies of multi-satellite collaborative inference force the processing satellite to remain idle while waiting for intermediate features to arrive. As illustrated in Fig.~\ref{fig_pipeline}, micro-batch pipelining mitigates this inefficiency by overlapping computation with communication, allowing a satellite to compute the next forward propagation while still transmitting the current one. This temporal overlap shifts the system bottleneck from the sum of computation and communication times to their maximum~\cite{hu2022pipeedge}. However, pipelining alone cannot protect against sudden link disruptions, which can cause in-flight micro-batches to be lost, wasting prior computation and incurring high energy costs. To prevent this, topology-aware admission control calculates the remaining link contact duration before admitting new tasks. A micro-batch enters the pipeline only if its predicted transmission window guarantees successful downstream delivery; otherwise, it is delayed until the orbital geometry provides a stable connection. Because physical links may still fail mid-transfer despite this admission control, a stateful checkpointing mechanism saves buffered activations in local memory, ensuring that execution can resume seamlessly once connectivity is restored~\cite{borzunov2023distributed}.

\subsubsection{Energy-Constrained Split-and-Exit Control}

The intermittent solar-energy provision and finite onboard battery capacity prevent a multi-satellite inference pipeline from sustaining the deepest execution path throughout an orbital cycle. Adaptive inference depth therefore provides an energy-aware control mechanism: when the participating satellites have sufficient energy, the inference pass can proceed through deeper layers to obtain higher inference accuracy; when a satellite enters eclipse, or its battery state becomes constrained, the pipeline can terminate at an earlier exit, reducing the computation and communication energy required by the remaining layers. This adjustment is coupled with model partitioning; the model split point determines how the executed layers are distributed across satellites, whereas the exit point determines where the forward propagation terminates. For dynamic task streams in LEO constellations, joint optimization of split and exit points balances inference accuracy against energy consumption under strict task deadlines~\cite{guan2024collaborative}. In multi-satellite LEO collaboration, coordinated scheduling aligns these two decisions to reduce average latency under energy constraints~\cite{zhang2025satcooper}. Although these existing methods account for energy consumption, they typically rely on simplified energy models rather than incorporating the fine-grained, predicted solar supply and battery state of participating LEO satellites.

\subsection{Satellite-GS Collaborative Inference}

Even after combining the computing and memory resources of multiple satellites through the multi-satellite collaboration in Section~\ref{subsec:multi_satellite}, some models may remain too demanding for any ISL-connected group of satellites to execute collectively. In such cases, inference can leverage the more powerful computing capacity of GS, whose hardware is not subject to the SWaP constraints of onboard platforms. This subsection first presents the workflow of
satellite-GS collaborative inference and then examines the design methodologies for operating it under the constraints of the SGLs.
Table~\ref{tab:satellite_ground_inference_summary} summarizes the key techniques for satellite-GS collaborative inference reviewed below.

\subsubsection{Workflow of Satellite-GS Collaborative Inference}

Recall Fig.~\ref{fig_three_infer}, satellite-GS collaborative inference partitions model execution between satellites and the GS. Given their asymmetric computing capacities, a typical deployment places a shallow sub-model at the satellite and assigns the remaining deeper layers to a GS. Given this model partitioning, inference proceeds through three steps. First, during satellite-side inference, the satellite preprocesses the sensor observations and executes the forward propagation for model layers up to the selected model split point, producing the intermediate features to be transmitted. Second, during downlink feature transmission, the satellite sends the intermediate features to the GS over an SGL during an available contact window, buffering them until a suitable window arises if immediate transmission is unavailable. Finally, during GS-side inference, the GS executes the remaining model layers with the received features and converts the predictions into task outputs. These outputs can be delivered directly to GS-side applications, or returned to the satellite for satellite-side decision-making.

While powerful GS can support models beyond the computing capacity of accessible satellites, exploiting this capacity requires transmitting features over constrained SGLs. Atmospheric attenuation can degrade high-frequency radio links, while cloud blockage can interrupt optical links, and for satellite-GS path lengths of 500 to 2,000 km, the one-way propagation delay alone is approximately 1.7 to 6.7~ms. The rapid orbital motion of satellites makes SGL availability intermittent: LEO Earth-observation satellites have only four to six contacts with a GS per day, with each lasting up to about ten minutes~\cite{tao2024serval}. These limited contact windows restrict the transmission timing and volume of intermediate features. We summarize three optimization approaches for improving inference efficiency under such SGL constraints in the following subsections.

\begin{table*}[!t]
\centering
\caption{Comparison of inference paradigms for SI}
\label{tab:inference_comparison}
\small

\resizebox{0.87\textwidth}{!}{%
\begin{tblr}{
  width=\textwidth,
  colspec={X[1.2,c,m] X[1.2,c,m] X[1.3,c,m] X[1.4,c,m]}, 
  colsep=4pt, rowsep=2.2pt, stretch=1.15,
  row{1}={font=\bfseries, bg=black!8, halign=c, m},
  column{1}={font=\bfseries}, 
  hline{1,Z}={0.8pt},
  vline{1,Z}={0.8pt},
  vline{2-4}={0.4pt},
  hline{2}={0.6pt},
  hline{3-Z}={wd=0.3pt, fg=black},
}
Dimension & Local & Multi-Satellite & Satellite-GS \\
Latency sources & Onboard processing & Processing + ISLs & Processing + SGLs + waiting \\
Model capacity & Single-satellite resources & Pooled orbital resources & Ground-assisted resources \\
Communication dependency & None during execution & ISLs & SGLs \\
Service continuity & Contact-independent & ISL-path dependent & Contact-window limited \\
Failure exposure & Local hardware & Participating nodes/links & Satellite/ground access \\
Accuracy factors & Model and compression & Model, depth, features & Model and received data \\
Onboard energy use & Local computation & Computation + ISLs & Onboard processing + SGLs \\
\end{tblr}%
}
\end{table*}

\subsubsection{Contact-Aware Model Partitioning}

\begin{figure}[t!]
\centering
\includegraphics[width=0.48\textwidth]{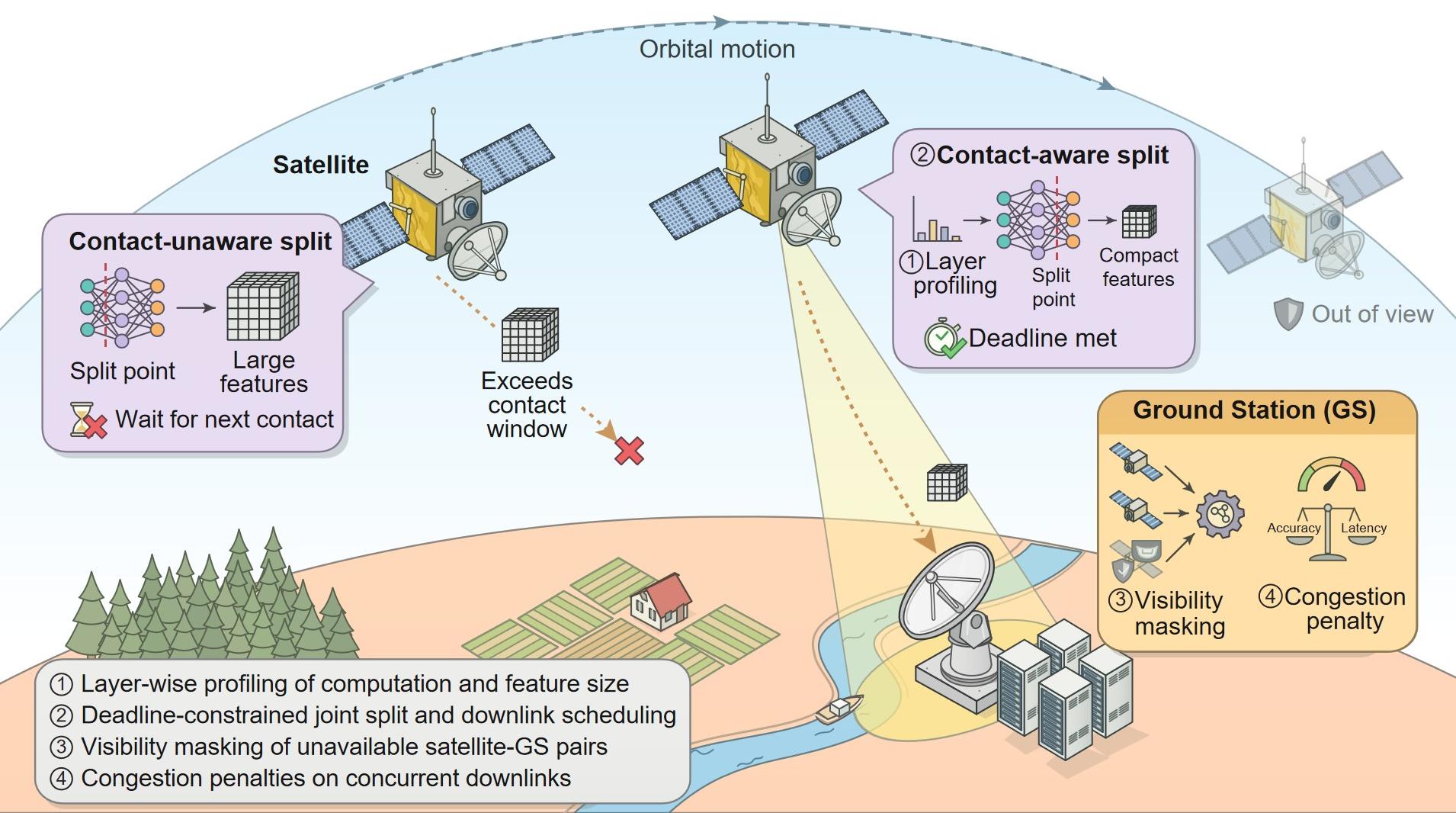}
\centering\caption{Contact-aware model partitioning in satellite-GS collaborative inference.}
\label{fig_contact_aware_split}
\end{figure}

Selecting a model split point without considering satellite–GS visibility risks wasting the contact windows. Specifically, satellite-executed layers chosen solely to minimize computation may produce intermediate outputs that exceed the capacity of a single transmission, or fail to reach the GS before the satellite moves out of view; in either case, the outcome forces the task to wait for the next contact and defeats the purpose of satellite–GS collaboration~\cite{nan2026robust,liu2025efficient}. As depicted in Fig.~\ref{fig_contact_aware_split}, contact-aware model partitioning and scheduling avoid this by coordinating satellite execution and downlink transmission with available contact windows. Rather than being fixed in isolation and later checked against latency constraints, the model split point is determined jointly with subsequent transmission and GS execution. For a single inference task, layer-specific computational profiles guide split point selection to satisfy end-to-end deadlines that account for inter-contact waiting times~\cite{chen2024energyaware}, balancing satellite processing workloads against the transmission delays of large intermediate features. When multiple satellites compete for a limited satellite-GS connection, this single-task formulation must be extended into a joint control problem: cross-satellite model split and downlink decisions are coordinated to balance inference accuracy against latency, excluding destinations without current visibility through masking and discouraging excessive concurrent transmissions through congestion penalties~\cite{song2026semantic}.

\subsubsection{Downlink-Efficient Compression}

While contact-aware scheduling optimizes transmission timing, the sheer data volume of each transmission poses an orthogonal bottleneck. Downlink-efficient compression mitigates this by minimizing the transmitted data volume within limited contact windows through two complementary paradigms: feature selection (optimizing content) and encoding design (optimizing representation). Feature selection filters out redundant information prior to encoding, preserving only task-critical semantics for GS execution. For instance, in hyperspectral classification, spectral-band selection extracts highly informative wavelength channels to balance classification accuracy against satellite processing overhead, ensuring the combined processing and downlink delay satisfies the contact-window duration~\cite{zhu2023collaborative}. Conversely, encoding design dictates the data representation format, with optimization objectives tailored to channel conditions. Over error-prone low-rate SGLs, robustness is prioritized: learned encoders map images into representations resilient to channel error patterns, preserving task-relevant semantics despite bit corruptions~\cite{gadre2024adapting}. Under reliable links where contact duration dictates the transmission bottleneck, the focus shifts to encoding efficiency: offline profiling of codec and tiling configurations determines the maximum imagery encodable and transmittable per contact, while inter-tile coding exploits spatial correlations to reduce payloads, balancing compression ratio against encoding throughput~\cite{furutanpey2025fool}. Ultimately, whether optimizing for robustness or efficiency, both encoding and feature selection converge on an identical objective: maximizing the delivery of task-relevant semantics within stringent contact windows.
For satellite-GS inference that fuses multiple observations, longer channel codewords improve decoding reliability but consume more transmission time, potentially reducing the number of observations delivered before the inference deadline. This reliability--observation tradeoff~\cite{wang2026ultralola} motivates jointly optimizing observation selection, feature compression, and transmission-time allocation within the satellite-GS contact window.

\subsubsection{GS-Access-Aware Task Coordination}

The intermittent satellite-GS links inherent to LEO networks prevent all features from being transmitted immediately, requiring GS-access-aware coordination to manage the resulting task streams dynamically: this approach addresses SGL constraints from a task-centric perspective, tailoring delivery paths and transmission priorities to available GS visibility windows and application-specific demands. Delivery path selection hinges on the remaining contact duration: by weighing transmission delays against available windows, routing mechanisms determine whether intermediate features should be directly transmitted, forwarded via inter-satellite links, or processed through a local fallback when GS access is temporarily infeasible.
For delay-tolerant tasks, mobile carrying provides an additional delivery option by exploiting satellite motion to transport buffered data~\cite{wang2021mobilecarrying}.
Applying this mechanism to inference features or results requires accounting for task deadlines and onboard storage limits.
For closed-loop tasks requiring GS feedback, this evaluation must be iterative: as visibility evolves, initial path assumptions are continuously updated to maintain optimal return routes despite rapid orbital motion~\cite{yao2025leoedge}. When downlink capacity cannot satisfy concurrent data demands, task prioritization resolves scheduling conflicts by leveraging either query urgency or inference confidence. For urgency-driven paradigms, priority queues deliver imagery or compact results matching latency-sensitive queries before other data during each contact~\cite{tao2024serval}. For confidence-driven paradigms, the reliability of satellite predictions instead dictates the necessity of GS refinement: after pruning redundant image tiles, a dynamic confidence threshold is calibrated against the downlink budget to select the remaining tiles for transmission, with the resulting GS-refined counts fused with the accepted satellite-generated counts to minimize overall counting error~\cite{zhang2024resource}.

\subsection{Paradigm Comparison and Selection}

\begin{figure*}[t!]
\centering
\includegraphics[width=0.70\textwidth]{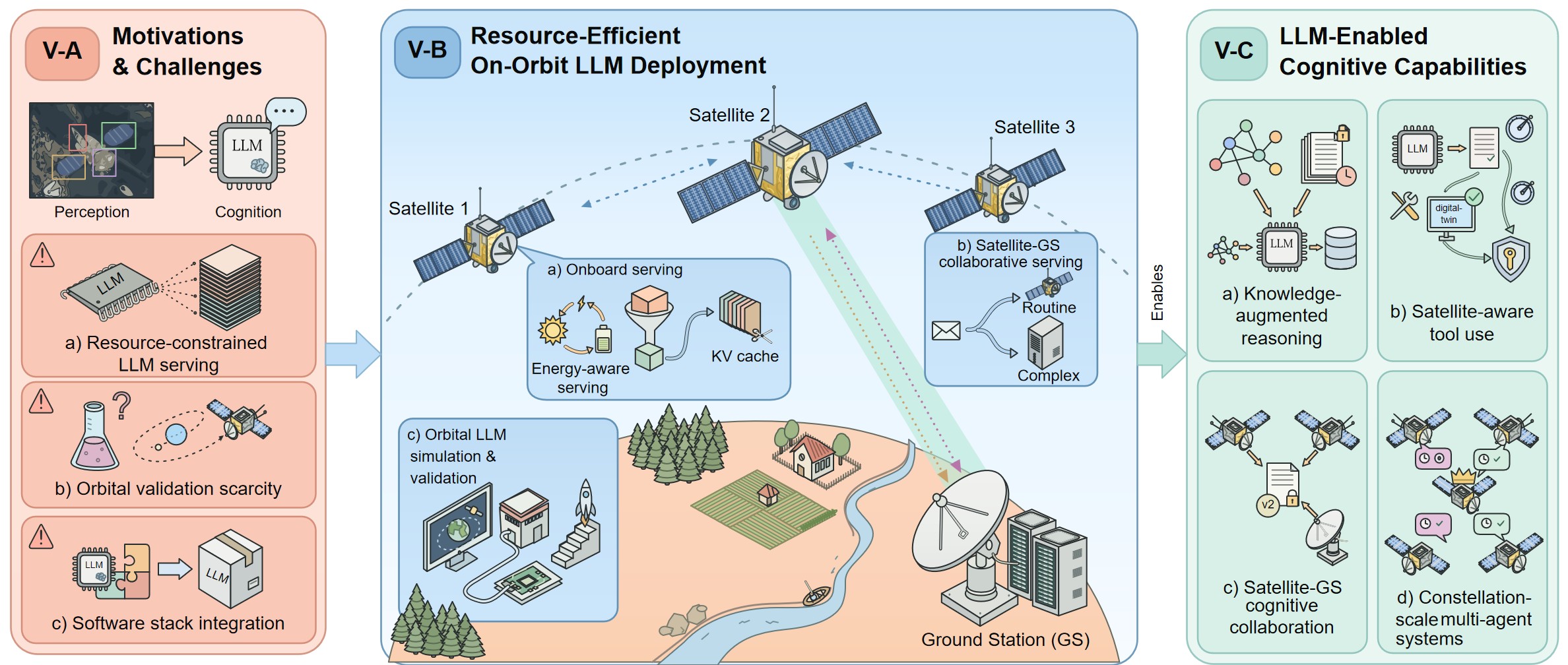}
\centering\caption{On-orbit LLM adaptation for SI, spanning resource-efficient deployment and LLM-enabled cognitive capabilities.}
\label{fig_LLM_overview}
\end{figure*}

The above three inference paradigms pursue the same fundamental objective: satisfying task requirements under constrained satellite resources. However, they differ in computing load distribution and communication reliance, as summarized in Table~\ref{tab:inference_comparison}. Consequently, the optimal paradigm is inherently dynamic, dictated by the deployed model, task workloads, and real-time communication conditions during execution. These differences motivate adaptive selection among the three paradigms rather than committing to one in advance, evaluating candidate deployments by jointly analyzing profiled model accuracy, resource overhead, and predicted communication availability, with latency estimates accounting for computation, communication, contact waiting, and any required result return. Local inference is optimal when the deployed model already meets the required accuracy and deadline within satellite resource budgets, avoiding unnecessary external communication delay; multi-satellite inference becomes preferable once additional computing or memory is needed and an ISL-connected satellite group can sustain the distributed execution within the task budget; and satellite-GS inference is preferable when ground resources support a model beyond what the constellation can host, or reduce satellite computation enough to justify the transmission and contact-waiting costs it incurs. However, selecting among these architectures incurs non-trivial overhead. While mechanisms like early exits and resource control can absorb resource variations within whichever architecture is already running, switching architectures requires verifying model availability and accommodating the associated loading, migration, or data-transfer costs. Therefore, the expected performance benefit must outweigh the costs while preserving task feasibility.

This adaptive selection follows the resource, topology, and energy
considerations in Section~\ref{subsec:design_principle}.
Sustained performance degradation under the selected deployment should instead motivate model updating through
the distributed training mechanisms in Section~\ref{sec:distributed_training}.

\section{On-Orbit Adaptation of LLMs}\label{sec:LLM_adaptation}

Sections~\ref{sec:distributed_training} and~\ref{sec:collaborative_inference} established distributed training and collaborative inference frameworks for conventional deep neural networks across LEO constellations. Future SI increasingly demands cognitive capabilities beyond task-specific models for emerging applications such as autonomous mission planning, onboard anomaly diagnosis, and collaborative constellation control. LLMs offer a unified foundation for these capabilities through semantic understanding, long-context knowledge integration, and multi-step reasoning~\cite{bai2025longbenchv2,wei2022chain}, and LLM-based agents extend this foundation toward autonomous operation through retrieval, tool invocation, memory, and workflow orchestration~\cite{wu2025agentic,yin2024lumos}. Nevertheless, deploying LLMs in orbit introduces challenges beyond those of conventional satellite AI. Their substantially higher demands on memory and computation, coupled with the stringent SWaP constraints of LEO satellites, motivate new paradigms for resource-efficient on-orbit LLM deployment and LLM-enabled SI.


As shown in Fig.~\ref{fig_LLM_overview}, in this section, we first motivate the shift from perception-oriented models toward LLM-enabled cognitive satellite systems and identify the key challenges for onboard LLM deployment, including resource-constrained LLM serving, orbital validation scarcity, and software stack integration (Section~\ref{subsec::moti_llm}). Then, we examine resource-efficient on-orbit LLM deployment, including onboard serving, satellite-GS collaborative serving, and orbital LLM simulation and validation (Section~\ref{subsec:llm_serving}). Finally, we investigate how deployed LLMs enable advanced SI through knowledge-augmented reasoning, satellite-aware tool use, satellite-GS cognitive collaboration, and constellation-scale multi-agent systems (Section~\ref{subsec::llm_power_si}).

\subsection{Motivations and Challenges}
\label{subsec::moti_llm}

\subsubsection{Why LLMs for SI}

Existing onboard AI systems~\cite{denby2023kodan,zhang2024resource,du2025earthplus} primarily support perception-oriented tasks, such as image classification, object detection, and change detection. Although these models can efficiently extract information from sensor data, they are typically designed for predefined tasks and provide limited support for knowledge-intensive reasoning, mission-level planning, and interaction with heterogeneous software tools. Future satellite systems are expected not only to perceive the environment but also to interpret telemetry and mission instructions, diagnose operational events, coordinate onboard resources, and adapt their actions to changing mission conditions. These emerging requirements naturally position LLMs as the cognitive foundation of SI. 
LLMs~\cite{lin2025pushing,zhang2026resource,lin2026redundancy,fang2026sibyl,lin2026hsplitlora} are not intended to replace specialized perception models, control algorithms, or verified aerospace software. Instead, they serve as a cognitive layer that interprets information, invokes appropriate tools, and coordinates task execution. The motivation for onboard LLMs can be summarized through the following three capability requirements:


\textit{a) Knowledge-Augmented Cognitive Reasoning.}
LEO satellite operations require decisions based on multiple heterogeneous sources of information, including orbital dynamics, satellite telemetry, mission objectives, spectrum regulations, and historical operational experience~\cite{tao2024serval,hundman2018detecting,chamberlain2024facilitating}. Conventional task-specific models are generally optimized for fixed input–output mappings and cannot easily integrate such heterogeneous information within a unified decision process. For example, anomaly diagnosis may require jointly analyzing telemetry changes with orbital conditions, subsystem dependencies, and maintenance records. Similarly, observation planning must jointly consider target priorities, visibility windows, payload constraints, communication opportunities, and available energy~\cite{tao2024serval,zhai2024seco,liu2024inorbit}. By combining domain adaptation with external knowledge retrieval, LLMs can interpret these information sources and generate context-aware recommendations. This capability allows onboard AI to move beyond isolated perception results toward mission-level understanding and decision support.

\textit{b) Satellite-Aware Tool Use and Workflow Orchestration.}
Satellite operations depend on specialized software tools, including orbit propagators, digital twins, payload schedulers, routing algorithms, spectrum management modules, and fault-diagnosis systems~\cite{lu2025opensn,liao2026goal,levinson2025optimal,deng2024satellites,muscettola1998remote}. These tools provide reliable domain-specific functions, but coordinating them often requires manually designed workflows or continuous intervention from ground operators. An LLM-based agent can translate high-level mission objectives into a sequence of executable subtasks, select appropriate tools, provide the required inputs, and integrate their outputs into a coherent operational plan~\cite{shen2023hugginggpt,schick2023toolformer}. For instance, an observation request may require the agent to determine orbital visibility, evaluate payload and energy availability, schedule data acquisition, select a communication route, and plan the subsequent downlink. Onboard LLMs can coordinate these steps locally when GS connectivity is unavailable. Combined with command validation, rule-based constraints, and human oversight for safety-critical operations, such tool-enabled orchestration can reduce operational workload while improving the flexibility of mission execution.

\textit{c) Autonomous Constellation Coordination.} 
Future LEO constellations may consist of thousands of satellites operating under time-varying connectivity, heterogeneous resource availability, and rapidly changing mission demands~\cite{basak2025leocraft,gao2025topology}, rendering continuous human supervision of every satellite and every decision difficult to scale. Onboard LLMs can support distributed coordination by allowing satellites to interpret shared mission goals and summarize their local states, then negotiate task assignments and exchange this high-level semantic information rather than raw observations or complete telemetry streams~\cite{nakatsuji2025act,wang2025beyond,zheng2024semantic}. This capability is particularly useful when satellites must coordinate observation, computation, routing, or energy use under limited ISLs and SGLs. A hierarchical architecture can further combine lightweight onboard LLMs for time-sensitive local decisions with more capable GS-based LLMs for global planning and long-term knowledge management. Such satellite-GS collaboration provides a practical path toward increasingly autonomous constellations while retaining supervision for complex or safety-critical decisions.

\subsubsection{Gaps Between LLMs and Satellite Operations}

While LLMs can offer unprecedented cognitive capabilities for SI, deploying them in orbit is substantially more challenging than conventional deep neural networks. Beyond the resource limitations of satellite platforms, LLMs also impose distinct requirements on inference serving, system validation, and integration with satellite software systems. Three research gaps are summarized as follows.

\begin{table*}[!t]
\centering
\caption{Summary of representative techniques for onboard LLM serving}
\label{tab:onboard_llm_serving}
\footnotesize
\setlength{\tabcolsep}{3pt}
\newlength{\onboardLLMServingTableWidth}
\setlength{\onboardLLMServingTableWidth}{\dimexpr\textwidth-10\tabcolsep-6\arrayrulewidth\relax}

\resizebox{0.92\textwidth}{!}{%
\begin{minipage}{\textwidth}
\begin{tblr}{
  colspec={Q[c,m,wd=0.15\onboardLLMServingTableWidth]
           Q[c,m,wd=0.05\onboardLLMServingTableWidth]
           Q[l,m,wd=0.20\onboardLLMServingTableWidth]
           Q[l,m,wd=0.22\onboardLLMServingTableWidth]
           Q[l,m,wd=0.38\onboardLLMServingTableWidth]},
  colsep=3pt, rowsep=2pt, stretch=1.15,
  row{1}={font=\bfseries, bg=black!8, halign=c, m},
  hline{1,Z}={0.8pt},
  vline{1,Z}={0.8pt},
  vline{2-5}={0.4pt},
  hline{2}={0.6pt},
  hline{8,13}={0.5pt}, 
  hline{3-7,9-12,14-16}={2-Z}{wd=0.3pt, fg=black}, 
}
\textbf{Approach} & \textbf{Ref.} & \textbf{Performance Metrics} & \textbf{Control Variables} & \textbf{Key Contribution} \\
\SetCell[r=6]{c,m} {\textbf{Hardware-Aware}\\\textbf{Model Compression}}
& \cite{frantar2023optq}
& Perplexity; Zero-shot accuracy; Speedup
& Weight bit-width; Quantization group size
& Second-order weight-only quantization reducing weight memory traffic during decoding \\
& \cite{lin2024awq}
& Perplexity; Accuracy; On-device speedup
& Per-channel scaling factors; Weight bit-width
& Activation-aware weight-only quantization protecting salient channels for on-device LLMs \\
& \cite{ling2024slimgpt}
& Perplexity; Zero-shot accuracy
& Layer-wise pruning ratio; Pruned heads and FFN channels
& Structured pruning of attention heads and FFN channels with layer-wise non-uniform ratios \\
& \cite{gu2024minillm}
& Response quality; Calibration; Exposure bias
& Student model size; Distillation objective
& Reverse-KL distillation transferring generative capability to a compact student model \\
& \cite{chen2025safety}
& Safety risk; Task utility
& Quantization method and bit-width; Safety-patching updates
& Quantization-aware safety patching restoring alignment degraded by quantization \\
& \cite{zhou2025qpruner}
& Accuracy; Memory usage
& Structured pruning ratio; Layer-wise bit-width
& Mixed-precision quantization of structurally pruned LLMs via layer-wise bit allocation \\
\SetCell[r=5]{c,m} {\textbf{KV-Cache and}\\\textbf{Long-Context Management}}
& \cite{liu2024kivi}
& Accuracy; Peak memory; Throughput
& Key/value bit-width; Per-channel key and per-token value grouping
& Asymmetric 2-bit KV-cache quantization reducing per-entry storage \\
& \cite{zhang2023h2o}
& Accuracy; Throughput; KV-cache memory
& KV-cache budget; Heavy-hitter/recent token ratio
& Heavy-hitter token eviction retaining influential KV states under a fixed budget \\
& \cite{xiao2024efficient}
& Streaming perplexity; Decoding speedup
& Number of attention-sink tokens; KV window size
& Attention sinks with a rolling KV window bounding cache growth for streaming inputs \\
& \cite{packer2023memgpt}
& Document QA accuracy; Multi-session consistency
& Main-context contents; External memory paging
& Virtual context management paging and summarizing older states beyond the context window \\
& \cite{zheng2024sglang}
& Throughput; Latency
& Prefix cache (radix tree); Cache-aware scheduling
& RadixAttention reusing KV caches of shared prompt prefixes across requests \\
\SetCell[r=4]{c,m} {\textbf{Energy-Aware}\\\textbf{Serving}}
& \cite{samsi2023words}
& Energy per token and per response; Inference latency
& Model size; Batch size; Generation length; GPU power cap
& Energy profiling of LLM inference across model sizes, generation lengths, and power caps \\
& \cite{patel2024power}
& Power consumption; Latency; Throughput
& GPU frequency; Power capping
& Phase-aware power capping and frequency locking reducing inference power with limited performance loss \\
& \cite{patel2024splitwise}
& Throughput; Cost; Power
& Prefill/decoding machine pools; Hardware and power per phase
& Phase-split execution exploiting the lower power demand of decoding to raise throughput under power budgets \\
& \cite{stojkovic2025dynamollm}
& Energy; SLO attainment
& Instance count; Model parallelism; GPU frequency
& Energy-minimizing reconfiguration of instances, parallelism, and GPU frequency under SLOs \\
\end{tblr}

\par\vspace{3pt}
\scriptsize
FFN: feed-forward network; KV: key-value; QA: question answering; SLO: service-level objective.
\end{minipage}
}
\end{table*}

\textit{a) Resource-Constrained LLM Serving.}
Unlike conventional deep neural networks, which typically perform a fixed forward pass, LLMs generate outputs autoregressively via repeated model execution and continuously growing key-value (KV) caches, incurring substantial runtime memory, computation, and energy costs~\cite{yu2022orca,kwon2023pagedattention,patel2024power}. Even lightweight LLMs typically contain billions of parameters, placing considerable pressure on the limited resources of onboard commercial off-the-shelf AI platforms, such as the NVIDIA Jetson AGX Orin~\cite{song2024powerinfer}. Moreover, rapidly varying orbital conditions and mission workloads necessitate adaptive onboard serving policies that jointly consider connectivity, available resources, latency requirements, and mission priorities~\cite{liu2026joint}. At the constellation level, model deployment and inference tasks must also be coordinated across satellites and GS infrastructure~\cite{liu2026joint,yao2025leoedge}. Supporting LLM inference under these constraints requires satellite-specific serving mechanisms that jointly optimize model execution and time-varying satellite-GS resources.

\textit{b) Orbital Validation Scarcity.}
Unlike terrestrial LLM deployment, on-orbit LLM systems operate under time-varying orbital dynamics, intermittent connectivity, heterogeneous onboard hardware, and dynamic mission workloads. These coupled factors make conventional LLM benchmarks, which typically evaluate model quality and serving performance on static computing platforms, insufficient for characterizing on-orbit behavior~\cite{zhong2024distserve,sun2024llumnix,wang2025rhone}. Reliable evaluation requires orbital-aware frameworks that jointly model satellite trajectories and connectivity, onboard hardware, mission workloads, and LLM serving behavior~\cite{lu2025opensn,wang2025rhone,zhong2024distserve}. However, integrated simulation and validation platforms with these capabilities remain lacking. In-orbit experiments are also costly, infrequent, and difficult to reproduce, limiting their use for systematic design exploration. This scarcity makes it difficult to identify system bottlenecks, compare deployment strategies, and assess operational reliability before actual deployment. Orbital simulation and hardware-in-the-loop testing are therefore essential for providing repeatable and representative validation of onboard LLM systems.

\textit{c) Software Stack Integration.} An LLM alone cannot support autonomous satellite operations. It must be integrated into a closed operational loop that connects perception, state estimation, reasoning, planning, tool execution, and feedback. This requires a satellite-specific LLM agent framework that manages mission context, provides structured access to onboard tools and application programming interfaces (APIs), monitors execution outcomes, and triggers replanning when intermediate results deviate from the intended mission objectives~\cite{wan2026compass,lu2025axis,wang2024imaginarium}. Integrating with existing flight software must also account for real-time scheduling, fault management, access control, telemetry, and command verification. The key challenge is therefore to establish a reliable interface between the LLM and the satellite software stack, supporting reliable, long-horizon task execution and progressively autonomous mission operations.

Overall, onboard LLM deployment is not simply a matter of transferring LLMs to satellite hardware. It requires a system-level design that addresses resource-constrained LLM serving, the scarcity of representative orbital validation, and reliable integration with the satellite software stack. Addressing these challenges is essential for moving from standalone onboard LLMs toward practical, reliable, and increasingly autonomous SI. The following sections examine resource-efficient LLM adaptation and deployment, followed by the mechanisms required to integrate LLMs into satellite operations.


\subsection{Resource-Efficient On-Orbit LLM Deployment}
\label{subsec:llm_serving}

Deploying LLMs in orbit requires substantially more than compressing a terrestrial model to fit onboard hardware. During inference, an LLM must accommodate model weights, intermediate activations, continuously growing KV caches, and runtime states while dynamically adapting execution to time-varying spacecraft conditions. These demands are further coupled with the distinctive physical and operational conditions of space: radiation-induced faults~\cite{wang2026radshield}, restricted heat dissipation in vacuum~\cite{li2024exploring}, periodically varying energy availability~\cite{liu2024inorbit}, and intermittent satellite-GS connectivity~\cite{ma2023network}. Therefore, resource-efficient on-orbit LLM deployment requires coordinated optimization on three fronts: efficient onboard serving, satellite-GS collaborative serving, and the orbital simulation and validation needed to evaluate deployment policies under these time-varying spacecraft conditions.

\begin{table*}[!t]
\centering
\caption{Summary of representative techniques for satellite-GS collaborative LLM serving}
\label{tab:satgs_llm_serving}
\footnotesize
\setlength{\tabcolsep}{3pt}
\newlength{\satLLMServingTableWidth}
\setlength{\satLLMServingTableWidth}{\dimexpr\textwidth-10\tabcolsep-6\arrayrulewidth\relax}

\resizebox{0.92\textwidth}{!}{%
\begin{minipage}{\textwidth}
\begin{tblr}{
  colspec={Q[c,m,wd=0.15\satLLMServingTableWidth]
           Q[c,m,wd=0.05\satLLMServingTableWidth]
           Q[l,m,wd=0.20\satLLMServingTableWidth]
           Q[l,m,wd=0.22\satLLMServingTableWidth]
           Q[l,m,wd=0.38\satLLMServingTableWidth]},
  colsep=3pt, rowsep=2pt, stretch=1.15,
  row{1}={font=\bfseries, bg=black!8, halign=c, m},
  hline{1,Z}={0.8pt},
  vline{1,Z}={0.8pt},
  vline{2-5}={0.4pt},
  hline{2}={0.6pt},
  hline{7}={0.5pt}, 
  hline{3-6,8-10}={2-Z}{wd=0.3pt, fg=black}, 
}
\textbf{Approach} & \textbf{Ref.} & \textbf{Performance Metrics} & \textbf{Control Variables} & \textbf{Key Contribution} \\
\SetCell[r=5]{c,m} {\textbf{Quality-Aware}\\\textbf{Routing}}
& \cite{chen2023frugalgpt}
& Accuracy; Inference cost
& LLM cascade order; Answer-scoring thresholds
& LLM cascade escalating a query to a larger model only when a scorer rejects the cheaper model's answer \\
& \cite{fan2025satellite}
& Total system cost of task inference; Delay; Energy consumption
& Inference location; Computing and communication resource allocation
& DRL-based choice of inferring each task on the device, the serving or a neighboring satellite, or the ground cloud \\
& \cite{ding2024hybrid}
& Response quality; Share of queries served by the small model
& Routing threshold
& Router predicting the small-large model quality gap, sending a query to the large model only when the gap is significant \\
& \cite{liao2025reward}
& Accuracy; FLOPs
& Process-reward acceptance threshold
& Step-level routing that invokes the target model only for draft reasoning steps with low process reward \\
& \cite{kadavath2022language}
& Calibration; Discrimination of correct answers
& Self-evaluation format; Number of sampled answers
& Self-evaluated probability that an answer is correct, usable as the confidence signal that triggers offloading \\
\SetCell[r=4]{c,m} {\textbf{Communication-Efficient}\\\textbf{Collaboration}}
& \cite{hao2024hybrid}
& Generation quality; Cost of cloud LLM invocation
& Token-level uncertainty threshold
& Edge SLM generation with the cloud LLM consulted only to verify and correct uncertain tokens, approaching LLM quality at near-SLM cost \\
& \cite{dhuliawala2024chain}
& Precision; F1; FactScore
& Verification execution variant (joint or factored)
& Draft-then-verify pipeline revising a provisional response through independently answered verification questions \\
& \cite{park2023generative}
& Believability of agent behavior
& Retrieval scoring (recency, importance, relevance); Reflection trigger
& Memory stream with retrieval and reflection, passing only relevant and summarized memories to the model instead of the full history \\
& \cite{jiang2023llmlingua}
& Task performance; Compression ratio; End-to-end latency
& Budget controller; Iterative token-level compression
& Budget-controlled coarse-to-fine prompt compression shrinking context volume with limited performance loss \\
\end{tblr}

\par\vspace{3pt}
\scriptsize
DRL: deep reinforcement learning; FLOPs: floating-point operations; SLM: small language model.
\end{minipage}
}
\end{table*}

\subsubsection{Efficient Onboard LLM Serving}\label{subsec:resource_effi_llm_serving}

LLMs execute the model repeatedly for every generated token via autoregressive decoding, and the KV cache underlying this process grows continuously with context length, leaving satellite platforms with the rigid onboard resource budgets in Section~\ref{subsec:SI_constraints} ill-equipped to sustain the resulting memory and computing demand. Efficient onboard serving must jointly reduce model and runtime overhead while adapting execution to time-varying spacecraft conditions. Therefore, we examine three complementary directions: hardware-aware model compression, KV-cache and long-context management, and energy-aware serving. Table~\ref{tab:onboard_llm_serving} summarizes the key techniques for efficient onboard LLM serving reviewed below.

\textit{a) Hardware-Aware Model Compression.}
Quantization, pruning, and knowledge distillation are key methods for reducing the resource requirements of onboard LLMs, but they must account for the characteristics of autoregressive inference rather than be applied as generic deep neural network compression techniques. During token-by-token decoding, LLM inference on resource-constrained satellites is often limited by memory bandwidth because model weights must be repeatedly read for each generated token. Weight-only quantization methods, such as GPTQ~\cite{frantar2023optq} and AWQ~\cite{lin2024awq}, directly reduce weight memory traffic while retaining higher-precision activations and largely preserving model capability. Structured pruning must consider the uneven contribution of individual attention heads and feed-forward components to different reasoning patterns, removing those with the least impact on task performance while preserving structures that the target hardware can directly accelerate~\cite{ling2024slimgpt,an2024flap,ashkboos2024slicegpt}. Knowledge distillation transfers selected capabilities from a larger teacher model to a compact onboard model~\cite{gu2024minillm}. However, compression should be evaluated against end-to-end serving performance rather than parameter count alone: irregular sparsity may yield little acceleration without matching hardware support~\cite{frantar2023sparsegpt}, and overly aggressive quantization can degrade both reasoning capability~\cite{okoshi2026qat} and safety alignment~\cite{chen2025safety}, which is particularly risky for mission-critical decisions. The selected configuration must also reserve sufficient memory for KV caches, retrieval indices, sensor-processing workloads, and system software, since compressing the weights alone can still leave the satellite short of memory once decoding begins. A practical strategy is therefore to use mixed precision with structured compression, retaining higher precision for capability-sensitive components while aggressively compressing less critical layers~\cite{zhou2025qpruner}.

\begin{table*}[!t]
\centering
\caption{Comparison of LLM serving paradigms for SI}
\label{tab:llm_serving_comparison}
\small

\resizebox{0.85\textwidth}{!}{%
\begin{tblr}{
  width=\textwidth,
  colspec={X[1.1,c,m] X[1.3,c,m] X[1.5,c,m]}, 
  colsep=4pt, rowsep=2.2pt, stretch=1.15,
  row{1}={font=\bfseries, bg=black!8, halign=c, m},
  column{1}={font=\bfseries}, 
  hline{1,Z}={0.8pt},
  vline{1,Z}={0.8pt},
  vline{2,3}={0.4pt},
  hline{2}={0.6pt},
  hline{3-Z}={wd=0.3pt, fg=black},
}
Dimension & Onboard Serving & Satellite-GS Collaborative Serving \\
Reasoning location & Single satellite & Satellite and GS \\
Model capacity & Compressed onboard LLM & Onboard LLM + GS-hosted LLM \\
Communication dependency & None during execution & SGLs at task/step boundaries \\
Latency sources & Prefill + decoding & Onboard + SGLs + contact waiting \\
Service continuity & Contact-independent & Contact-window limited \\
Dominant bottleneck & Memory (weights, KV cache), energy & SGL bandwidth, contact windows \\
Suited tasks & Routine, time-critical requests & Knowledge-intensive analysis, planning \\
\end{tblr}%
}
\par\vspace{3pt}
\end{table*}

\begin{figure}[t!]
\centering
\includegraphics[width=0.48\textwidth]{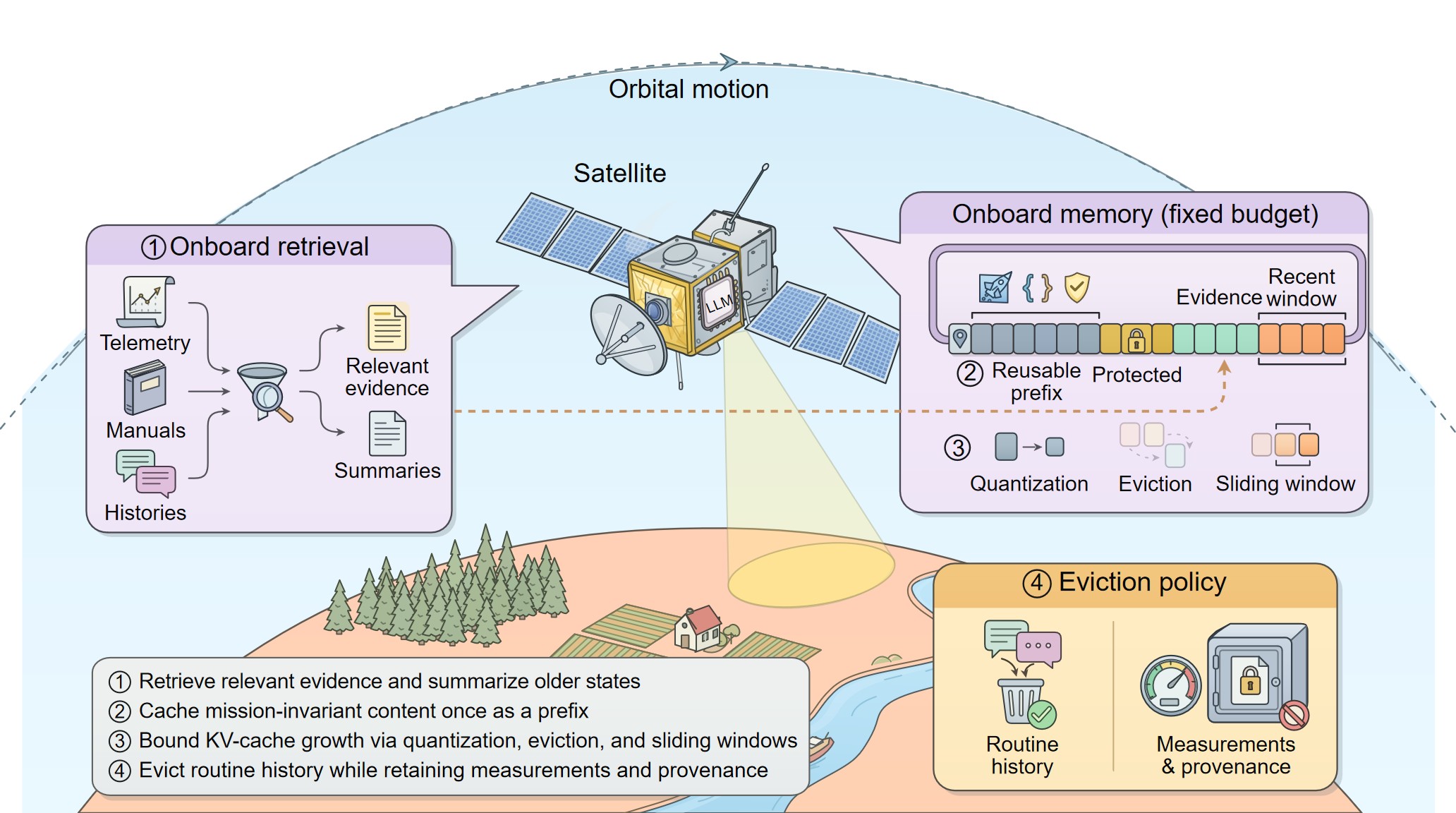}
\centering\caption{KV-cache and long-context management for onboard LLM serving.}
\label{fig_KV_cacge_mag}
\end{figure}

\textit{b) KV-Cache and Long-Context Management.}
The KV cache constitutes a primary bottleneck because it grows without bound, and this growth is far more consequential on a satellite than in a data center, where additional memory can simply be provisioned when the cache runs long. As shown in Fig.~\ref{fig_KV_cacge_mag}, this pressure can be relieved at two points: controlling what enters the cache and compressing what it already holds. On the input side, mission prompts routinely draw on telemetry archives, operational manuals, and interaction histories that a memory-constrained satellite cannot afford to re-insert in full on every query; an onboard retrieval module therefore selects only the task-relevant evidence per query~\cite{lewis2020retrieval} and summarizes older states~\cite{packer2023memgpt}, while content that remains valid across an entire mission, such as spacecraft specifications, tool schemas, and safety constraints, is cached once as a reusable prefix rather than re-encoded each time~\cite{zheng2024sglang,gim2024prompt}. Within the cache, KV-cache quantization reduces the storage cost of individual entries~\cite{liu2024kivi,hooper2024kvquant}, importance-aware token eviction retains only influential historical states~\cite{zhang2023h2o,li2024snapkv}, and sparse or sliding-window attention bounds cache growth by combining recent tokens with a small set of persistent tokens~\cite{xiao2024efficient}, each extending the effective context length on the fixed onboard memory budget. This compression must nonetheless be applied with care under such a tight budget, since discarding the wrong entry can silently remove the evidence a later decision depends on: eviction must therefore target routine conversational history and never the authoritative measurements and decision provenance a mission may need to audit.

\textit{c) Energy-Aware Serving.}
A static serving configuration cannot remain efficient throughout an orbit, since the energy available to sustain it varies with the sunlit-eclipse cycle. This variation affects the two phases of LLM inference differently~\cite{patel2024splitwise,zhong2024distserve,patel2024power}: Prefill processes the input prompt in a single batched pass and can therefore be delayed, batched, or rescheduled with limited impact on service continuity. In contrast, decoding generates tokens sequentially and requires sustained power for its duration. Although it can resume from a preserved KV cache, pausing or migrating an active generation incurs additional state-management overhead. Onboard runtimes should therefore schedule non-urgent prefill, cache construction, and report generation during energy-rich sunlit phases, reserve battery capacity for latency-sensitive decoding requests, and dynamically adjust model size, quantization level, and context and generation lengths as eclipse approaches~\cite{stojkovic2025dynamollm,samsi2023words}. Because the sunlit-eclipse cycle is predictable well in advance, these adjustments need not wait until energy reserves become low. Request- or token-level energy budgets and early stopping can tighten progressively as the eclipse nears, rather than triggering only once a hard limit is reached~\cite{han2025token,muennighoff2025s1}. If available energy nevertheless approaches this limit, the runtime can reduce concurrency, pause or checkpoint active decoding~\cite{miao2024spotserve}, switch to a smaller model~\cite{ahmad2024proteus}, or defer pending requests for execution by the GS when communication opportunities permit~\cite{liu2026joint,yao2025leoedge}.

\subsubsection{Satellite-GS Collaborative LLM Serving}\label{subsubsec:sate_GS_serving}

While the techniques examined in Section~\ref{subsec:resource_effi_llm_serving} expand the serving capacity of individual satellites, they cannot circumvent the fundamental physical limits of satellite hardware. Even a heavily compressed and energy-aware LLM cannot match the knowledge capacity or reasoning depth of a model with hundreds of billions of parameters running on the GS. Satellite-GS collaborative serving bridges this gap by integrating the low latency of satellite-side execution with the superior cognitive capacity of GS-hosted LLMs. However, this paradigm introduces the intermittent and bandwidth-limited SGL as a bottleneck absent in local-only serving. Consequently, orchestrating this collaboration necessitates strategically routing requests between the two computing tiers and sustaining their interaction over restricted links. Table~\ref{tab:satgs_llm_serving} summarizes the key techniques for satellite-GS collaborative LLM serving reviewed below.

\textit{a) Quality-Aware Routing.}
Relying solely on satellite execution or offloading every task to the GS undermines satellite-GS collaboration. The former lacks the reasoning capability of terrestrial LLMs, while the latter degrades onboard latency and exhausts limited contact windows~\cite{chen2023frugalgpt,fan2025satellite}. Effective quality-aware routing collaboration instead routes each task to the tier best suited to it~\cite{ding2024hybrid}. Routine command interpretation, local retrieval, and time-critical responses are typically executed on satellites, whereas knowledge-intensive analysis and complex planning are offloaded to the GS. Because an LLM often reasons through a task as a sequence of intermediate steps rather than a single-shot prediction~\cite{lightman2024let,yao2023tree}, this offloading can occur either as a complete request or at the level of the specific step that requires it, without discarding steps already completed onboard~\cite{liao2025reward}. Routing must further weigh task criticality, determined by the mission context, against model confidence, read from the token-level probabilities the model assigns to its own output~\cite{kadavath2022language,kuhn2023semantic}, since the two are not always aligned. A highly critical task may warrant offloading even at reasonable confidence, since an undetected error costs far more than the transmission delay it avoids, whereas a low-confidence but routine task may still be resolved onboard if its deadline cannot accommodate a contact window~\cite{tao2024serval}. Therefore, task deadlines and predicted contact durations dictate the feasibility of GS execution, while the required data volume determines whether the link can carry it. Orchestrating these factors together selects the least costly tier that still satisfies the required quality and reliability, rather than offloading simply because GS connectivity happens to be available.

\textit{b) Communication-Efficient Collaboration.}
Satellite-GS collaborative LLM serving relies on continuous data exchange over intermittent and bandwidth-constrained SGLs, which can easily saturate these links~\cite{zhang2026communication,xing2024deciphering,furutanpey2025fool}. For multi-step reasoning, querying the GS after every intermediate step wastes limited contact windows and forces the satellite to remain idle while waiting for responses. Instead, the satellite and the GS should interact strictly at task or reasoning-step boundaries. The satellite-side model can generate provisional responses or compact intermediate results, which the GS-side LLM asynchronously verifies or refines once a connection is established~\cite{hao2024hybrid,dhuliawala2024chain,madaan2023self}. This decoupling allows on-satellite reasoning to progress during orbital occlusion rather than stalling until the next contact. Furthermore, this boundary-level interaction allows the satellite to transmit a structured semantic state, comprising the current objective, selected observations, completed tool calls, and decision provenance, rather than the entire interaction history~\cite{park2023generative,jiang2023llmlingua}. Beyond saving bandwidth, this approach avoids LLM context limits, where forwarding full histories increases computational cost and risks truncating critical information as the context window fills~\cite{liu2024lost,hsieh2024ruler}, whereas a structured state remains compact regardless of how long the task runs. Subsequent contacts then synchronize only the changes to this state, ensuring that transmission volume scales with actual reasoning progress rather than accumulated context.

Table~\ref{tab:llm_serving_comparison} summarizes the differences between onboard and satellite-GS collaborative LLM serving.

\begin{figure}[t!]
\centering
\includegraphics[width=0.48\textwidth]{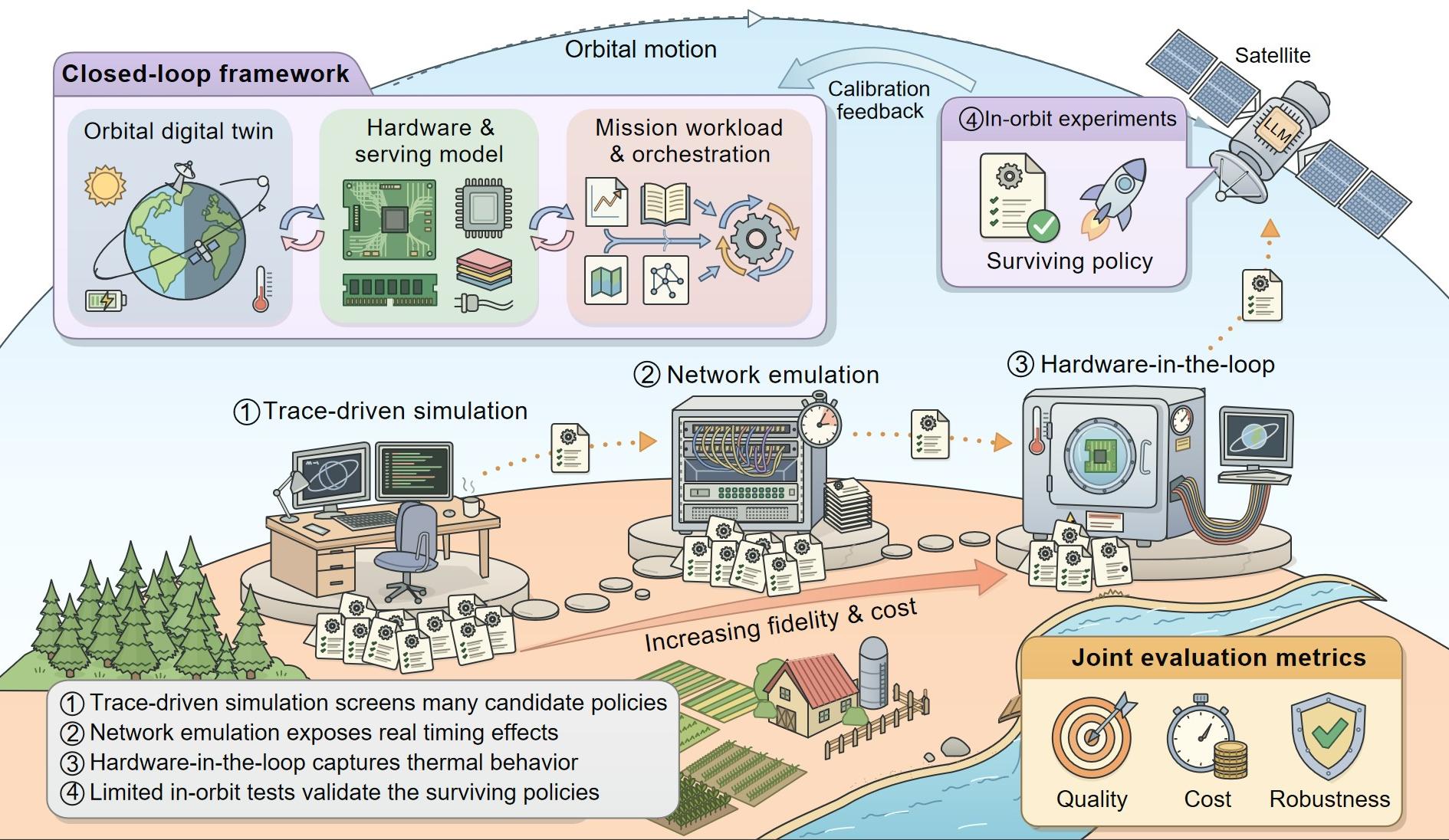}
\centering\caption{Closed-loop simulation and validation framework for orbital LLM deployment.}
\label{fig_orbit_LLM_simu_valid}
\end{figure}

\subsubsection{Orbital LLM Simulation and Validation}

Directly evaluating LLM deployment policies on operational satellites is prohibitively expensive and poses severe operational risks~\cite{denby2020orbital,furano2020towards}, while static terrestrial testbeds cannot reproduce how orbital motion, connectivity, spacecraft state, and LLM serving affect one another in practice. Orbital motion governs when a link exists and how much it can carry; that link in turn governs how an LLM can be served, and the resulting serving delay governs how long the satellite remains exposed to conditions that continue to evolve throughout~\cite{mohan2024multifaceted,bhattacherjee2019network}. Existing satellite network simulators and digital twins capture one half of this loop, modeling trajectories, dynamic topologies, and link performance~\cite{kassing2020exploring,lai2023starrynet}, but reduce inference itself to a fixed computation delay. LLM-serving simulators capture the other half, modeling model execution, batching, and KV-cache allocation~\cite{agrawal2024vidur,cho2024llmservingsim}, but assume the stable conditions of a datacenter and treat the network as continuously available. Because each tool omits exactly what the other represents, evaluating deployment policies with either alone, or combining their results, cannot reproduce this coupling; rigorous orbital LLM evaluation instead requires a single integrated architecture that bridges both simulators.

Fig.~\ref{fig_orbit_LLM_simu_valid} presents such a closed-loop framework, coupling orbital dynamics, hardware and serving behavior, and mission workloads across progressively higher-fidelity evaluation stages: i) an orbital digital twin reproduces trajectories, satellite-GS visibility, inter-satellite topology, link capacity, illumination, battery state, thermal evolution, payload schedules, and representative fault events; ii) a hardware and serving model captures processor and memory characteristics, model configuration, prefill and decoding performance, KV-cache behavior, and power consumption; iii) a mission workload and orchestration layer then generates representative requests, such as telemetry diagnosis, knowledge retrieval, mission planning, and constellation coordination, and executes the same model-selection, routing, admission-control, and recovery policies used by the target system.

The validation process proceeds through four stages of increasing fidelity and cost. First, trace-driven simulation replays orbital traces (e.g., SGP4-propagated trajectories~\cite{vallado2006revisiting}) and real LLM workload traces~\cite{qin2025mooncake} through the coupled models to screen many candidate policies quickly and cheaply. Second, network emulation tests the strongest candidates against a real, timed network stack, exposing timing effects that the simulation abstracted away~\cite{lantz2010network,netravali2015mahimahi}. Third, hardware-in-the-loop testing narrows the candidate set further by running actual onboard hardware under emulated orbital and network conditions to capture effects, such as thermal behavior, that no software model fully reproduces~\cite{kiesbye2019hardware,lentaris2018high}. Finally, limited in-orbit experiments are reserved for the few policies that survive this narrowing, since flight time is far too scarce to spend on anything not already validated on the ground~\cite{mateo2023orbit,mankins2009technology}. At each stage, evaluation should jointly report output quality (task quality, tool-use correctness), cost (latency, memory, energy per request or token, communication volume), and robustness (thermal throttling, service availability, deadline violations, recovery from link or hardware failures), since a policy strong on one dimension can still fail on another~\cite{liang2022holistic}. Measurements from the later, more expensive stages then feed back into the digital twin, calibrating the earlier ones against what the system actually did~\cite{tao2018digital}. This progression supports reproducible comparison of deployment policies and surfaces rare but critical operating conditions before they affect an operational mission.

\subsection{LLM-Enabled Cognitive Capabilities for SI}
\label{subsec::llm_power_si}

The mechanisms examined in Section~\ref{subsec:llm_serving} render onboard LLM execution feasible within satellite resource constraints, but execution alone does not constitute the objective: resource-efficient deployment provides only the computational foundation, and an LLM built on this foundation should serve as a cognitive interface linking satellite observations, mission knowledge, domain tools, and distributed decision processes. Rather than replacing specialized perception, communication, or control algorithms, an LLM can organize these components around mission objectives and support reasoning across heterogeneous information and different operational timescales. Thus, an LLM can integrate mission knowledge and specialized tools while supporting satellite-GS cognitive collaboration and constellation-scale multi-agent coordination. Table~\ref{tab:llm_cognitive_comparison} compares these four LLM-enabled capabilities reviewed below.

\subsubsection{Satellite Knowledge-Augmented Reasoning}

\begin{table*}[!t]
\centering
\caption{Comparison of LLM-enabled cognitive capabilities for SI}
\label{tab:llm_cognitive_comparison}
\footnotesize
\setlength{\tabcolsep}{3pt}
\newlength{\llmCognitiveTableWidth}
\setlength{\llmCognitiveTableWidth}{\dimexpr\textwidth-10\tabcolsep-6\arrayrulewidth\relax}

\resizebox{0.92\textwidth}{!}{%
\begin{minipage}{\textwidth}
\begin{tblr}{
  colspec={Q[c,m,wd=0.14\llmCognitiveTableWidth]
           Q[l,m,wd=0.215\llmCognitiveTableWidth]
           Q[l,m,wd=0.215\llmCognitiveTableWidth]
           Q[l,m,wd=0.215\llmCognitiveTableWidth]
           Q[l,m,wd=0.215\llmCognitiveTableWidth]},
  colsep=3pt, rowsep=2pt, stretch=1.15,
  row{1}={font=\bfseries, bg=black!8, halign=c, m},
  column{1}={font=\bfseries}, 
  hline{1,Z}={0.8pt},
  vline{1,Z}={0.8pt},
  vline{2-5}={0.4pt},
  hline{2}={0.6pt},
  hline{3-Z}={wd=0.3pt, fg=black}, 
}
Dimension & Satellite Knowledge-Augmented Reasoning & Satellite-Aware Tool Use & Satellite-GS Cognitive Collaboration & Constellation-Scale Multi-Agent Systems \\
Scope
& Single LLM
& Single LLM
& Satellite--GS
& Constellation \\
Core challenge
& Time-bounded evidence validity
& Physical-model compliance
& Differential observability under intermittent links
& Partial observability and stale information \\
LLM role
& Query generation; Evidence validation
& Objective interpretation; Action proposal
& Satellite: Provisional planning; GS: Deep analysis
& Objective decomposition; Role assignment \\
Key mechanism
& Spatio-temporal knowledge graph; Hybrid retrieval
& Propose--verify--authorize loop
& Versioned shared mission state
& Structured messaging; Event-triggered belief sharing \\
Information unit
& Metadata-tagged knowledge items
& Typed tool calls; Simulator feedback
& Versioned state writes
& Time-bounded structured messages \\
Non-LLM safeguard
& Metadata filtering; Time decay
& Digital-twin verification; Predefined authorization
& Version checks; Directive/advisory distinction
& Message expiration; Quorum-based decisions \\
Risk mitigated
& Stale evidence treated as current
& Unverified command execution
& Stale state overwrites
& Inconsistent world models \\
\end{tblr}
\end{minipage}
}
\end{table*}

Satellite missions rely on highly dynamic knowledge spanning telemetry archives, spacecraft manuals, orbital data, and mission plans~\cite{teng2025efficient,thangavel2024artificial}. Encoding this continuously evolving information directly into model parameters restricts the LLM to a fixed training-time snapshot, inevitably producing outdated responses as system states change~\cite{kasai2023realtime}. While knowledge-augmented reasoning addresses this by retrieving task-relevant evidence at inference time, standard retrieval-augmented generation paradigms remain inherently inadequate for satellite operations. Spaceborne evidence is strictly governed by orbital physics and rapid temporal degradation~\cite{yates2021age}. For instance, a Keplerian state vector becomes invalid immediately following a trajectory maneuver; thermal mitigation procedures depend strictly on dynamic sunlit-eclipse phases; and telemetry cross-linked from neighboring satellites may lag behind current satellite-side readings, demanding strict freshness evaluations. Consequently, the LLM must evaluate the precise spatio-temporal validity boundaries of retrieved evidence rather than merely its semantic content. Each knowledge item must encapsulate rich metadata, such as precise epoch timestamps, coordinate reference systems, and topological provenance, allowing the model to differentiate verified real-time measurements from outdated estimates and explicitly flag unresolved data conflicts~\cite{teng2025efficient,jensen2013provenance}.

To realize this knowledge augmentation, system implementations must move beyond conventional dense vector retrieval. One prominent approach is the integration of spatio-temporal knowledge graphs to explicitly encode hardware dependencies and telemetry lifespans as dynamic graph edges~\cite{radtke2026rdf}. Instead of relying solely on semantic similarity, the retrieval pipeline employs hybrid search architectures: the LLM acts as a query generator (e.g., producing SQL or Cypher scripts) to extract deterministic real-time states from time-series databases while simultaneously retrieving unstructured manuals from vector stores~\cite{pourreza2023dinsql,lewis2020retrieval}. Finally, deterministic pre-filtering and time-decay functions can automatically prune stale telemetry before it enters the LLM's context window. By coupling semantic search with structured graph traversal and metadata filtering, this architecture transforms retrieval-augmented generation from a generic query mechanism into an auditable foundation for mission-critical reasoning~\cite{patel2024acorn,sun2024think}.

\subsubsection{Satellite-Aware Tool Use}

Autoregressive language generation is insufficient for critical satellite operations, such as orbital propagation, link-budget calculation, and collision assessment, whose outputs must strictly adhere to physical models~\cite{dziri2023faith,mirzadeh2025gsm}. Instead of relying on heuristic reasoning, the LLM must function as a semantic orchestrator that interprets mission objectives, invokes specialized deterministic solvers, and synthesizes numerical results~\cite{gao2023pal,qin2024toolllm}. However, standard function-calling mechanisms are unreliable because they accept model proposals without validating physical constraints. Reliable integration demands a decoupled execution loop comprising three strict stages: the LLM proposes an action sequence; deterministic tools verify its physical feasibility; and local flight software, or ground operators, grants final authorization based on safety classifications determined in advance, rather than the LLM's runtime judgment~\cite{xiang2024guardagent}. To implement this architecture, developers must integrate the LLM orchestration layer with established aerospace middleware, such as NASA's Core Flight System~\cite{mccomas2016core} or Space ROS~\cite{probe2023space}. Each domain solver must be wrapped in strictly typed and machine-readable schemas that enforce input validation, explicit validity boundaries, and predefined side effects~\cite{dong2025xgrammar,geng2023grammar}. To operationalize the verification stage, the orchestrator should route the LLM's proposed command sequences through an isolated digital twin or physics-based simulation environment prior to live execution~\cite{ruan2024identifying,alhassan2026digital}. If a proposed maneuver violates thermal limits or trajectory constraints within this environment, the simulator returns explicit error logs and execution statuses, and the orchestrator feeds this structured feedback back to the LLM for iterative refinement~\cite{xinyun2023teaching,silver2024generalized}. This closed-loop implementation programmatically guarantees that only physically verified, schema-compliant commands can reach the final authorization and execution layer.

\subsubsection{Satellite-GS Cognitive Collaboration}

Satellite-GS collaborative serving, examined in Section~\ref{subsubsec:sate_GS_serving}, determines where model computation is executed; cognitive collaboration instead determines which side holds authority over a given unit of knowledge, reasoning, or decision-making, a division that follows directly from the differential observability of the satellite and GS. A satellite-side LLM possesses direct access to current observations and can therefore respond immediately to local events, whereas the GS retains broader historical knowledge, constellation-wide state, mission objectives, and human expertise that no individual satellite accumulates independently~\cite{yao2022edge,zhang2025satellite}. Consequently, the satellite is positioned to perform time-critical situation assessment and provisional planning, whereas the GS is better positioned to conduct deeper analysis, reconcile state across satellites, and update longer-term strategy~\cite{li2026enabling,lin2023swiftsage}. This division supports autonomy without requiring either the satellite or GS to maintain a complete and continuously synchronized view of the mission.

\begin{figure}[t!]
\centering
\includegraphics[width=0.48\textwidth]{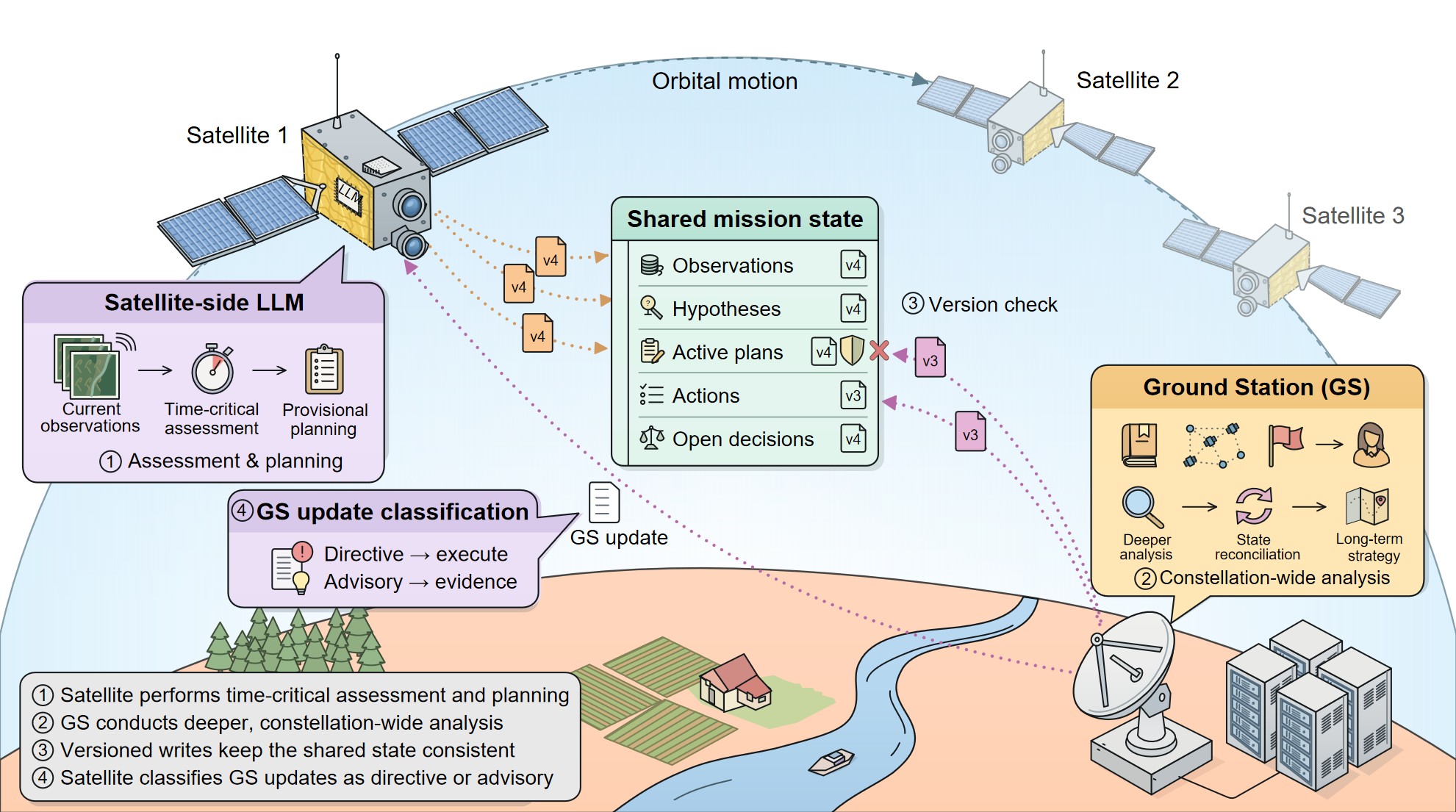}
\centering\caption{Satellite-GS cognitive collaboration via a versioned shared mission state under intermittent connectivity.}
\label{fig_satellite_gs_colla_cognitive}
\end{figure}

Preserving this division without the satellite or GS operating on stale state demands more than correct request routing. It necessitates a shared mission state, independently readable and writable by both the satellite and the GS, as illustrated in Fig.~\ref{fig_satellite_gs_colla_cognitive}. This store must record verified observations, current hypotheses, active plans, completed actions, and unresolved decisions~\cite{xu2026mem}. Because the satellite and the GS update this state asynchronously, each write should carry a monotonically increasing version identifier and be annotated with the operational context assumed at write time. This enables subsequent operations to explicitly determine whether the preconditions underlying an earlier write remain valid~\cite{zhang2026patchboard}. This scheme effectively implements optimistic concurrency control. Prior to overwriting a plan, the GS must verify its version, deferring to the satellite's locally committed state rather than applying a stale write. Conversely, the satellite must classify each incoming GS update as either a mandatory directive or an advisory inference result. The former must be executed immediately, while the latter is incorporated merely as evidence for local reasoning rather than as a binding instruction~\cite{wallace2024instruction}. Explicit authority, conflict-resolution, and expiration semantics associated with each write render this arbitration tractable. Satellite-GS cognitive collaboration can accordingly be formalized as asynchronous shared-state reasoning over a versioned, eventually consistent store, wherein the satellite and the GS contribute complementary capabilities while maintaining decision consistency under delayed and intermittent connectivity~\cite{loven2026meld,margalit2026governed}.

\subsubsection{Constellation-Scale Multi-Agent Systems}


As a constellation grows to hundreds or thousands of satellites, routing every coordination decision through a single GS becomes a bottleneck, and many tasks, such as cooperative target tracking or distributed sensing, require satellites to coordinate with one another rather than through the GS at all. This motivates designing each satellite as an autonomous agent, maintaining its own local observations, tools, resources, and mission responsibilities. This differs fundamentally from distributed model serving, where multiple satellites jointly execute a single model: constellation agents instead reason and act independently while coordinating toward shared objectives~\cite{cao2025exploring,lowe2017multi}. Such systems can support distributed sensing, cooperative target tracking, communication-resource allocation, anomaly response, and constellation reconfiguration~\cite{liu2024multi,guo2025resilience}. LLMs provide a flexible interface for this coordination, decomposing high-level objectives, assigning semantic roles to individual satellites, exchanging summarized observations, and adapting task plans as satellites enter or leave the available communication topology~\cite{guo2024large,li2023camel}.

However, coordinating agents is harder at constellation scale than between a single satellite and the GS: no agent observes the full state of the system, resources and links vary across satellites, and information reaching one agent may already be outdated by the time it acts on it~\cite{foerster2016learning,kaul2012real}. Under these conditions, unrestricted natural-language interaction between agents can produce inconsistent world models, or a consensus that no agent can actually verify~\cite{cemri2026multi}. Multi-agent coordination should therefore constrain what agents can interact with each other rather than leaving communication open-ended: each message can follow a fixed pattern identifying its source, timestamp, confidence, the specific action it requests, and the interval during which it remains valid, so that a receiving agent can discard it once that interval has passed rather than acting on outdated information~\cite{hong2024metagpt,chen2024agentverse}. Decisions whose consequences extend beyond a single agent, such as reallocating shared communication resources or reconfiguring the constellation's coverage pattern, should be resolved through a predefined procedure rather than open negotiation, in which a designated coordinator, or a quorum of agents meeting a minimum threshold, must agree before the action is taken, and any agent excluded from that quorum defers to its outcome~\cite{ongaro2014search}. To keep this exchange within what limited ISLs and SGLs can support, agents can share only a compact summary of their current beliefs rather than complete observation histories, transmitting only when that belief changes materially rather than on a fixed schedule~\cite{tabuada2007event,singh2018learning}. Therefore, designing a constellation-scale multi-agent system must account not simply for the number of agents involved, but for keeping their collective behavior communication-efficient, interpretable, and consistent with constellation-level safety constraints as that number grows.

\section{Open Challenges and Future Directions}
\label{sec:challenges}

SI remains a largely unexplored direction. This section discusses several pressing concerns and potential solutions for preserving system security, data compliance, and data efficiency, toward practical and efficient SI.

\subsection{Secure and Reliable SI}

While the SI frameworks established in Sections~\ref{sec:distributed_training}-\ref{sec:LLM_adaptation}, spanning distributed training, collaborative inference, and satellite-GS LLM serving, avoid downlinking raw telemetry data, these workflows remain vulnerable to adversarial exploitation. Constellations are increasingly operated through multi-tenant GS-as-a-service infrastructures and multi-vendor inter-satellite relays, expanding the attack surface beyond traditional point-to-point links~\cite{koisser2024orbital}. Consequently, intermediate representations, ranging from model updates and smashed data to LLM KV caches and semantic prompts, can still be intercepted and exploited to reconstruct sensitive orbital imagery or deduce critical mission intents.

Securing SI workflows requires defending against two distinct threat models. First, SI must mitigate inference attacks from honest-but-curious parties, such as third-party GSs or relay satellites, which follow the protocol but attempt to infer private information from the data they receive. One potential defense is to add noise based on differential privacy (DP)~\cite{sun2024efficient}. However, in satellite networks, injected DP noise is inherently compounded by channel bit-errors, aggressive quantization, and feature sparsification~\cite{lang2023joint}. Across all SI frameworks, privacy is deeply coupled with resource allocation. In particular, offloading more computation to the GS, whether via shallower cut layers in split architectures or delegating LLM reasoning steps, exposes lower-level features and richer semantic contexts that are significantly easier to invert~\cite{lin2025pushing}. Therefore, privacy-preserving mechanisms must be jointly optimized with network conditions, determining noise thresholds that bound information leakage without destroying the fragile signal-to-noise ratio of compressed orbital transmissions.

Second, SI can enforce Byzantine resilience against malicious satellites attempting to poison global models, alter training results, or inject malicious prompts into LLM workflows. Conventional robust coordination schemes assume synchronous participation, which fails in LEO topologies where asynchronous execution is necessary to handle frequent stragglers~\cite{zhou2024decomposition}. Under these conditions, adversarial training updates or manipulated inference features easily masquerade as delayed, honest contributions. Furthermore, techniques like PEFT and retrieval-augmented generation concentrate the attack surface into a minimal set of adapter weights or retrieved tokens, altering the statistical footprint of malicious perturbations~\cite{sun2025peftguard}. Future SI architectures therefore demand lightweight, topology-aware robust verification protocols capable of distinguishing legitimate orbital staleness from deliberate adversarial drift without exceeding onboard compute limits. Beyond security, SI must also be reliable when driving autonomous operations, since an erroneous decision, such as a flawed collision-avoidance maneuver, may cause catastrophic outcomes like satellite collisions.

\subsection{Data-Compliance-Aware SI}

LEO constellations inherently traverse multiple jurisdictions within a single orbital period, subjecting collected telemetry data to diverse and often conflicting regional data regulations~\cite{tricco2025protection}. Unlike terrestrial data centers where data remains geographically localized, distributed SI frameworks aggregate orbital knowledge into globally shared models, ranging from conventional neural networks to LLMs. When a regulatory body enforces data withdrawal, such as invoking the right to erasure or regional remote sensing restrictions, the algorithmic influence of that local data must be deterministically excised from every model artifact it has reached~\cite{xu2024machine}.

Resolving this compliance requirement necessitates both proactive architectural isolation and reactive algorithmic unlearning. Proactively, SI deployments must enforce structural isolation during training and adaptation. By confining region-specific updates to dedicated parameter-efficient modules (e.g., LoRA for LLMs) atop a frozen and shared backbone, forgetting a jurisdiction simply requires dropping its associated module, trading cross-regional knowledge sharing for strict regulatory compliance~\cite{bourtoule2021machine}. Reactively, when targeted forgetting is required within a shared parameter space, approximate machine unlearning (such as targeted knowledge erasure in LLMs) must be employed to estimate and subtract the data's influence without triggering a computation-prohibitive full-model retraining~\cite{yao2024large}. Executing this unlearning over LEO topologies introduces severe systemic complexities. First, the targeted data's influence has already propagated through multiple aggregation rounds, compounding the difficulty of cleanly isolating its gradients~\cite{romandini2024federated}. Second, the unlearned state must be consistently broadcast to every satellite across the constellation despite intermittent contact windows and frequent topological partitions. Crucially, regulators demand cryptographic or algorithmic proof of compliance; establishing such verifiable unlearning on resource-constrained satellite hardware, without interrupting time-critical mission operations, remains an open challenge. Beyond regulatory compliance, cross-constellation collaboration further calls for standardized model formats, software interfaces, and communication protocols, since satellites from different operators integrate heterogeneous hardware, software, and communication systems from multiple vendors.

\subsection{Data-Efficient SI}

SI confronts an inversion of the terrestrial AI paradigm. Terrestrial models are trained with ample datasets and computing resources, whereas SI operates in an environment where raw observations are abundant, but onboard compute, energy, and ground-truth labels are scarce. The majority of orbital data is both unlabeled and highly redundant, dominated by cloud-obscured scenes, open oceans, and unchanged landscapes. Training on such data exhausts limited onboard resources and may even degrade model performance. Existing onboard filtering mechanisms~\cite{aybar2024onboard} are designed to save downlink bandwidth and remain decoupled from the training pipeline. Therefore, automated data valuation is needed to drive on-satellite learning.

There are two key research directions for improving the data efficiency of SI. First, SI can shift from training on all data to targeted data selection~\cite{sorscher2022beyond}. Lightweight coreset selection and importance sampling can identify high-value samples, though calibrating these scores with computationally constrained or stale models remains challenging. At the constellation scale, data valuation must guide orchestration; the GS should dynamically allocate aggregation bandwidth and training rounds to satellites possessing the most globally informative updates~\cite{amiri2021convergence}. However, aggressively filtering for informative samples risks severely skewing local datasets toward rare edge cases, exacerbating the non-IID distribution across the constellation. Future mechanisms therefore need to balance data efficiency against this distribution skew. Second, SI can overcome the scarcity of orbital labels~\cite{wang2022self}. Since labels are only available on the ground, active learning can prioritize downlinking the highly uncertain samples whose annotation would most improve the model, while the remaining unlabeled data can be exploited in orbit via self-supervised and semi-supervised learning~\cite{cong2022satmae}. Notably, LEO constellations provide a supervisory signal rarely available terrestrially: multi-view spatio-temporal consistency. As multiple satellites revisit the same region from different orbits and viewing angles, the consistency among their observations can serve as a training signal~\cite{manas2021seasonal}. The key challenge is to distinguish genuine land-cover changes from observational discrepancies caused by clouds, illumination, or seasons. Once this is addressed, orbital redundancy can be transformed from a computational burden into a source of supervision.

\section{Conclusion}
\label{sec:conclusion}

In recent years, the proliferation of LEO constellations and onboard sensing has driven an unprecedented demand for intelligent spaceborne processing. In this article, we advocated the paradigm shift from terrestrial-centric computation to SI. We first identified the key deployment challenges that arise from rigid SWaP constraints, highly dynamic topologies, and intermittent energy provision. To address these challenges, we established constraint-aware design principles and then elaborated on several frameworks to enable efficient distributed training, collaborative inference, and on-orbit LLM adaptation under the resource-constrained space environment. Finally, we outlined critical open challenges spanning security, data compliance, and data efficiency. We hope this article can inspire more researchers from both the AI and satellite communication communities to explore the deployment of SI and further advance this emerging space-native AI field.

\ifCLASSOPTIONcaptionsoff
  \newpage
\fi

\bibliographystyle{IEEEtran}
\bibliography{reference}

\end{document}